\documentclass[aps, prd, 10pt,twocolumn,superscriptaddress,nofootinbib]{revtex4-2}
\usepackage{mathrsfs, amssymb, amsmath, mathtools}  
\usepackage{cancel, comment}
\usepackage{footmisc}
\usepackage{latexsym}
\usepackage{natbib}
\usepackage{url}
\usepackage{dcolumn}
\usepackage{multirow}
\usepackage{color}
\usepackage{soul}
\usepackage[normalem]{ulem}
\usepackage{amsfonts,amssymb,amsmath, txfonts}
\usepackage{graphicx,epsfig}
\usepackage{psfrag}
\usepackage{hyperref}
\usepackage{mathtools}
\usepackage{enumitem}
\usepackage{float}
\usepackage[dvipsnames]{xcolor}
\usepackage{xcolor}
\hypersetup{linktoc=all,
    colorlinks=true, linkcolor={brightpink},
    citecolor={blue}, urlcolor={blue}
}
\definecolor{rosy}{RGB}{230,235,252}
\definecolor{myframetitle}{RGB}{90,89,170}
\definecolor{myblocktitle}{RGB}{140,185,249}
\definecolor{mytitle}{RGB}{10,80,26}

\definecolor{darkgreen}{RGB}{27,130,45}
\definecolor{darkblue}{rgb}{0,0,0.3}
\definecolor{darkred}{rgb}{0.7,0,0}

\definecolor{light gray}{RGB}{220,220,220}
\definecolor{dark purple}{RGB}{108,0,217}
\definecolor{pink}{RGB}{190,20,100}
\definecolor{orang}{RGB}{193,63,0}
\definecolor{green}{RGB}{11,98,17}
\definecolor{darkpink}{RGB}{153,0,76}
\definecolor{bluegreen}{RGB}{0,102,102}
\definecolor{greenlagan}{RGB}{0,102,0}
\definecolor{redgreen}{RGB}{102,102,0}
\definecolor{Redgreen}{RGB}{153,76,0}
\definecolor{vividviolet}{rgb}{0.62, 0.0, 1.0}
\definecolor{amaranth}{rgb}{0.9, 0.17, 0.31}
\definecolor{palatinateblue}{rgb}{0.15, 0.23, 0.89}
\definecolor{brightpink}{rgb}{1.0, 0.0, 0.5}
\definecolor{cornflowerblue}{rgb}{0.39, 0.58, 0.93}
\definecolor{deepcarminepink}{rgb}{0.94, 0.19, 0.22}
\definecolor{radicalred}{rgb}{1.0, 0.21, 0.37}

\def\H0{{\text{H}\hspace*{-2.05mm}\text{H} 0\hspace*{-1.35mm}0\ }}

\def\be{\begin{equation}}
\def\ee{\end{equation}}
\def\beq{\begin{equation}}
\def\eeq{\end{equation}}
\def\bea{\begin{eqnarray}}
\def\eea{\end{eqnarray}}
\newcommand{\dd}{\textrm{d}}
\newcommand{\nn}{\nonumber \\}

\begin{document}

\title{Do high redshift QSOs and GRBs corroborate JWST?}

\author{Eoin \'O Colg\'ain}
\affiliation{Atlantic Technological University, Ash Lane, Sligo F91 YW50, Ireland}
\author{M. M. Sheikh-Jabbari}
\affiliation{ School of Physics, Institute for Research in Fundamental
Sciences (IPM), P.O.Box 19395-5531, Tehran, Iran}

\author{Lu Yin}
\affiliation{Department of Physics, Shanghai University, Shanghai, 200444,  China}
\affiliation{Asia Pacific Center for Theoretical Physics, POSTECH, Pohang 37673, Korea}

\begin{abstract}
The James Webb Space Telescope (JWST) is reporting massive high redshift galaxies that appear challenging from the $\Lambda$CDM perspective. Interpreted as a cosmological problem, this necessitates the Planck collaboration underestimating either matter density $\Omega_m$ or physical matter density $\Omega_m h^2$ at higher redshifts. Through standard frequentist profile likelihoods, we identify corroborating quasar (QSO) and gamma-ray burst (GRB) data sets where $\Omega_m$ increases with effective redshift $z_{\textrm{eff}}$ and remains anomalously large at higher redshifts. We relax the traditional priors by allowing for $\Omega_m > 1$, consistent with negative dark energy density, to locate profile likelihood peaks where possible. While the variation of $\Omega_m$ with $z_{\textrm{eff}}$ is at odds with the $\Lambda$CDM model, demarcating frequentist confidence intervals through differences in $\chi^2$ in profile likelihoods, the prevailing technique in the literature, points to $3.9 \sigma$ and $7.9 \sigma$ tensions between GRBs and QSOs, respectively, and Planck-$\Lambda$CDM. We explain the approximations inherent in the existing profile likelihood literature, and highlight fresh methodology that generalises the prescription. We show that alternative methods, including Bayesian approaches, lead to similar tensions. Finally, in the large sample limit, we show that Feldman-Cousins prescription for frequentist confidence intervals in the presence of a boundary (prior) leads to confidence intervals that are bounded above by Wilks' theorem. 

\end{abstract}

\maketitle

\section{Introduction}
Preliminary James Webb Space Telescope (JWST) data has revealed a bevy of high-redshift galaxies with intriguingly large stellar masses \cite{10.1093/mnras/stac3347, Labbe:2022ahb, Castellano:2022ikm, Finkelstein:2023, 2023MNRAS.519.1201A, 2023MNRAS.518.6011D, 2022ApJ...940L..14N, 2023ApJ...942L...9Y}. This unforeseen finding has seeded investigations into observational systematics \cite{Lovell:2022bhx, Fujimoto:2022aco, Prada:2023dix, Chen:2023ugq, Forconi:2023izg, Steinhardt:2023oow, Vikaeus:2023cyi} and speculations on new astrophysics. The latter musings range from more mundane changes to the star formation rate \cite{Robertson:2022gdk, Parashari:2023cui, Sabti:2023xwo, Pallottini:2023yqg, Tkachev:2023acf, Wang:2023xmm, Wang:2023gla} to exotic physics that could explain the origin of the galaxies, such as accelerated growth through massive seed black holes \cite{Inayoshi:2024xwv, Jeon:2024iml} and primordial black holes \cite{Liu:2022bvr, Huang:2023chx, Gouttenoire:2023nzr}; invoking primordial magnetic fields \cite{Ralegankar:2024ekl}; primordial density fluctuations \cite{Hirano:2023auh}; cosmic string loops \cite{Jiao:2023wcn, Wang:2023len, Jiao:2024rcr};  changes to the dark matter (DM)  paradigm \cite{Maio:2022lzg, Gong:2022qjx, Hutsi:2022fzw, Haslbauer:2022vnq, Domenech:2023afs, Lin:2023ewc, Bird:2023pkr, Davari:2023tam}. While DM represents a cornerstone in the Lambda-Cold Dark Matter ($\Lambda$CDM) cosmological model, one could also consider relaxing other $\Lambda$CDM assumptions, notably dark energy described by the cosmological constant $\Lambda$ \cite{Santini:2022bib, Menci:2022wia, Wang:2022jvx, Wang:2023ros, Adil:2023ara, Menci:2024rbq}, in order to alleviate any tension. More generally, one may radically rethink cosmology to allow more time for galaxies to form \cite{Melia:2023dsy, Binici:2024smk, lopezcorredoira2024age}. In the big picture, it is imperative to tease apart the astrophysical and cosmological implications of JWST observations, otherwise JWST can \textit{never} challenge $\Lambda$CDM. Tellingly, a host of studies inspired by JWST reaffirm that $\Lambda$CDM is fine \cite{Keller:2022mnb, Desprez, Sun:2023ocn, McCaffrey_2023}.

Separately, $\Lambda$CDM is troubled by persistent, independent anomalies, most notably $H_0$ and $S_8$ tensions (reviewed in \cite{DiValentino:2021izs, Perivolaropoulos:2021jda, Abdalla:2022yfr}). Admittedly, $\Lambda$CDM is a cornerstone of modern cosmology, and JWST anomalies \textit{on their own} may not be conclusive enough to bring an end to the normal ($\Lambda$CDM) science cycle. For this reason, it is instructive to establish that independent anomalies are indeed consistently pointing to qualitatively similar deviations from Planck-$\Lambda$CDM \cite{Planck:2018vyg} behaviour. To that end, it is noteworthy that popular resolutions to $H_0$ tension, which introduce pre-recombination physics to adjust the BAO scale \cite{Knox:2019rjx, Poulin:2018cxd, Agrawal:2019lmo, Lin:2019qug, Niedermann:2019olb, Ye:2020btb, Poulin:2023lkg}, typically lead to larger values of physical matter density $\Omega_m h^2$ relative to Planck \cite{Planck:2018vyg}, thereby alleviating tension with JWST \cite{Boylan-Kolchin:2022kae, McGaugh:2023nkc,  Liu:2024yan, Forconi:2023izg, Forconi:2023hsj} (however see \cite{Jedamzik:2020zmd}). See also \cite{Forconi:2023hsj, vanPutten:2023ths} for implications of JWST data for other models claiming to alleviate $H_0$ tension.  

In this paper, we showcase a synergy between an existing $\Lambda$CDM anomaly and JWST observations. Concretely, it has been proposed that quasars (QSOs) are standardisable through a non-linear UV, X-ray flux relation popularised by Risaliti \& Lusso \cite{Risaliti:2015zla}. This begets an anomaly, whereby the resulting QSO data set \cite{Risaliti:2018reu, Lusso:2020pdb} favours smaller luminosity distances than Planck-$\Lambda$CDM \cite{Planck:2018vyg}, especially at higher redshifts, $z \gtrsim 1.5$ \cite{Risaliti:2018reu, Lusso:2020pdb}. Translated into the flat $\Lambda$CDM cosmology, this implies a larger value of matter density $\Omega_m$ relative to Planck \cite{Yang:2019vgk}. As can be seen from Fig.~1 of \cite{Forconi:2023hsj}, not only does JWST data prefer larger values of physical matter density $\Omega_m h^2$, as claimed in \cite{Boylan-Kolchin:2022kae, McGaugh:2023nkc,  Liu:2024yan, Forconi:2023izg, Forconi:2023hsj},\footnote{Note that $h:=H_0/100$ is simply a multiplicative factor of the Hubble constant $H_0$.} JWST also prefers larger values of matter density today $\Omega_m$. The point of this paper is that the larger values of $\Omega_m$ and $\Omega_m h^2$ preferred by the Risaliti-Lusso QSOs are also preferred by JWST, at least at the level of qualitative comparison. Since the standardisability of the Risaliti-Lusso QSOs has been openly challenged \cite{Khadka:2020vlh, Khadka:2020tlm, Khadka:2021xcc, Khadka:2022aeg, Singal:2022nto, Petrosian:2022tlp, Zajacek:2023qjm}, we identify a  gamma-ray burst (GRB) data set, which apparently passes standardisable candle tests failed by QSOs \cite{Cao:2024vmo}, that exhibits the same trend. Along the way, we explain the limitations of prevailing methods for determining frequentist confidence intervals from profile likelihoods and comment on fresh methodology \cite{Gomez-Valent:2022hkb} (see also \cite{Herold:2021ksg, Colgain:2023bge}) that generalises existing methods.    

\section{JWST Anomaly}
\label{sec:JWST}
We begin by reviewing JWST observations of massive high redshift galaxies that appear anomalous from the $\Lambda$CDM perspective \cite{10.1093/mnras/stac3347, Labbe:2022ahb, Castellano:2022ikm, Finkelstein:2023, 2023MNRAS.519.1201A, 2023MNRAS.518.6011D, 2022ApJ...940L..14N, 2023ApJ...942L...9Y}. The point of this section is to serve as an appetiser for later QSO and GRB anomalies that appear to be pointing to consistent deviations from the Planck-$\Lambda$CDM cosmology \cite{Planck:2018vyg}. Concretely, we will demonstrate that larger $\Omega_m$ and $\Omega_m h^2$ values relative to the Planck-$\Lambda$CDM model help alleviate tensions with JWST. Our findings in this section are expected since a host of papers claim that tension between Planck-$\Lambda$CDM and JWST can be reduced by increasing either matter density today $\Omega_m$ \cite{Forconi:2023hsj} or physical matter density $\Omega_m h^2$ \cite{Boylan-Kolchin:2022kae, McGaugh:2023nkc,  Liu:2024yan, Forconi:2023hsj}. 

We quickly review how one arrives at these conclusions. Basically, one needs to compute the comoving number or mass density of haloes. To begin, the dark matter (DM) halo mass function $ \dd n(M, z)/ \dd M$ quantifies the number of DM haloes of mass $M$ per unit mass per unit co-moving volume at redshift $z$: 
\begin{equation}
    \frac{\dd n}{\dd M} = - \frac{\rho_{m}^{(0)}}{M} \frac{\dd \ln \sigma}{\dd M} f(\sigma), 
\end{equation}
where $\rho_{m}^{(0)}$ is the matter energy density today, $\sigma(R, z)$ is the variance of matter density fluctuations in a sphere of comoving radius $R$ at redshift $z$, and $f(\sigma)$ is defined through the Sheth-Tormen prescription for DM haloes \cite{Sheth:1999mn}: 
\begin{equation}
f(\sigma) = A \sqrt{ \frac{2 a}{\pi} } \left[ 1 + \left( \frac{\sigma^2}{a \delta_c^2}\right)^p\right]
\frac{\delta_c}{\sigma} \exp \left( - \frac{\delta_c^2 a}{2 \sigma^2} \right).  
\end{equation}
Here we follow \cite{Adil:2023ara} and adopt the numbers $A = 0.322, a = 0.707, p = 0.3$ and $\delta_c = 1.686$. As is the norm, the mass $M$ within a sphere of comoving radius $R$ is simply the volume of the sphere times matter density today, $M = \frac{4}{3} \pi R^3 \rho_m^{(0)}$, and the variance $\sigma (R,z)$ takes the form, 
\begin{equation}
    \sigma^2(R, z) = \frac{1}{2 \pi^2} \int_0^{\infty} \dd k \, k^2 \, P(k, z) \, W^2( k R), 
\end{equation}
where $P(k, z)$ is the power spectrum of matter density fluctuations and $W(x) = 3 (\sin x - x \cos x)/x^3$. 

\begin{figure}[htb]
   \centering
\includegraphics[width=80mm]{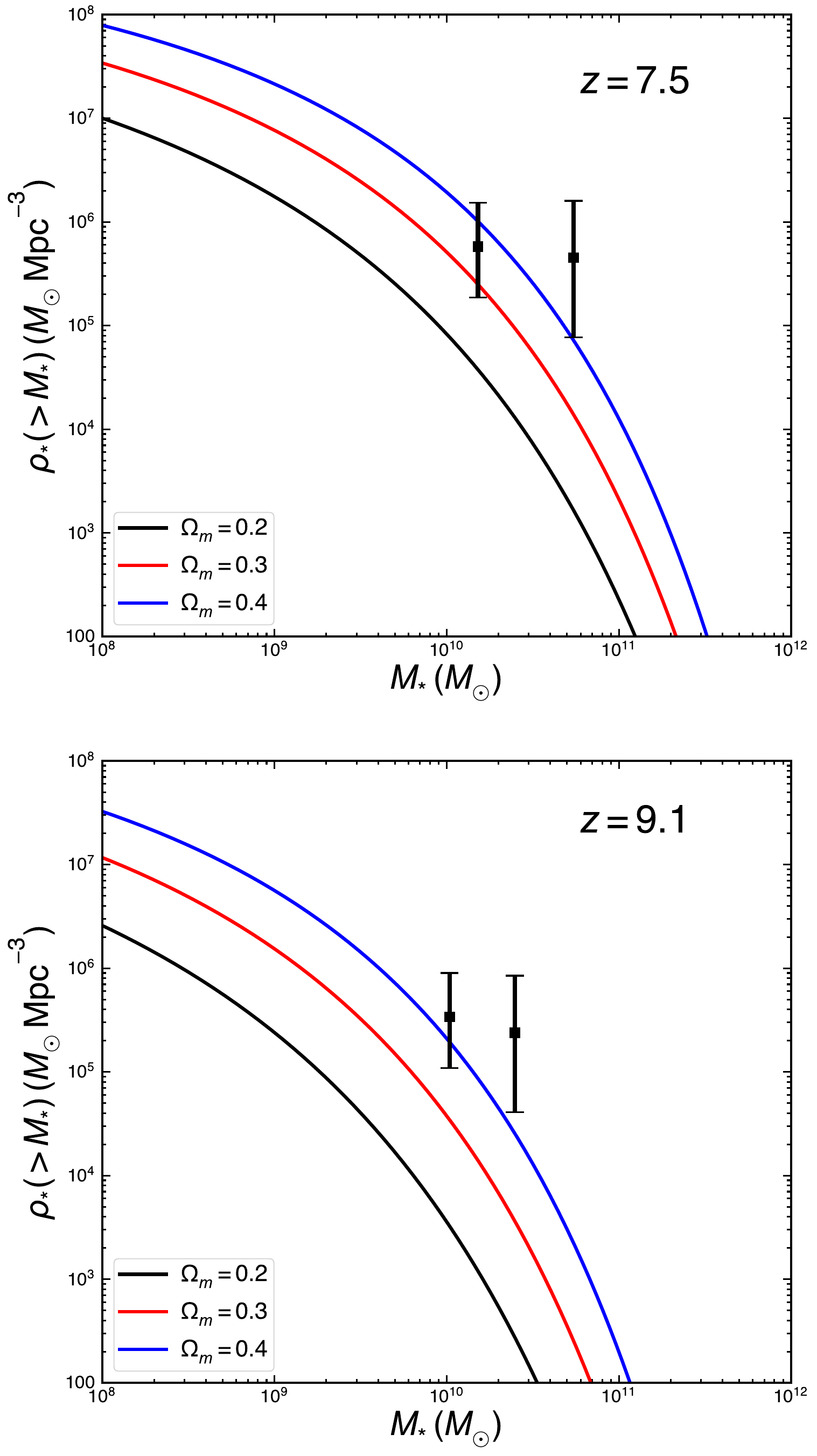} 
\caption{The comoving stellar mass density contained within galaxies more massive than $M_{\star}$ at $z \approx 7.5$ (above) and $z \approx 9.1$ (below). With fixed cosmic baryon fraction, $f_b$, and fixed efficiency of converting gas into stars, $\epsilon$, increasing $\Omega_m$ alleviates tensions with the Labb\'e et al. \cite{Labbe:2022ahb} data points.}
\label{fig:JWST_OM}
\end{figure}

\begin{figure}[htb]
   \centering
\includegraphics[width=80mm]{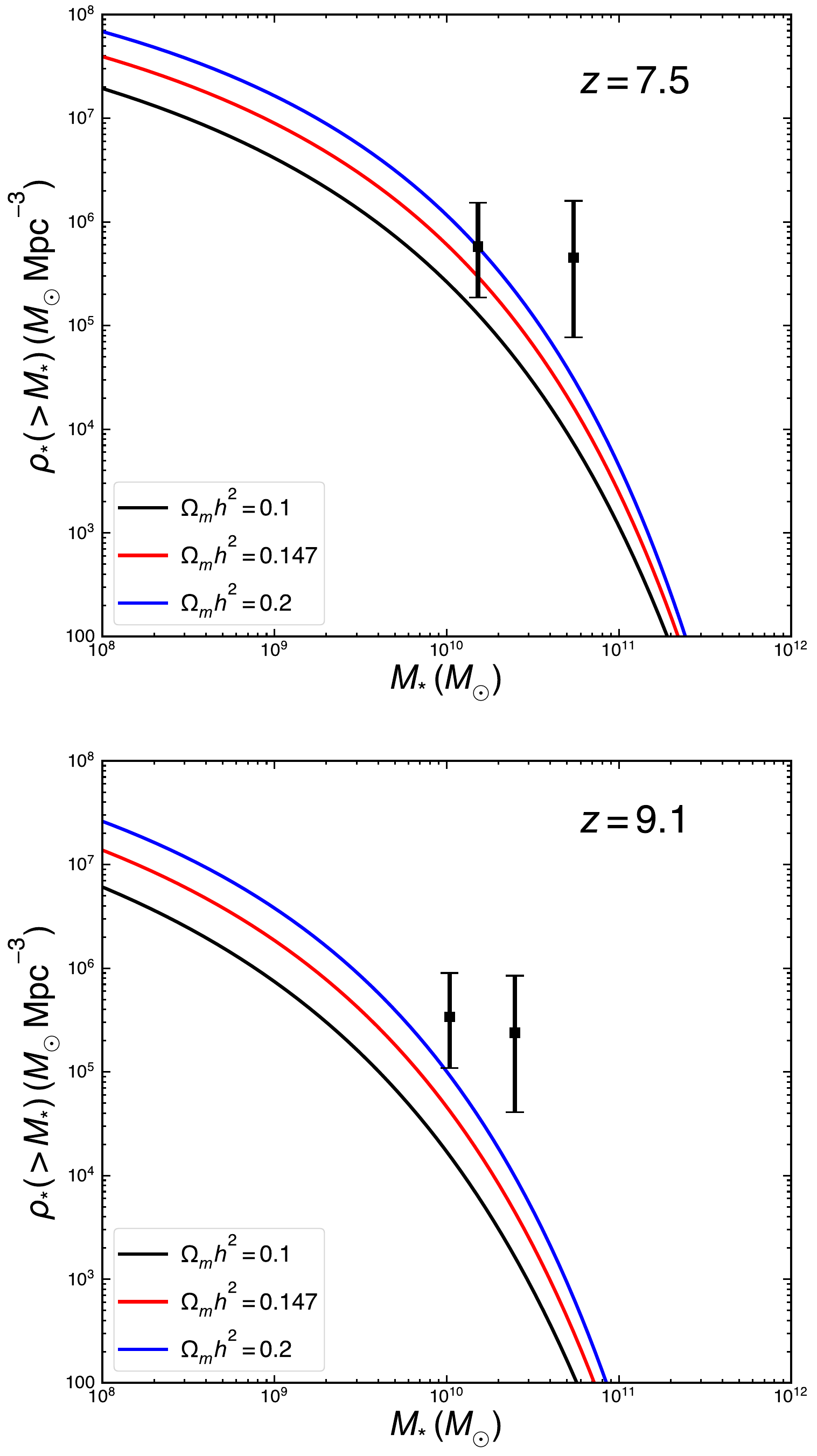} 
\caption{The comoving stellar mass density contained within galaxies more massive than $M_{\star}$ at $z \approx 7.5$ (above) and $z \approx 9.1$ (below). With fixed cosmic baryon fraction, $f_b$, and fixed efficiency of converting gas into stars, $\epsilon$, increasing $\Omega_m h^2$ alleviates tensions with the Labb\'e et al. \cite{Labbe:2022ahb} data points.}
\label{fig:JWST_OMH2}
\end{figure}

At this point, one has all the ingredients to calculate the cumulative comoving mass density of stars contained in galaxies more massive than $M_{\star}$ \cite{Boylan-Kolchin:2022kae}: 
\bea
    \rho (>M_{\star}, z) = \epsilon f_b  \int_{M_{\textrm{halo}}}^{\infty} \dd M M \frac{\dd n (M, z)}{\dd M}, 
\eea
where $f_b := \Omega_b/\Omega_m$ is the cosmic baryon fraction and $\epsilon$ is the efficiency of converting gas into stars. We assume $f_b$ and $\epsilon$ to be cosmology-independent constants and fix them to $f_b = 0.156$ and $\epsilon = 0.32$. {Finally, we fix $A_s$ and $n_s$ to their Planck values.} One is now in a position to produce Fig. \ref{fig:JWST_OM} and Fig. \ref{fig:JWST_OMH2}. What they demonstrate simply backs up observations made elsewhere in the literature, namely that increasing both matter density $\Omega_m$ and physical matter density $\Omega_m h^2$ relative to the Planck values alleviates tension with observations \cite{Boylan-Kolchin:2022kae, McGaugh:2023nkc,  Liu:2024yan, Forconi:2023hsj}. {Our analysis here follows \cite{Boylan-Kolchin:2022kae} and only compares the cumulative comoving mass density at specific redshifts $z=7.5$ and $z=9.1$ (to compare with \cite{Labbe:2022ahb}). However, in \cite{Forconi:2023hsj} the authors extend this analysis to a greater number of JWST observations and redshift ranges. From Fig. 1 of \cite{Forconi:2023hsj} it is clear that increasing $\Omega_m$ and $H_0$ while keeping the remaining $\Lambda$CDM parameters fixed to their Planck values, improves the fit to JWST observations.\footnote{{Fig. 1 \cite{Forconi:2023hsj} allows for two possibilities when varying $H_0$, either fixing $A_s$ or fixing $\sigma_8$. Only when $A_s$ is fixed, the scenario we consider here, does one find that increasing $H_0$ improves the fit to JWST observations.}}  Varying $\Omega_m$ has a more pronounced effect than varying $H_0$ and it is evident that increasing physical matter density $\Omega_m h^2$ will also improve the fit to JWST observations.}

Note that $\Omega_m$ and $h$ are constant fitting parameters, alternatively cosmological parameters, in the $\Lambda$CDM model, which cannot vary without running into an inconsistency between the $\Lambda$CDM model and observations. While this subtelty is typically glossed over in Bayesian analysis, from the perspective of physics \footnote{As a sub-branch of physics it is imperative that the cosmology community conducts consistency tests that the physics community routinely performs elsewhere.} it is important to check that the $\Lambda$CDM cosmological parameters are not redshift dependent \cite{akarsu2024lambdacdm}.   

It should be stressed that the purpose of this section is to make the paper as self-contained as possible. Evidently, there appears to be some tension between the Planck-$\Lambda$CDM model amd JWST. Of course, this tension may be overcome through new astrophysics. For example, it is evident that increasing the efficiency $\epsilon$ can also alleviate the problem. What we will now show is that high redshift cosmological probes have a preference for the larger $\Omega_m$ values that would help alleviate JWST anomalies. This may be a coincidence, and neither QSOs nor GRBs may make sense as cosmological probes, but given the synergy, further study is warranted. We expect that Type Ia supernovae (SNe) samples will be able to confirm or refute these observations in QSOs/GRBs once we have higher redshift samples.

\section{Frequentist Confidence Intervals}
\label{sec:methods}
In the literature frequentist confidence intervals typically approximate likelihoods as Gaussian distributions. A case in point is the Fisher matrix. As is evident from section 2.6 of \cite{Trotta:2017wnx}, one can Taylor expand the likelihood $\mathcal{L} (\theta)$ about the maximum likelihood estimator (MLE) $\bar{\theta}$ dropping terms beyond quadratic order. This truncation of the expansion evidently throws information away, and as a result, one is \textit{approximating} the likelihood about its peak as a Gaussian.
The second frequentist method highlighted in \cite{Trotta:2017wnx} is profile likelihoods. Concretely, one constructs the profile likelihood ratio,  
\begin{equation}
\label{eq:Ratio}
    R(\theta) = \frac{\mathcal{L}(\theta, \hat{\hat{\psi}})}{\mathcal{L}(\hat{\theta}, \hat{\psi})}, 
\end{equation}
where $\theta$ is the parameter of interest, $\psi$ denotes auxiliary parameters, $\hat{\hat{\psi}}$ is the value of $\psi$ that maximises the likelihood $\mathcal{L} (\theta, \psi)$ for fixed $\theta$, and $(\hat{\theta}, \hat{\psi})$ denote the values of $(\theta, \psi)$ that globally maximise the likelihood $\mathcal{L} (\theta, \psi)$. We note that the denominator is a constant so that the numerator $\mathcal{L}(\theta, \hat{\hat{\psi}})$, called a profile likelihood, and the profile likelihood ratio are identical up to a constant. For this reason, we will use the terminology profile likelihood and profile likelihood ratio interchangeably; we hope the meaning is clear from the context. 

In our work, we will study profile likelihoods with one degree of freedom, $\theta = \Omega_m$. The logic then goes that one invokes Wilks' theorem \cite{Wilks}, which states for large enough samples profile likelihoods are close to Gaussian:
\begin{equation}
\label{eq:PL}
\mathcal{L} (\theta, \hat{\hat{\psi}}) \propto e^{- \frac{c}{2} (\theta- \bar{\theta})^2} (1+ O(1/\sqrt{n})), 
\end{equation}
where $c$ is a positive constant, $\bar{\theta}$ is the MLE for the parameter $\theta$, $n$ is the size of the sample, and we have specialised to the case of interest with one degree of freedom $\theta$. Equation (\ref{eq:PL}) is essentially equation (7) from \cite{Wilks}. It should be evident that Wilks' theorem makes assumptions so that the profile likelihood converges to a Gaussian distribution in the large sample limit $n \rightarrow \infty$. As a result, $-2 \ln R(\theta)$ converges to a chi-squared distribution in the same limit.  

Making use of Wilks' theorem \cite{Wilks}, the prescription is that $100 \,\alpha \%$ confidence intervals correspond to the values of $\Delta \chi^2$ satisfying the equation \cite{Trotta:2017wnx}: 
\begin{equation}
\label{eq:alpha}
    \alpha = \int_{y=0}^{y = \Delta \chi^2} \frac{1}{\sqrt{2 \pi y}} e^{-\frac{1}{2} y} \textrm{d} y, 
\end{equation}
where we have specialised to the chi-squared distribution with one degree of freedom and employed $\Gamma(\frac{1}{2}) = \sqrt{\pi}$. Integrating the right hand side to $\Delta \chi^2 = 1$ and $\Delta \chi^2 = 3.9$ one finds $\alpha = 0.6827$ and $\alpha = 0.9517$, corresponding to $68 \%$ and $95 \%$ confidence intervals, respectively. A key point here is that if $X$ is a random variable with a  standard normal distribution with probability density function (PDF),  
\begin{equation}
f_{X}(x) = \frac{1}{\sqrt{2 \pi}} e^{-\frac{1}{2} x^2}, \quad \int_{-\infty}^{+\infty} f_{X} (x) \dd x = 1, 
\end{equation}
then it is easy to prove that $Y = X^2$ has the chi-squared distribution PDF (see appendix for the  derivation) 
\begin{equation}
\label{eq:fYy}
f_{Y}(y) = \frac{1}{\sqrt{2 \pi y}} e^{-\frac{1}{2} y}, \quad \int_{-\infty}^{+\infty} f_{Y} (y) \dd y = 1.  
\end{equation}
Thus, one has the result 
\begin{equation}
\label{eq:alpha_gaussian}
    \alpha = \int_{0}^{\Delta \chi^2} \frac{1}{\sqrt{2 \pi y}} e^{-\frac{1}{2} y} \textrm{d} y = \int_{-\sqrt{\Delta \chi^2}}^{+\sqrt{\Delta \chi^2}} \frac{1}{\sqrt{2 \pi}} e^{-\frac{1}{2} x^2}. 
\end{equation}
In other words, integrating the chi-squared distribution with one degree of freedom between $\Delta \chi^2 = 0$ and $\Delta \chi^2 = 1$ is equivalent to integrating a standard normal distribution between $-\sqrt{\Delta \chi^2}$ and $+\sqrt{\Delta \chi^2}$. Specialising to the region of parameter space with $\Delta \chi^2 \leq 1$, we recognise from equation (\ref{eq:alpha_gaussian}) that the $68 \%$ confidence intervals one recovers are precisely the $1 \sigma$ confidence interval of a Gaussian. So, in summary, equation (\ref{eq:alpha}) \textit{approximates} confidence intervals and this approximation becomes \textit{exact} in the $n = \infty$ limit whereby the profile likelihood (\ref{eq:alpha_gaussian}) is Gaussian. 

As an aside, there is one further subtlety worth discussing with the $\Delta \chi^2 \leq 1$ and $\Delta \chi^2 \leq 3.9$ prescription for $68\%$ and $95 \%$ confidence intervals. Wherever the confidence intervals are impacted by a boundary, one should make use of the Feldman-Cousins prescription \cite{Feldman:1997qc}. We remark that the prescription is employed in  \cite{Herold:2021ksg}, but since the profile likelihood is Gaussian to a good approximation, and the $\Delta \chi^2 \leq 1$ interval is not impacted by the boundary, the methodology is redundant for $68\%$ confidence intervals. More concretely, from Fig. 2 of \cite{Herold:2021ksg} it is evident that the boundary starts to impact results at $95 \%$ confidence interval corresponding to $\Delta \chi^2 \leq 3.9$. From Table X of \cite{Feldman:1997qc} we see that there is no correction to the Wilks' $68\%$ confidence interval when the boundary is $\gtrsim 2 \sigma$ away from the MLE. In \cite{2024arXiv240504455G} the Feldman-Cousins prescription is employed to greater effect, but we observe that the authors assume that their profile likelihoods follow equations (4.2) and (4.3) from the Feldman-Cousins paper \cite{Feldman:1997qc}, i. e. Gaussian profile likelihoods are assumed.\footnote{We thank Giacomo Galloni for correspondence on this point.} In the appendix we take a closer look at the effect of a boundary on Gaussian profiles. We demonstrate a key corollary of Table X of \cite{Feldman:1997qc} that the Feldman-Cousins prescription results in confidence intervals that are typically smaller than a naive application of Wilks' theorem.

We have hopefully convinced the reader that frequentist confidence intervals in cosmology rest heavily on Gaussian distributions.\footnote{It is often argued in the literature that one can reparameterise to get a Gaussian profile likelihood, e. g. \cite{Karwal:2024qpt}. This is true, but one needs to make sure that reparamaterisations are physical. For example, the degeneracies or banana-shaped contours in MCMC posteriors in the $(H_0, \Omega_m)$-plane in \cite{Colgain:2023bge} can be removed by redefining $(H_0, \Omega_m)$ as $(H_0, \Omega_m h^2)$. Since $\Omega_m h^2$ is well constrained, this should lead to a Gaussian profile in $\Omega_m h^2$. The question is, what does one do about $H_0$, a universal constant in all FLRW models? How does one reparametrise $H_0$? It is also worth noting that $H_0$ may be directly and definitively measured by local distance ladder measurements in a cosmological model-independent way assuming homogeneity and isotropy at cosmological scales.} This begs the question, how does one treat generic profile likelihoods? To that end  \cite{Gomez-Valent:2022hkb} (see also \cite{Herold:2021ksg, Colgain:2023bge}) suggests the following prescription. To extract the $68\%$ confidence interval for the profile likelihood, one normalises the profile likelihood ratio by the total area under the curve, 
\be
\label{eq:w}
w(\theta) = \frac{R(\theta)}{\int R(\theta) \, \dd \theta}, 
\ee
and solves the equation 
\be
\label{eq:conf}
\int_{\theta^{(1)}}^{\theta^{(2)}} w(\theta) \, \dd \theta = 0.68, \quad w(\theta^{(1)}) = w(\theta^{(2)}). 
\ee
To get $95 \%$ ($2 \sigma$) and $99.7 \%$ ($3 \sigma$) confidence intervals one simply changes the number on the right hand side of equation (\ref{eq:conf}). It should be noted that equations (\ref{eq:w}) and (\ref{eq:conf}) are more or less equations (2) and (3) from \cite{Herold:2021ksg}. The content of equation (\ref{eq:conf}) is that one builds $68\%$ confidence intervals outwards from the peak of the profile likelihood ratio, $R(\hat{\theta}) = 1$, by steadily including points in the $\theta$ parameter space that are steadily less likely until one reaches $68 \%$ of the area under the profile likelihood curve. It should be clear from (\ref{eq:PL}) and (\ref{eq:alpha_gaussian}) that for a Gaussian profile likelihood we can expect the $68 \%$ confidence interval to agree with the $\Delta \chi^2 \leq 1$ confidence interval one gets by exploiting Wilks' theorem. Nevertheless, it should be stressed that Wilks' theorem is at best an approximation. As a further comment, one may worry that integrating the likelihood, suitably normalised, over the parameter is difficult to justify. However, here one can invoke Bayes' theorem, where assuming uninformative uniform priors, as we do here, the posterior probability of the parameter $\theta$ given the data $d$, $P(\theta | d)$, must reduce to the likelihood $\mathcal{L}(\theta)$ up to a normalising constant, $P(\theta | d) \propto \mathcal{L}(\theta)$ (see equation (74) of \cite{Trotta:2017wnx}). It is worth stressing here that in the large sample limit, employing Wilks' theorem is mathematically equivalent to integrating the normalised profile likelihood over the parameter. This is the content of equation (\ref{eq:alpha_gaussian}). This prescription holds for Gaussian profile likelihoods, but from the perspective of mathematics there is no obstacle to extending it to smaller samples with non-Gaussian profile likelihoods.

\section{QSO Anomaly}
We begin by studying changes in the best fit $\Lambda$CDM parameters ($H_0, \Omega_m$) in QSO data \cite{Risaliti:2015zla, Risaliti:2018reu, Lusso:2020pdb}. Trends of increasing $\Omega_m$ with effective redshift have been reported \cite{Colgain:2022nlb, Colgain:2022rxy, Pasten:2023rpc, Malekjani:2023dky} across a number of observables, including Type Ia SNe, and here we revisit findings with fresh methodology. Previously, the observation that $\Omega_m$ increased with effective redshift rested upon i) extremising the log-likelihood to identify best fits and ii) Markov Chain Monte Carlo (MCMC) posteriors \cite{Colgain:2022nlb}. Both analyses imposed an admittedly physical but restrictive prior on matter density, $0 \leq \Omega_m \leq 1$. 

Here we relax this prior to allow $\Omega_m > 1$ values while exploiting frequentist analysis introduced in section \ref{sec:methods}. Relaxing the prior $\Omega_m \leq 1$ allows the QSO data to constrain the $\Lambda$CDM model without theoretical preconceptions. One should take special care that the priors used for Bayesian analysis have strong and well-motivated theoretical grounds, and if not, physics demands that the chosen priors should not impact the final outcome. On the first point, we observe that there is no theoretical requirement from General Relativity that $\Lambda > 0$ (recall that in $\Lambda$CDM, $\Lambda \propto (1 - \Omega_m)$). Moreover,  a small positive $\Lambda$ as the dominant part of dark energy causes problems for both quantum field theory \cite{Weinberg:1988cp} and string theory \cite{Dvali:2014gua, Dvali:2018fqu, Obied:2018sgi}, which are routinely swept under the rug. For these reasons, the $\Omega_m \leq 1$ bound is not strongly motivated by theory, certainly not to the point that it is beyond question.

On the second point that priors should not impact the final result, in frequentist analysis, the focus is not on priors but on identifying the MLE; priors should be generous enough to facilitate the identification of the MLE. At this point the reader may object that we have no right comparing QSO data that prefer $\Omega_m > 1$ with Planck data, because the Planck collaboration has imposed the prior $\Omega_m \leq 1$ in their data analysis. To dispel this concern, one can check the Planck baseline MCMC chains and confirm that the $\Omega_m$ values never venture outside of the range $0.29 \lesssim \Omega_m \lesssim 0.34$. In fact, the Planck value $\Omega_m = 0.315 \pm 0.007$ is $98 \sigma$ from $\Omega_m =1$, so a narrow range is expected. What we learn from this is that the Planck collaboration could have imposed any upper bound on $\Omega_m$ beyond the maximum $\Omega_m$ value from the MCMC chain, e. g. $\Omega_m \leq 0.4, \Omega_m \leq 1$, $\Omega_m \leq 10$, $\Omega_m < \infty$, etc, and it would have no bearing on the Planck result \footnote{Note, at each step of MCMC, the \textit{derived} parameter $\Omega_m$ needs to be determined from the parameters fitted by Planck, $\Omega_b h^2, \Omega_c h^2$, etc, before this additional prior is imposed. The point we are making is that for a sufficiently large bound, i. e. $\Omega_m \lesssim 0.4$ or larger, one cannot impact the Planck MCMC chains and therefore one cannot change the Planck $\Omega_m$ constraint. Ultimately, any sufficiently large $\Omega_m$ bound is just a redundant piece of code in the data analysis. }. As explained above, this ensures that none of the ``physics" is prior dependent. In summary, from either the theoretical or data analysis perspective, we see no obstacle to relaxing the traditional $\Omega_m \leq 1$ upper bound.

Even if the reader refuses to engage with these arguments, we should still not lose sight of context. Our paper attempts to argue that the profile likelihoods are shifting to larger $\Omega_m$ values as the effective redshift increases. One way to check this is to compare to an $\Omega_m$ ``yardstick", which we choose to be the Planck $\Omega_m$ value. Doing so allows us to verify that the Planck value is moving further into the lower tail of our profile likelihoods as the effective redshift increases. There may be some exceptions, which we comment upon later, but there is utility in the Planck $\Omega_m$ constraint as a yardstick. One could choose another dataset for this yardstick, for example the Pantheon+ SNe dataset \cite{Brout:2022vxf}, where it is easy to relax the priors and make sure that they have no bearing on the results. 

While one typically constructs profile likelihoods by isolating a fitting parameter of interest and extremising the log-likelihood with respect to the remaining parameters, here we will also employ methodology that recycles and bins the MCMC chain \cite{Gomez-Valent:2022hkb, Colgain:2023bge}.\footnote{See \cite{Trotta:2017wnx} for a discussion on binning the MCMC chain to construct profile likelihoods.} We remark that one traditionally maximises the log-likelihood, but it allows a more direct comparison between Bayesian and frequentist methods if one starts from information in the same MCMC chain. As we shall see, whether one maximises the log-likelihood or bins the MCMC chain, the difference is negligible and it can be improved by running longer MCMC chains, e. g. \cite{Colgain:2023bge}.

Recall the Risaliti-Lusso procedure for standardising QSOs \cite{Risaliti:2015zla}. At its heart, one assumes a UV, X-ray luminosity relation that is intrinsic to QSOs: 
\be
\label{eq:lum}
\log L_{X}  = \beta + \gamma \log L_{UV},  
\ee
where $\beta$ and $\gamma$ are constant fitting parameters. Through the standard luminosity-flux relation, $L = 4 \pi D_{L}(z)^2 F$ with luminosity distance $D_{L}(z)$, (\ref{eq:lum}) is equivalent to the UV, X-ray flux relation: 
\bea
\label{eq:flux}
\log F_{X} = \tilde \beta + \gamma \log F_{UV} + 2(\gamma-1) \log D_{L}(z),  
\eea
where $\tilde \beta=\beta+ (\gamma-1) \log 4 \pi$. 
Equation (\ref{eq:lum}) is empirically motivated, but a physical backstory has been provided in terms of an interaction between the accretion disc of the Active Galactic Nucleus (AGN) and the X-ray Corona. This would allow photons to inverse Compton scatter from UV to X-ray without hot electrons losing their energy. The precise physical mechanism is unclear, but models exist to explain the physics \cite{Lusso:2017hgz, Kubota:2018cuj, Arcodia:2019oyq}. Returning to the empirical result, the key point is that given observations of the X-ray flux $F_{X}$ and UV flux $F_{UV}$, one can constrain the cosmological model Hubble parameter $H(z)$ through the luminosity distance 
\be
\label{eq:DL}
D_{L}(z) = c (1+z) \int_{0}^{z} \frac{1}{H(z^{\prime})} \dd z^{\prime}. 
\ee
Throughout, we will assume the (flat) $\Lambda$CDM model with 
\be
\label{eq:lcdm}
H(z) = H_0 \sqrt{1-\Omega_m + \Omega_m (1+z)^3}. 
\ee
The parameters $(H_0, \Omega_m)$ arise as integration constants in either the Friedmann equation or the matter continuity equation.\footnote{$H_0$ is an integration constant in the Friedmann equations and $\rho_{m} (z=0) \propto H_0^2 \Omega_m$ is an integration constant in the matter continuity equations. As a result, $\Omega_m$ must be constant.} Alternatively put, \textit{mathematically} neither parameter can change with effective redshift, however it remains to be seen if this is the case \textit{observationally}. These consistency checks have been overlooked in cosmology, but they are well motivated \cite{akarsu2024lambdacdm}. We will see that, as highlighted initially in \cite{Risaliti:2018reu, Lusso:2019akb, Yang:2019vgk}, there is a strong tension between the QSO data and the $\Lambda$CDM model. However, in contrast to earlier work, the inconsistency need not be relative to a specific value of $\Omega_m$ preferred by a different data set, e. g. Planck CMB data \cite{Planck:2018vyg}, but may be regarded as an inconsistency driven by the non-constancy of integration constants. The contradiction between data and model is difficult to ignore.   

Evidently, (\ref{eq:lum}) is a working assumption. It is the subject of a growing body of work in the cosmology literature either questioning \cite{Khadka:2020vlh, Khadka:2020tlm, Khadka:2021xcc, Khadka:2022aeg, Singal:2022nto, Petrosian:2022tlp, Zajacek:2023qjm} or refining the Risaliti-Lusso prescription \cite{Dainotti:2022rfz, Dainotti:2024aha}.\footnote{Despite the corrections imposed on the QSO data, a residual increasing trend of the $\Omega_m$ parameter is still reported \cite{Dainotti:2024aha}.} {In particular, it has been argued in  \cite{Zajacek:2023qjm} that the Risaliti-Lusso QSOs are impacted by extinction, but this effect may not be large enough to explain the discrepancy seen in the Hubble diagram \cite{Trefoloni:2024dei}.} Here, it is worth stressing that QSOs represent emerging cosmological probes \cite{Moresco:2022phi} and considerable work remains to turn the proposal into standardisable candles on par with Type Ia SNe, where developments span 3 decades. See \cite{Czerny:2022xfj} for a historical account of the quest to construct a QSO Hubble diagram. Nevertheless, in defence of the Risaliti-Lusso methodology, it should be noted that the logarithms of X-ray and UV fluxes show an apparent correlation (see for example Figs. 3 and 4 of \cite{Risaliti:2015zla}), and to first approximation it is valid to fit a line to any empirical relation.\footnote{Unsurprisingly, the slope $\gamma$ is robust to changes in redshift \cite{Vignali:2002ct, Just:2007se, Lusso_2010, Salvestrini:2019thn, Bisogni:2021hue}. However, it is imperative from  (\ref{eq:flux}) that the $y$-intercept is not a constant in the $\log F_X-\log F_{UV}$ plane and that it decreases (note $\gamma \approx 0.6 < 1$) with increasing redshift/luminosity distance for any cosmology. Such a trend is visible in Fig. 4 of \cite{Risaliti:2015zla}, but a greater number of redshift bins is warranted. 
} Ultimately, our goal here is not to question (\ref{eq:lum}), but to maintain an open mind and treat it as a working assumption in order to study the implications. The point of the work is to draw a parallel between the preference of Risaliti-Lusso QSOs \footnote{Similar trends are also seen in Type Ia SNe and observational Hubble data \cite{Colgain:2022rxy, Pasten:2023rpc, Malekjani:2023dky}, but high redshift subsamples are smaller than QSO subsamples.} and JWST observations for larger $\Omega_m$ and larger $\Omega_m h^2$ values at higher redshifts. Along the way, we will comment on the tension between the QSO data set and the Planck-$\Lambda$CDM model using the frequentist methods in section \ref{sec:methods}. 

The QSO data set we employ consists of 2421 QSOs spanning the redshift range $0.009 \leq z \leq 7.5413$ \cite{Lusso:2020pdb}. The data set comprises the redshifts $z_i$, X-ray and UV fluxes ($F_{X, i}, F_{UV, i}$) and their corresponding errors $(\sigma_{F_{X, i}}, \sigma_{F_{UV, i}})$. There is considerable scatter in the QSO data and this is absorbed through an intrinsic error or intrinsic dispersion parameter $\delta$ \cite{Risaliti:2015zla}. Thus, restricting our attention to the $\Lambda$CDM model (\ref{eq:lcdm}), there are three nuisance parameters $(\beta, \gamma, \delta)$ in addition to the cosmological parameters $(H_0, \Omega_m)$. As is the case with Type Ia SNe data sets, where the absolute magnitude $M_B$ is degenerate with $H_0$, here $\beta$ and $H_0$ are degenerate. To overcome this degeneracy, one typically assumes a canonical value of $H_0 = 70$ km/s/Mpc and fits $(\Omega_m, \beta, \gamma, \delta)$ to the data. To do so, we consider the log-likelihood \cite{Risaliti:2015zla, Risaliti:2018reu, Lusso:2020pdb}, 
\bea
\label{eq:L1}
\ln \mathcal{L} &=& -\frac{1}{2} \left[ \chi^2 + \sum_{i=1}^N \ln (2 \pi s_i^2) \right],  \nn 
&=& - \frac{1}{2} \sum_{i=1}^{N} \left[ \frac{ \left(\log F^{\textrm{obs}}_{X, i} - \log F^{\textrm{model}}_{X,i}\right)^2}{s^2_{i}} + \ln (2 \pi s_i^2) \right], 
\eea
where $\log F^{\textrm{model}}_{X,i}$ is defined in (\ref{eq:flux}). Here, $s_i^2 = \sigma_{\log F^{\textrm{obs}}_{X,i}}^2+ \delta^2$ in (\ref{eq:L1}) contains the measurement error on the observed flux $\log F^{\textrm{obs}}_{X,i}$ and the intrinsic dispersion. Following the Risaliti-Lusso prescription, the $F_{UV}$ errors can be safely ignored \cite{Risaliti:2015zla} on the grounds that both $F_{X}$ and $F_{UV}$ errors are considerably smaller than $\delta$.

\subsection{\texorpdfstring{$\Omega_m$}{Omegam} trend}

\begin{figure}[htb]
   \centering
\includegraphics[width=90mm]{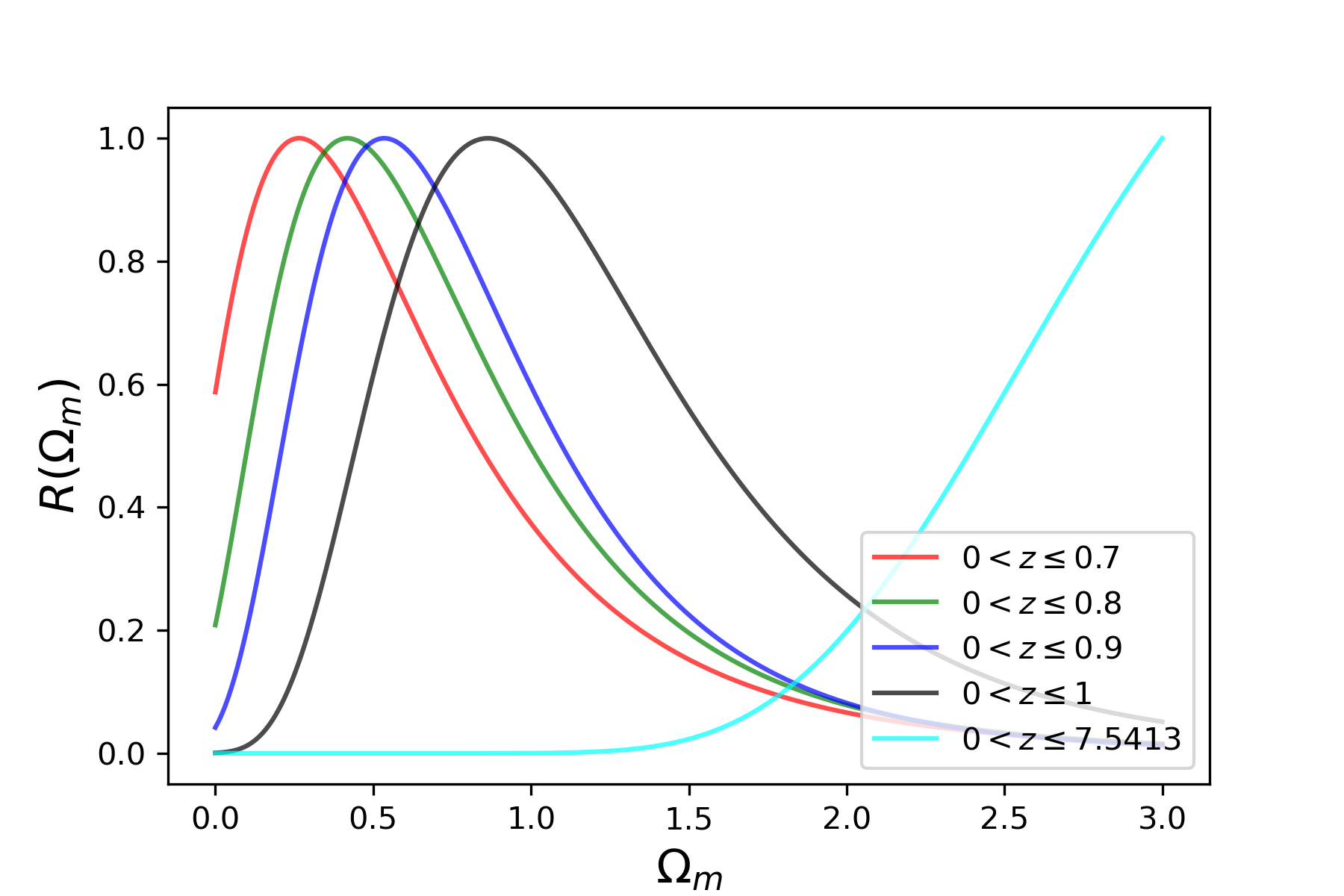} 
\caption{Variation of profile likelihoods $R(\Omega_m)$ with increasing effective redshift of the QSO sample \cite{Lusso:2020pdb}. The $\Lambda$CDM parameter $\Omega_m$ is not a constant when confronted to the QSO data.}
\label{fig:QSO_OM_ZEFF}
\end{figure}

In Table \ref{tab:QSOvsZ} we record frequentist $68\%$ confidence intervals for $\Omega_m$ as the redshift range of the sample is increased. The corresponding profile likelihood ratios are shown in Fig. \ref{fig:QSO_OM_ZEFF}, where the cyan curve corresponds to the full sample. As explained above, we fix $H_0$ to a canonical value, scan over $\Omega_m$ values in the range $\Omega_m \in [0, 3]$, and for each value of $\Omega_m$ we identify the maximum value of the likelihood at each value of $\Omega_m$, $\mathcal{L}_{\textrm{max}}(\Omega_m)$. Note, our choice of $\Omega_m$ bound allows us to track movements in the profile likelihood peak and confirm that the profile likelihood is well constrained, i. e. falls off at smaller and larger values of $\Omega_m$. That being said, it is evident that the bounds are not generous enough to accommodate the profile likelihood for the full sample (cyan curve). 

\begin{table}[htb]
\centering 
\begin{tabular}{c|c|c|c}
 \rule{0pt}{3ex} Redshift & \# QSOs & $ \Omega_{m}$ ($68\%$) & $\Omega_m$ ($ \Delta \chi^2 \leq 1$) \\
\hline 
\rule{0pt}{3ex} $0 < z \leq 0.7$ & $398$ & $< 0.79$ ($0.26$) & $ 0.26^{+0.46}_{-0.25}$ \\
\hline
\rule{0pt}{3ex} $0 < z \leq 0.8$  & $543$ & $0.42^{+0.54}_{-0.39}$ & $0.42^{+0.40}_{-0.28}$ \\
\hline
\rule{0pt}{3ex} $0 < z \leq 0.9$ & $680$ & $0.53^{+0.53}_{-0.31}$ & $0.53^{+0.45}_{-0.28}$ \\
\hline
\rule{0pt}{3ex} $ 0 < z \leq 1$ & $826$ & $0.86^{+0.64}_{-0.39}$ & $0.86^{+0.57}_{-0.36}$ \\
\hline
\rule{0pt}{3ex} $ 0 < z \leq 7.5413$ & $2421$ & $2.42 <$ $(3)$ & $2.55<$ ($3$) \\
\end{tabular}
\caption{The number of QSOs in redshift bins with $68\%$ frequentist confidence intervals for $\Omega_m$ determined using the two methods outlined in section \ref{sec:methods}. $\Omega_m$ best fits and confidence intervals increase with effective redshift. In the absence of an upper or lower bound we present the profile likelihood peak (MLE) in brackets.}
\label{tab:QSOvsZ}
\end{table}


Note, it is observed in \cite{Khadka:2020tlm}  that the variations in cosmological parameters across different models become more severe at higher redshifts when one fits QSO data. This was interpreted as evidence for the non-standardisability of the Risaliti-Lusso QSOs. However, as explained in \cite{Luongo:2021nqh}, the biggest jumps in cosmological parameters happen when the curvature parameter $\Omega_k$ is introduced, but this can be traced to the added flexibility that $\Omega_k < 0$ gives to lower the distance modulus or luminosity distance at higher redshits. The key point here is that the full QSO data set prefers $\Omega_m > 1$ values, so if one constrains $\Omega_m$ to the usual range, $0 \leq \Omega_m \leq 1$, one can get a better fit to the data through a curvature parameter $\Omega_k$. In short, the $\Omega_m$ bounds prevent the QSO data from finding the $\Omega_m > 1$ MLE values. This should come as no surprise as a pronounced fall off in the distance modulus relative to Planck-$\Lambda$CDM is glaringly obvious at higher redshifts in Fig. 2 of  \cite{Risaliti:2018reu}. Evidently, great care is required with QSOs over extensive redshift ranges. Thus, the point of Fig. \ref{fig:QSO_OM_ZEFF} is to focus on evolution in the QSO sample at more accessible and conservative lower redshifts $z \lesssim 1$, while demonstrating that the increasing $\Omega_m$ trend with effective redshift persists to the full sample. {We remark that once the profile likelihood peak enters $\Omega_m > 1$ parameter space, one can interpret this as negative dark energy density in the $\Lambda$CDM model.}

Scanning over $\Omega_m$ and maximising the log-likelihood (\ref{eq:L1}) leads to an array of $(\Omega_m, \mathcal{L}_{\textrm{max}}(\Omega_m))$ values. Inevitably, there will be a global maximum $\mathcal{L}_{\textrm{max}}$ in the range $\Omega_m \in [0, 3]$. This allows us to construct the profile likelihood ratio,  
\bea
\label{eq:R}
R(\Omega_m) = \frac{\mathcal{L}_{\textrm{max}}(\Omega_m)}{\mathcal{L}_{\textrm{max}}} &=& \exp \left( -\frac{1}{2} \Delta \chi^2 \right), \nn &=& \exp \left( - \frac{1}{2} (\chi^2_{\textrm{min}}(\Omega_m) - \chi^2_{\textrm{min}})\right),  
\eea
which by construction peaks at an $\Omega_m$ value where $R(\Omega_m) = 1$. The profile likelihoods $R(\Omega_m)$ for different redshift ranges are presented in Fig. \ref{fig:QSO_OM_ZEFF}. The cyan distribution has evidently been curtailed by our bounds on $\Omega_m$ and the distribution peaks at a larger $\Omega_m > 3$ value. What the cyan curve demonstrates is that there is a strong tension between the full QSO sample and canonical Planck values of $\Omega_m \sim 0.3$. However, as is clear from the plot, and earlier results from \cite{Colgain:2022nlb}, QSOs restricted to lower redshifts prefer Planck $\Omega_m$ values. Even if QSOs are non-standardisable, this is a surprising coincidence. To extract the $\Omega_m$ $68\%$ confidence intervals for each curve, one normalises by the total area under the curve (\ref{eq:w}) and solves equation (\ref{eq:conf}). 

To be more concrete about our methodology, we break the $\Omega_m \in [0, 3]$ interval up into 300 intervals of $\Delta \Omega_m \approx 0.01$ and maximise the log-likelihood for the auxiliary parameters $(\beta, \gamma, \delta)$ at each discrete value of $\Omega_m$. Throughout, we make sure that the values of $(\beta, \gamma, \delta)$ that maximise (\ref{eq:L1}) are not impacted by our bounds on those parameters. Even without interpolating the discrete values, this led to the smooth curves in Fig. \ref{fig:QSO_OM_ZEFF}. To perform integration, we made use of Simpson's rule. The resulting confidence intervals are presented in the third column in Table \ref{tab:QSOvsZ}. It is worth noting that the profiles have long non-Gaussian tails in the direction of larger $\Omega_m$ values. This behaviour is expected when one confronts the $\Lambda$CDM model to binned luminosity distance data and removes the low redshift observables \cite{Colgain:2022nlb}. Note, we have not binned the data, but the Risaliti-Lusso QSO samples become sparse at low redshift, so the redshift distribution of the observables is the same. 

One interesting feature of our analysis is that the $\Delta \chi^2 \leq 1 \Leftrightarrow R( \Omega_m) \geq 0.607$ prescription leads to smaller, less conservative $68 \%$ confidence intervals in column 4 in Table \ref{tab:QSOvsZ}. The $68 \%$ confidence intervals for the full sample need to be treated with caution as the peak of the profile likelihood exceeds our upper bound $\Omega_m = 3$.  
Nevertheless, it is a given that one can treat $\Omega_m = 3$ as a lower bound on the position of the peak. We can then get a \textit{lower bound} on the tension with the $1 \sigma$ Planck upper bound $\Omega_m \leq 0.322$ \cite{Planck:2018vyg}. Translated into the profile likelihood ratio, one has $R(\Omega_m) \leq 2.4 \times 10^{-14}$ corresponding to $\Delta \chi^2 \geq 62.7$ and a $\sim 7.9 \sigma$ disagreement between the QSOs and Planck. Once again bearing in mind the obvious caveat that the peak of our profile likelihood is beyond $\Omega_m = 3$, one may employ  (\ref{eq:w}) and (\ref{eq:conf}). Doing so, one finds that the Planck value is disfavoured at $99.9999999999994\%$ confidence level or $\sim 7.8 \sigma$. It is interesting that both our methodologies agree. Given the close agreement, we can infer that cyan curve bears a close approximation to a Gaussian in the redshift range of interest. Let us emphasise again that this is a lower bound because as is clear from Fig. \ref{fig:QSO_OM_ZEFF} the bulk of the profile likelihood distribution is beyond $\Omega_m = 3$. Relaxing that prior will mean that the tail of the profile likelihood corresponding to the Planck value will correspond to an even smaller percentage of the area under the curve. To find such a strong tension with standard frequentist methods is a little surprising. However, let us stress here that our profile likelihood is unobstructed as far as the Planck value, so there is no problem inferring the confidence level of lower bounds based on the $\Delta \chi^2$ prescription. One could employ the Feldman-Cousins prescription \cite{Feldman:1997qc}, but $\Omega_m=3$ is not a physically motivated boundary. Even if one ignores this point, as we show in the appendix, the prescription leads to smaller confidence intervals than the naive Wilks' theorem approach, which would only exacerbate a tension that is already very large.

Moving along, our profile likelihood analysis to this point has been relatively standard frequentist analysis. This provides a complementary perspective on earlier results based on best fits and MCMC posteriors \cite{Colgain:2022nlb}. Previously, it was evident that $\Omega_m$ increased from a Planck value $\Omega_m \approx 0.3$ to larger values, but our one-dimensional $\Omega_m$ MCMC posteriors were impacted by restrictive $\Omega_m \in [0, 1]$ priors \cite{Colgain:2022rxy}. Here, we have relaxed the priors, and it is evident that the $68\%$ confidence intervals for QSOs in the redshift range $0 < z \leq 1$ have shifted to the extent that the canonical $\Omega_m \approx 0.3$ falls outside the confidence intervals, irrespective of the method used to identify confidence intervals. The analysis here provides confirmation that the $\Omega_m$ best fits and corresponding confidence intervals shift to larger $\Omega_m$ values as one increases the QSO redshift range. Noting that $\Omega_m$ is by definition a constant, since it is related to an integration constant, this underscores the tension between the QSO sample and the $\Lambda$CDM model. In short, the tension is evident throughout the sample.   

\begin{figure}[htb]
   \centering
\includegraphics[width=90mm]{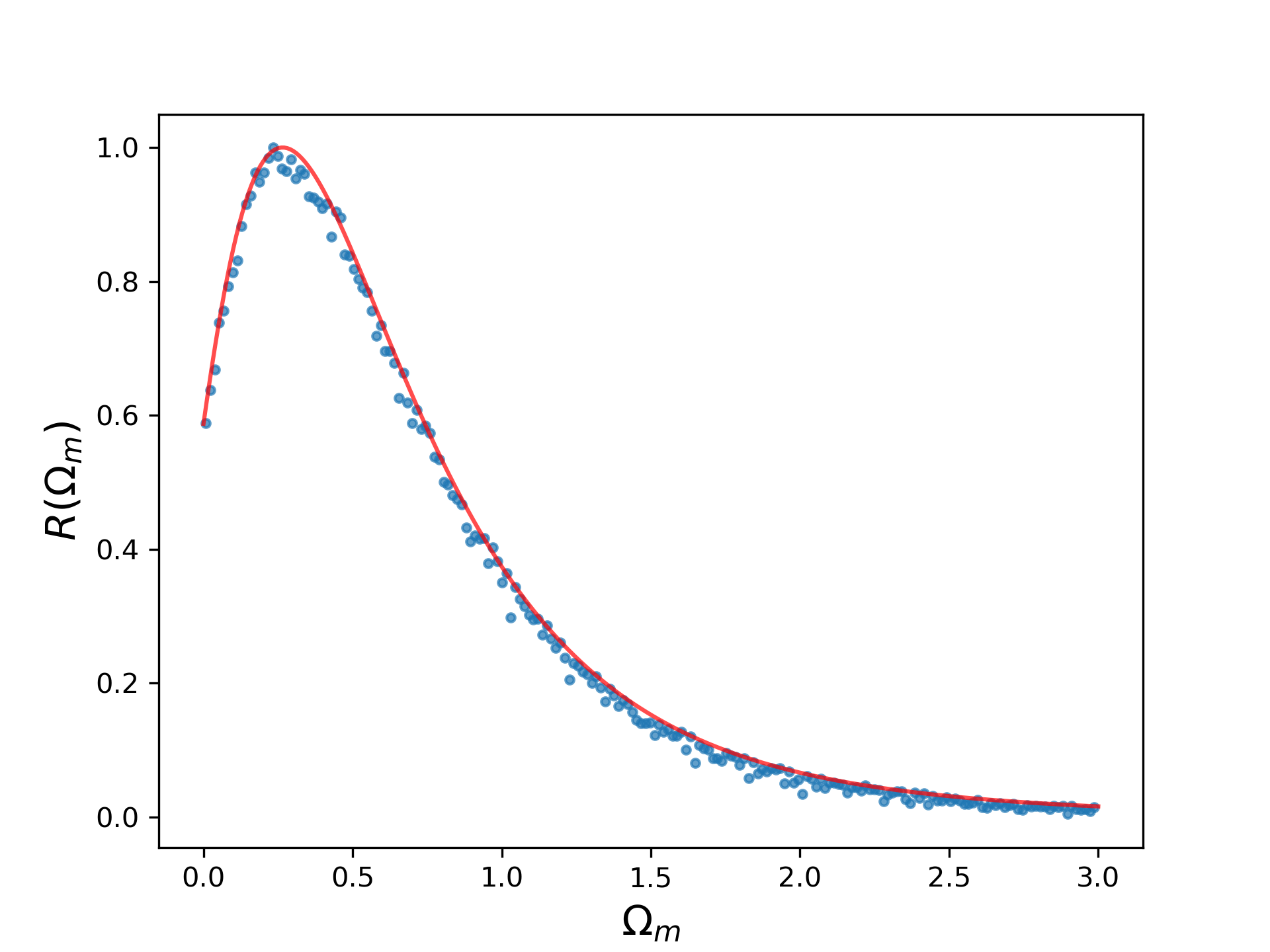} 
\caption{$R(\Omega_m)$ from binned, converged MCMC chain in blue dots versus maximisation of the log-likelihood in red. Scatter in blue dots can be reduced by running a longer MCMC chain. The blue dots are expected to converge to the red curve from below.}
\label{fig:opt_vs_mcmc}
\end{figure}

Before moving to study $\Omega_m h^2$ in the next subsection, it is interesting to compare profile likelihoods from maximising the log-likelihood, i. e. optimisation or gradient descent, with the corresponding result from binning MCMC chains \cite{Gomez-Valent:2022hkb, Colgain:2023bge}. While the latter allows a better comparison between Bayesian and frequentist methods, as one is exploiting the MCMC chain throughout, one worry is that MCMC may poorly identify $\mathcal{L}_{\textrm{max}}(\Omega_m)$. Focusing on QSOs in the lowest redshift range, $0 < z \leq 0.7$, in Fig. \ref{fig:opt_vs_mcmc} we present $R(\Omega_m)$ from a binned, converged MCMC chain in blue dots alongside the optimised $R(\Omega_m)$ in red. There is noticeable scatter in the blue dots and they visibly underestimate $\mathcal{L}_{\textrm{max}}(\Omega_m)$ and $R(\Omega_m)$. This is the expected outcome because the goal is to maximise $\mathcal{L}_{\textrm{max}}(\Omega_m)$, but any MCMC algorithm prioritises exploring parameter space over optimisation. That being said, any difference to the inferred $68\%$ confidence interval is negligible. One can evidently improve the agreement by running a longer MCMC chain, for example as in \cite{Colgain:2023bge}.   

\subsection{\texorpdfstring{$\Omega_m h^2$}{Omegamh2}  trend}
In the previous section we confirmed that there is an inconsistency between the $\Lambda$CDM model and standardisable QSOs precisely because $\Omega_m$ is not observationally a constant. Moreover, QSOs prefer larger $\Omega_m$ values than the Planck value \cite{Planck:2018vyg}, thereby alleviating the tension with independent JWST observations as explained in section \ref{sec:JWST}. In section \ref{sec:JWST} we also confirmed that increasing $\Omega_m h^2$ relative to Planck values helps alleviate the JWST anomaly. Thus, here we also look for an increasing $\Omega_m h^2$ trend in QSO data.  

Previously, we fixed $H_0$ to the nominal value $H_0 = 70$ km/s/Mpc and allowed $\beta$ to vary in order to break a degeneracy between the parameters. In this section, we fix $\beta$ to a nominal value and allow $H_0$ to vary. We do this to ascertain whether only $\Omega_m$ varies with effective redshift or whether the combination $\Omega_m h^2$ varies. Note, while $\Omega_m$ increases from Fig. \ref{fig:QSO_OM_ZEFF}, it is plausible that $h := H_0/100$ decreases so that the combination $\Omega_m h^2$ remains constant. In some sense, this question is already partially addressed in Table I in \cite{Colgain:2022nlb}, where it was observed that both $\Omega_m$ and $\beta$ MLEs increase with effective redshift. Since $\gamma \approx 0.6 < 1$, from equation (\ref{eq:flux}) it follows that increases in $\beta$ with fixed $H_0$ are mapped to increases in $H_0$ with fixed $\beta$. Given that the data is the same here and there, it is easy to guarantee that $H_0$ increases with effective redshift, and therefore that $\Omega_m h^2$ must increase with effective redshift.

\begin{figure}[htb]
   \centering
\includegraphics[width=90mm]{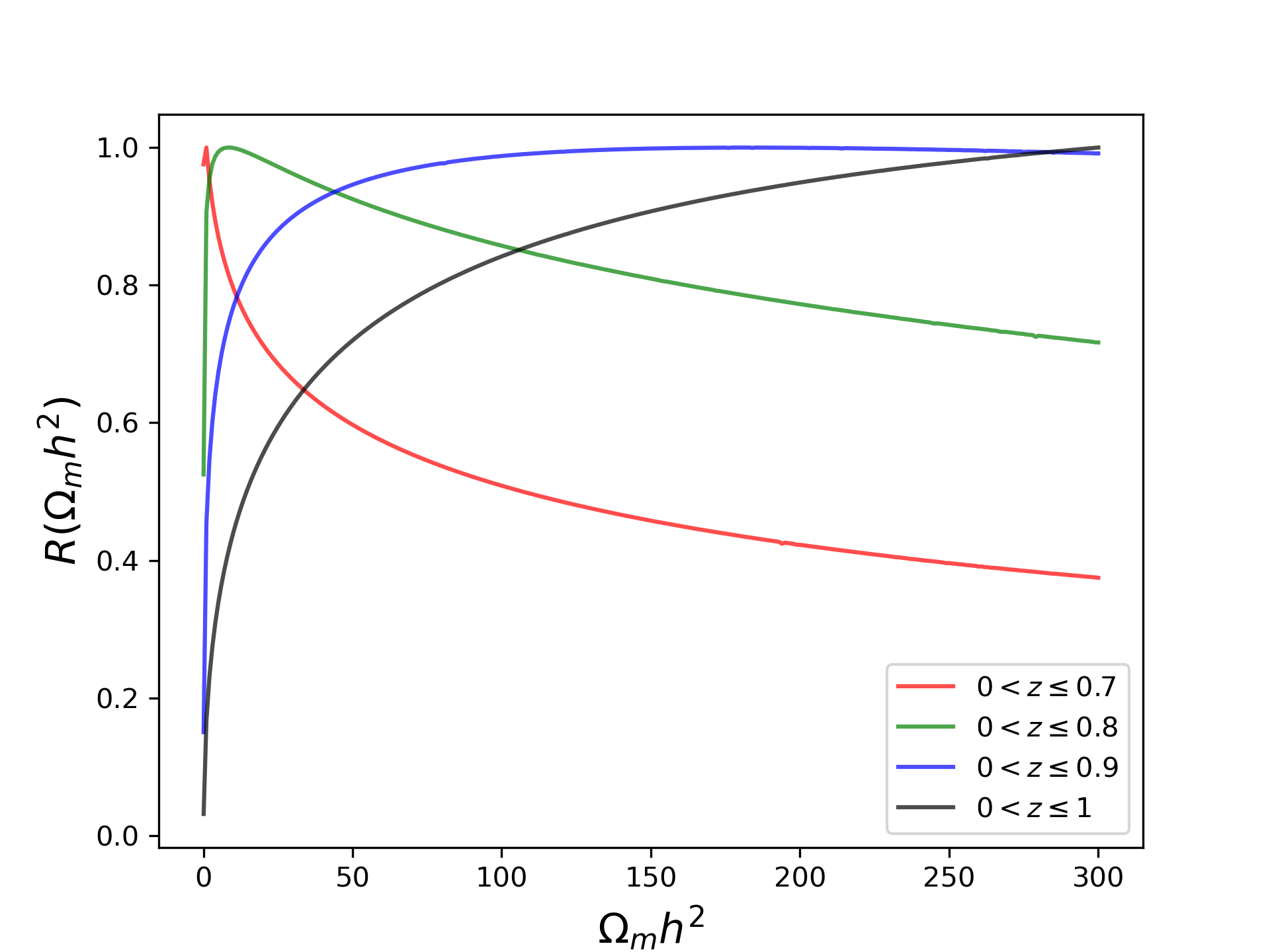} 
\caption{Variation of profile likelihoods $R(\Omega_m h^2)$ with increasing effective redshift of the QSO sample. The $\Lambda$CDM parameter $\Omega_m h^2$ is not a constant when confronted to the QSO data. }
\label{fig:QSO_OMH2_ZEFF}
\end{figure}

In Fig. \ref{fig:QSO_OMH2_ZEFF} we show the profile likelihoods for $\Omega_m h^2$. To get this plot, we fixed $\beta$ to the nominal value $\beta = 6.1$, so that $H_0$ adopts a canonical value, i. e. $H_0 \approx 70$ km/s/Mpc in the lowest redshift bin, $0 < z \leq 0.7$. Note that in this range $\Omega_m$ has a best fit value of $\Omega_m = 0.261$ from Table \ref{tab:QSOvsZ}. In turn, this implies $\Omega_m h^2 \approx 0.13$. In Fig. \ref{fig:QSO_OMH2_ZEFF} we scan over $\Omega_m h^2$ in intervals of $\Delta (\Omega_m h^2) = 1$ in the range $\Omega_m h^2 \in [0, 300]$. Unfortunately, this means that there is only one point below $\Omega_m h^2  = 1$, but nevertheless if one looks closely at the plot, a peak is evident in the red curve. Throughout, to get the curves we fixed $\beta$, but maximised the log-likelihood with respect to the parameters $(H_0, \Omega_m, \gamma, \delta)$ while imposing a constraint on the combination in $\Omega_m h^2$ through the likelihood. By scanning over the $\Omega_m h^2$ constraint in the range $\Omega_m h^2 \in [0, 300]$ we produced Fig. \ref{fig:QSO_OMH2_ZEFF}. 

There are a number of lessons. First, we confirm that $\Omega_m h^2$ increases with effective redshift. This is evident in the shift in the peaks of the profiles. In particular, the red, green and blue curves are all peaked in the range $0 < \Omega_m h^2 < 300$, whereas the black curve peaks at $\Omega_m h^2 > 300$. It is evident from the profiles that the confidence intervals are large, and that the profiles may fall away to $R(\Omega_m h^2) = 0$ very gradually even for profile likelihoods peaked at larger $\Omega_m h^2$ values, despite blue and black curves disfavouring the peak of the red curve strongly. It is clear from the constrained profiles in Fig. \ref{fig:QSO_OM_ZEFF} that the broad profiles in Fig. \ref{fig:QSO_OMH2_ZEFF} are driven by an inability to constrain both $H_0$ and $\Omega_m$. This is presumably due to the considerable scatter in the QSO data and the fact that this scatter is absorbed through a relatively large intrinsic dispersion $\delta$. The sparseness of QSOs at lower redshifts is also expected to lead to poorer constraints on $H_0$. Ultimately, QSO data on its own may be good enough to constrain $\Omega_m$ but the constraints on $H_0$ are poor. At best, Fig. \ref{fig:QSO_OMH2_ZEFF} appears consistent with the combination $\Omega_m h^2$ increasing with effective redshift, but the statistical significance of the trend is low. Nevertheless, it is clear that as effective redshift increases, the expected Planck value $\Omega_m h^2 \sim O(1)$ is disfavoured and this is the takeaway message from the analysis. Next, we turn our attention to GRB data sets, which have fewer observables and larger intrinsic scatter, so for this reason, we focus exclusively on $\Omega_m$, and not $\Omega_m h^2$, where already results are less conclusive in a better quality QSO data set.     

\section{GRB Anomaly}
\label{sec:GRB}
In this section we repeat the analysis with GRB samples. Concretely, we study a compilation of 220 long GRBs \cite{Cao:2024vmo} \footnote{This sample is the same as the 221 GRB sample from \cite{Jia:2022mwb} modulo the removal of a single GRB with an unreliable redshift and updating 8 GRBs.} in the redshift range $0.034 \leq z \leq 8.2$ and 118 long GRBs in the redshift range $0.34 < z < 8.2$ \cite{Khadka:2021vqa}. Our interest here is exploring high redshift observables that have different systematics to QSOs. In addition, while the standardisability of QSOs has been called into question \cite{Khadka:2020vlh, Khadka:2020tlm, Khadka:2021xcc, Khadka:2022aeg, Singal:2022nto, Petrosian:2022tlp, Zajacek:2023qjm}, similar tests on GRBs standardised through the Amati correlation \cite{Amati:2002ny} have been presented with a relatively clean bill of health \cite{Cao:2024vmo}. More precisely, the authors of \cite{Cao:2024vmo} claim that despite the GRBs being standardisable, the full sample of 220 GRBs leads to a value of the $\Lambda$CDM parameter $\Omega_m$ that is discrepant at the $> 2 \sigma$ level with the Planck-$\Lambda$CDM cosmology. We revisit this disagreement with profile likelihoods.    

Here, beginning with the 220 GRB sample \cite{Cao:2024vmo}, we ask a number of questions. First, as confirmed by earlier analysis, there is a low redshift subsample of the Risaliti-Lusso QSOs \cite{Lusso:2020pdb} that recovers canonical Planck values, as claimed originally in \cite{Colgain:2022nlb} in the absence of external data sets, notably Type Ia SNe, \textit{cf.} \cite{Risaliti:2018reu}. Thus, is there a low redshift subsample of the 220 GRB sample that recovers canonical Planck values $\Omega_m \sim 0.3$? Second, \cite{Cao:2024vmo} imposes the conventional $\Omega_m$ bounds, $ 0 \leq \Omega_m \leq 1$. Following the QSO analysis, we relax the bounds to identify the MLEs from the log-likelihoods. This allows us to confirm the preference of the GRB sample for larger $\Omega_m$ values. Lastly, we investigate if the profile likelihood changes with the effective redshift of the GRB sample. 

\begin{figure}[htb]
   \centering
\includegraphics[width=90mm]{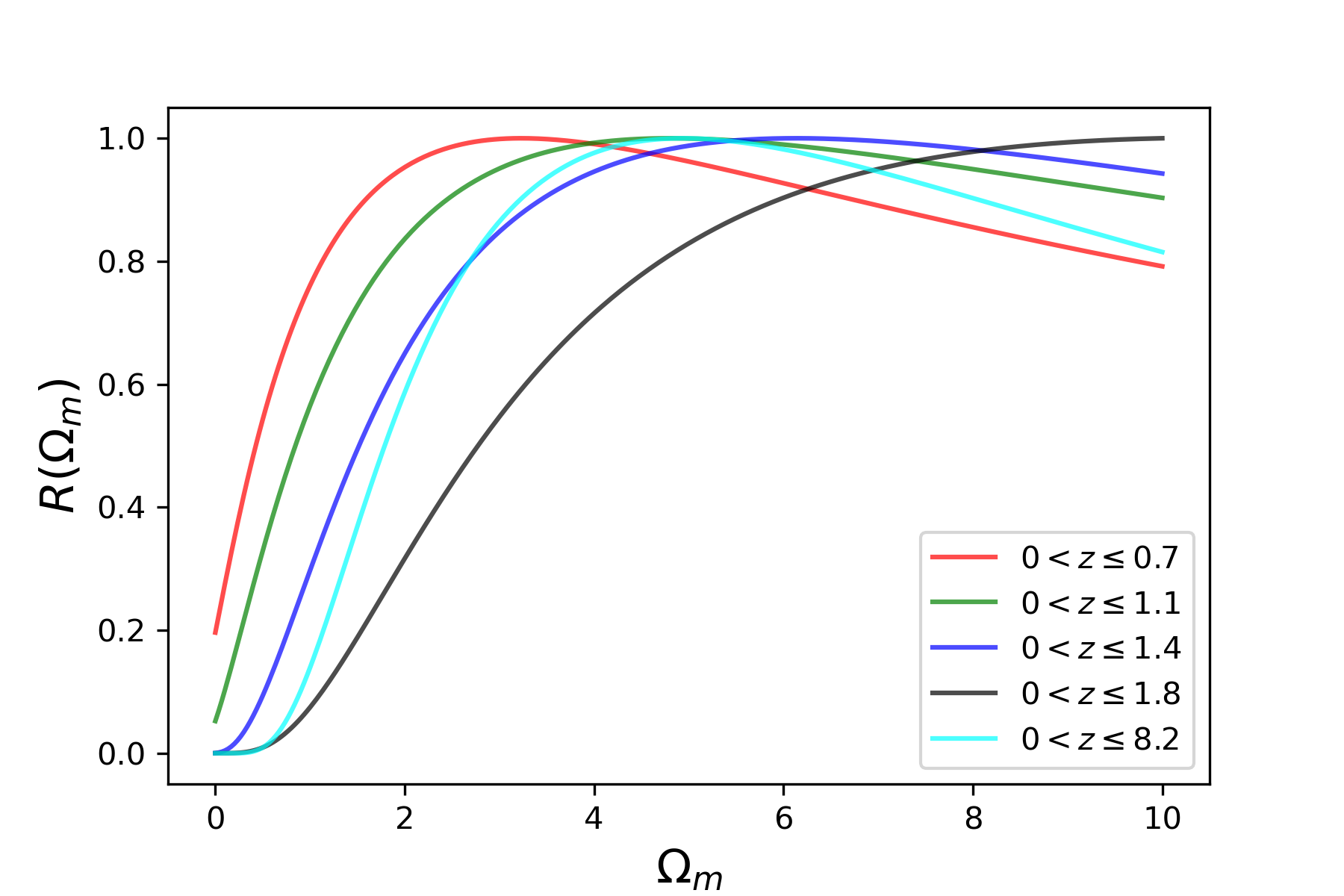} 
\caption{Variation of profile likelihoods $R(\Omega_m)$ with increasing effective redshift of the 220 GRB sample \cite{Cao:2024vmo}. The $\Lambda$CDM parameter $\Omega_m$ is not a constant when confronted to the GRB data.}
\label{fig:GRB_OM_ZEFF}
\end{figure}

We consider the log-likelihood 
\begin{equation}
\label{eq:L2}
\ln \mathcal{L} = -\frac{1}{2} \sum_{i=1}^{N} \left[ \frac{ \left(\log E_{\textrm{iso},i} - (\beta + \gamma \log E_{\textrm{p},i}) \right)^2}{s^2_{i}} + \ln (2 \pi s_i^2) \right]
\end{equation}
where $ E_{\textrm{p},i}$ is the spectral peak energy and $ E_{\textrm{iso},i}$ is the isotropic equivalent radiated energy of the $i^{\textrm{th}}$ GRB, and $s_i$ denotes the associated error, 
\begin{equation}
s_i^2 = \delta^2 + \left( \frac{\sigma_{S_{\textrm{bolo},i}}}{S_{\textrm{bolo},i} \ln (10)} \right)^2 + \beta^2 \left( \frac{\sigma_{E_{\textrm{p},i}}} {E_{\textrm{p},i} \ln (10)}\right)^2. 
\end{equation}
The data set \cite{Cao:2024vmo} provides redshifts, $z_i$, $E_{\textrm{p},i}$ and its error $\sigma_{E_{\textrm{p},i}}$, along with the measured bolometric fluence $S_{\textrm{bolo}, i}$ and its error $\sigma_{S_{\textrm{bolo}, i}}$. The $\Lambda$CDM cosmological parameters ($H_0, \Omega_m$) enter through the luminosity distance (\ref{eq:DL}), which together with $S_{\textrm{bolo}, i}$ allows one to reconstruct $E_{\textrm{iso}, i}:$
\begin{equation}
    E_{\textrm{iso},i} = 4 \pi D_L(z_i)^2 S_{\textrm{bolo},i} (1+z_i)^{-1}. 
\end{equation}
This leaves us with 5 parameters, two of which are cosmological, and another 3 nuisance parameters $(\beta, \gamma, \delta)$. Once again $H_0$ is degenerate with $\beta$, so we set $H_0 = 70$ km/s/Mpc to remove this degeneracy and this leaves 4 parameters to be constrained. Our first check of the data set, especially the $S_{\textrm{bolo},i}$ entries, is to adopt Planck values $(H_0, \Omega_m) = (67.34, 0.315)$ and confirm that one recovers Table 1 of \cite{Jia:2022mwb}. Having performed this step, we move onto constructing the profile likelihood ratios. 

Given that our GRB data set comprises a factor of 10 fewer observables compared to the QSO data, yet spans a comparable redshift range, we expect larger errors. Translated into confidence intervals, one then expects broader profile likelihoods, which necessitates relaxing our uniform prior on $\Omega_m$ to $\Omega_m \in [0, 10]$. We next break this interval up into $300$ evenly spaced values of $\Omega_m$ and for each value of $\Omega_m$ we maximise the log-likelihood (\ref{eq:L2}) with respect to $(\beta, \gamma, \delta)$. Throughout, we make sure that the values of $(\beta, \gamma, \delta)$ that maximise (\ref{eq:L2}) are not impacted by our bounds on those parameters. For each value of $\Omega_m$ we record the maximum value of the likelihood $\mathcal{L}_{\textrm{max}}(\Omega_m)$, determine the global maximum of the likelihood $\mathcal{L}_{\textrm{max}}$ in the range $\Omega_m \in [0, 10]$, and construct the profile likelihood ratio for $R(\Omega_m)$ in equation (\ref{eq:R}). 

The result of the exercise is shown in Fig. \ref{fig:GRB_OM_ZEFF}, where we have constructed the profile likelihood ratios $R(\Omega_m)$ in bins of increasing effective redshift. Once again, similar to Fig. \ref{fig:QSO_OM_ZEFF}, we see that as the effective redshift of the sample increases, the peak of the likelihood shifts to larger $\Omega_m$ values. However, in contrast to Fig. \ref{fig:QSO_OM_ZEFF}, the profile likelihoods are broader so the shifts in the peak are less pronounced. That being said, inspection of the bottom left corner of the plot confirms that smaller $\Omega_m$ values, including the canonical Planck $\Omega_m \sim 0.3$ value, become steadily more disfavoured as the effective redshift of the sample increases. In Table \ref{tab:GRBvsZ} we provide an estimate of the $68 \%$ confidence intervals, {where it is also obvious that all profile likelihood peaks inhabit the $\Omega_m > 1$ regime corresponding to negative dark energy density.} We do this by integrating under the curve following equations (\ref{eq:w}) and (\ref{eq:conf}), and in addition we document values of $\Omega_m$ where $\Delta \chi^2 \leq 1$. Since the curves do not exceed $\Delta \chi^2 =1$, corresponding to $R(\Omega_m)$ values below $R(\Omega_m) = 0.607$, we are unable to determine upper bounds. It is evident that there is a considerable difference in the results, but this is expected, since the profile likelihoods are far from Gaussian, so this is a reasonable outcome. Nevertheless, the main takeaway message is that the frequentist confidence intervals for $\Omega_m$ shift to larger values with increasing effective redshift and this can be seen from either methodology for frequentist confidence intervals. This trend continues through to $0 < z \leq 1.8$, but since the peak of the profile likelihood shifts back to smaller $\Omega_m$ values for the full sample, it is clear that the trend is not universal and it only persists at lower redshifts $z \lesssim 2$. We remind the reader that $\Omega_m$ is theoretically a constant, so its variation with effective redshift in the GRB sample \cite{Cao:2024vmo} represents an inconsistency between the data set and the $\Lambda$CDM model.      

\begin{table}[htb]
\centering 
\begin{tabular}{c|c|c|c}
 \rule{0pt}{3ex} Redshift & \# GRBs & $ \Omega_{m}$ ($68 \%$) & $\Omega_m$ ($\Delta \chi^2 \leq 1$) \\
\hline 
\rule{0pt}{3ex} $0 < z \leq 0.7$ & $32$ & $3.2^{+4.4}_{-1.8}$ & $0.6 <$ $(3.2)$ \\
\hline
\rule{0pt}{3ex} $0 < z \leq 1.1$  & $57$ & $4.8^{+4.0}_{-2.1}$ & $1.1<$ $(4.8)$ \\
\hline
\rule{0pt}{3ex} $0 < z \leq 1.4$ & $90$ & $6.1^{+3.6}_{-2.1}$ & $1.9<$ ($6.1$) \\
\hline
\rule{0pt}{3ex} $ 0 < z \leq 1.8$ & $122$ & $5.1<$ ($10$) & $3.3<$ ($10$)  \\
\hline
\rule{0pt}{3ex} $ 0 < z \leq 8.2$ & $220$ & $5.0^{+3.6}_{-1.9}$ & $2.1<$ ($5.0$)  \\
\end{tabular}
\caption{The number of GRBs in redshift bins with $68\%$ confidence intervals for $\Omega_m$ from the 220 GRB sample \cite{Cao:2024vmo}. $\Omega_m$ best fits and confidence intervals increase with effective redshift. In the absence of an upper bound we present the profile likelihood peak in brackets. }
\label{tab:GRBvsZ}
\end{table}


We now attempt to assess the inconsistency between the full data set and the Planck $\Omega_m$ value. The profile likelihood ratio for the full sample appears as the cyan curve in Fig. \ref{fig:GRB_OM_ZEFF}. It is clear that small values of $\Omega_m$ are unlikely. To ascertain how unlikely is the Planck $1 \sigma$ upper bound $\Omega_m \leq 0.322$ \cite{Planck:2018vyg}, we identify the $R(\Omega_m)$ values. We find $R(\Omega_m) \leq 0.00052$. Translated into a difference in $\Delta \chi^2$ in  (\ref{eq:R}), we ascertain that the Planck values correspond to $\Delta \chi^2 \geq 15.1$. Given the GRB data, this excludes the Planck value at $99.99 \%$ confidence level corresponding to $\sim 3.9 \sigma$. We can get another perspective on this number by employing our methodology from (\ref{eq:w}) and (\ref{eq:conf}), where the reader should bear in mind that we have cut off the profile likelihood at $\Omega_m = 10$. We find that the Planck value is disfavoured at $99.9994 \%$ confidence interval corresponding to $\sim 4.5 \sigma$. The large difference between $\sim 3.9 \sigma$ and $\sim 4.5 \sigma$ can be attributed to the fact that the profile likelihood is far from Gaussian. The key point here is that irrespective of the methodology, \textit{provided one resorts to frequentist profile likelihood ratios}, there is a strong tension $> 3 \sigma$ between the 220 GRB data set \cite{Cao:2024vmo} and the Planck-$\Lambda$CDM model. In the appendix we confirm a $>4 \sigma$ tension with complementary Bayesian methods.

\begin{figure}[htb]
   \centering
\includegraphics[width=90mm]{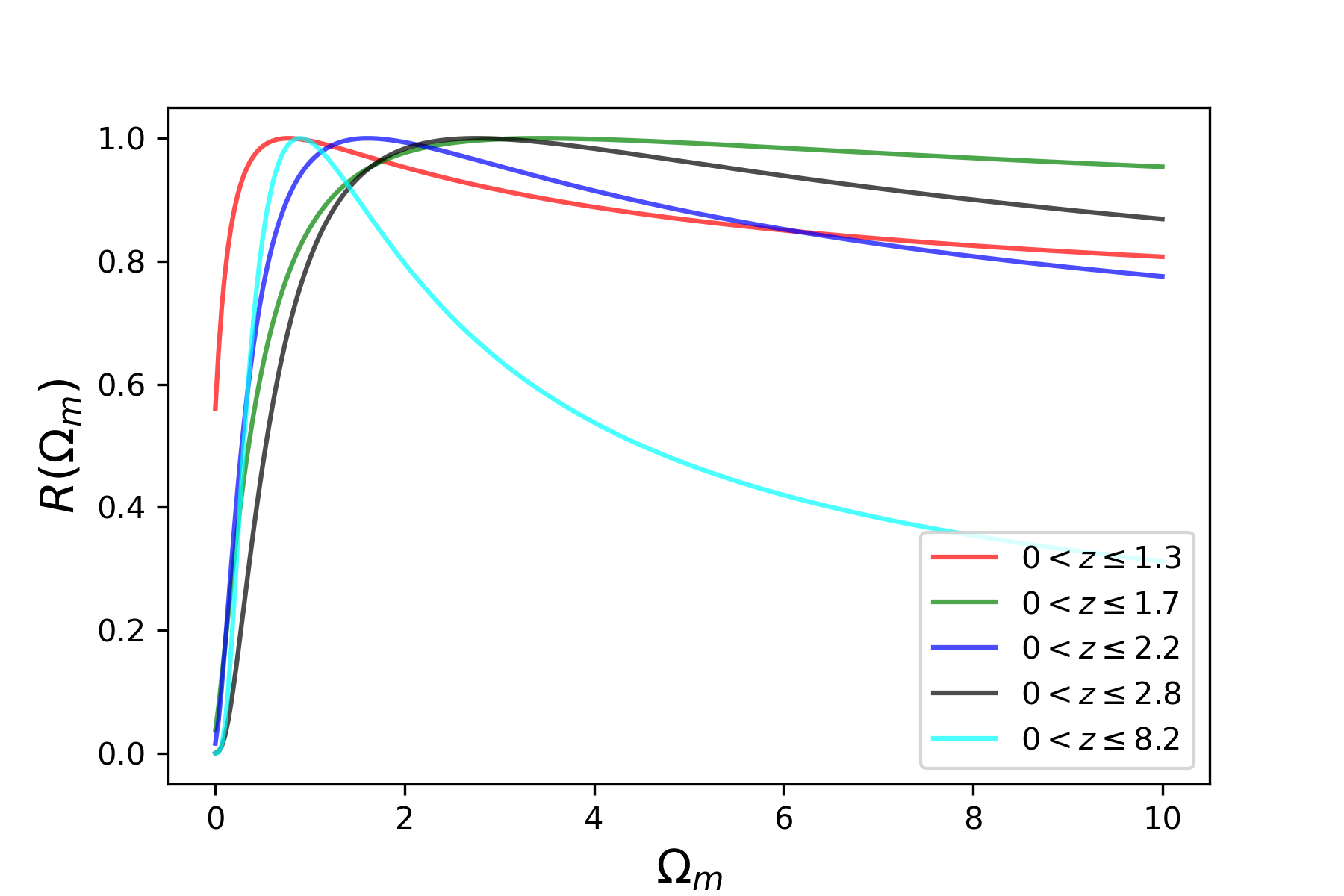} 
\caption{Variation of profile likelihoods $R(\Omega_m)$ with increasing effective redshift of the 118 GRB sample \cite{Khadka:2021vqa}. While the data prefers large $\Omega_m \gtrsim 1$ values, in contrast to Fig. \ref{fig:QSO_OM_ZEFF} and Fig. \ref{fig:GRB_OM_ZEFF}, $\Omega_m$ does not vary with effective redshift.}
\label{fig:GRB_118_OM_ZEFF}
\end{figure}

One puzzling outcome of the analysis is that even when we restrict GRBs to redshift ranges where Type Ia SNe and QSOs (see Fig. \ref{fig:QSO_OM_ZEFF}) lead to profile likelihoods peaked close to the Planck value $\Omega_m \sim 0.3$, in the sample of 220 GRBs studies in \cite{Cao:2024vmo} this is not the case. Thus, it is interesting to change the GRB sample, and repeat the exercise. To that end, we focus on the smaller sample of 118 GRBs from  \cite{Khadka:2021vqa}. The price one pays is halving the size of the GRB sample, while marginally increasing the intrinsic dispersion parameter $\delta$, a measure of the scatter in the sample, from $\delta \sim 0.38$ to $\delta \sim 0.4$.  

In Table \ref{tab:GRB_118vsZ} and Fig. \ref{fig:GRB_118_OM_ZEFF} we show the results of the same exercise with the 118 GRB sample \cite{Khadka:2021vqa}. Despite halving the number of GRBs, once again we see that the data set has a preference for larger $\Omega_m \gtrsim 1$ values. However, in contrast to the QSO and 220 GRB samples, we no longer see a definite trend of increasing $\Omega_m$ with increasing effective redshift. However, one similarity with the 220 GRB sample in Fig. \ref{fig:GRB_OM_ZEFF} is that including the high redshift GRBs $z \gtrsim 2-3$ pulls the $\Omega_m$ value back to lower values. This similarity is expected as GRB samples presumably possess a considerable number of high redshift GRBs in common. Thus, as is evident from Table \ref{tab:GRB_118vsZ}, GRB subsamples at intermediate redshifts show the greatest discrepancy with the Planck $\Omega_m$ value, however across the full sample, any disagreement is lower. Finally, we remark that frequentist confidence intervals based on complementary techniques lead to wildly different results, but where the profile likelihood is most Gaussian, namely for the full sample, we see that the confidence intervals show the best agreement. However, once again it is worth noting that the methodology based on  (\ref{eq:w}) and (\ref{eq:conf}) leads to larger confidence intervals and thus more conservative results.   

\begin{table}[htb]
\centering 
\begin{tabular}{c|c|c|c}
 \rule{0pt}{3ex} Redshift & \# GRBs & $ \Omega_{m}$ ($68 \%$) & $\Omega_m$ ($\Delta \chi^2 \leq 1$) \\
\hline 
\rule{0pt}{3ex} $0 < z \leq 1.3$ & $19$ & $0.8^{+6.0}_{-0.6}$ & $0.0<$ ($0.8$) \\
\hline
\rule{0pt}{3ex} $0 < z \leq 1.7$  & $39$ & $3.4^{+4.8}_{-1.6}$ & $0.5<$ ($3.4$) \\
\hline
\rule{0pt}{3ex} $0 < z \leq 2.2$ & $60$ & $1.6^{+5.6}_{-1.0}$ & $0.4<$ ($1.6$) \\
\hline
\rule{0pt}{3ex} $ 0 < z \leq 2.8$ & $82$ & $2.8^{+4.9}_{-1.4}$ & $0.7<$ ($2.8$)  \\
\hline 
\rule{0pt}{3ex} $ 0 < z \leq 8.2$ & $118$ & $0.9^{+4.6}_{-0.6}$ & $0.9^{+2.3}_{-0.5}$ \\
\end{tabular}
\caption{The number of GRBs in redshift bins with $68\%$ confidence intervals for $\Omega_m$ from the 118 GRB sample \cite{Khadka:2021vqa}. $\Omega_m$ best fits and confidence intervals increase with effective redshift. In the absence of an upper bound we present the profile likelihood peak in brackets. }
\label{tab:GRB_118vsZ}
\end{table}


Before leaving this section, it is good to drill down on the discrepancy between the full 118 GRB sample \cite{Khadka:2021vqa} and the $1 \sigma$ Planck upper bound $\Omega_m \leq 0.322$ \cite{Planck:2018vyg}. We find that the resulting $R(\Omega_m)$ values are $R(\Omega_m) \leq 0.502$ corresponding to $\Delta \chi^2 \geq 1.38$, which disfavours the Planck value at $76\%$ confidence level or $\sim 1.2 \sigma$. In contrast, using our methodology from  (\ref{eq:w}) and (\ref{eq:conf}), the $1 \sigma$ Planck $\Omega_m$ upper bound appears at $58 \%$ confidence level or $\sim 0.8 \sigma$. In contrast to the larger 220 GRB sample \cite{Cao:2024vmo}, where a significant $\gtrsim 4 \sigma$ tension arises with Planck, in the smaller 118 GRB sample, despite large $\Omega_m$ values being favoured, the disagreement with Planck is negligible at the $\sim 1 \sigma$ level.

\section{Conclusions}
In this paper we have explored the synergies between existing JWST \cite{10.1093/mnras/stac3347, Labbe:2022ahb, Castellano:2022ikm, Finkelstein:2023, 2023MNRAS.519.1201A, 2023MNRAS.518.6011D, 2022ApJ...940L..14N, 2023ApJ...942L...9Y} and QSO \cite{Risaliti:2018reu, Lusso:2019akb} anomalies. Interpreted as a problem of cosmological origin, the former anomaly can be alleviated if matter density $\Omega_m$ \cite{Forconi:2023hsj} or physical matter density $\Omega_m h^2$ \cite{Boylan-Kolchin:2022kae, McGaugh:2023nkc,  Liu:2024yan, Forconi:2023hsj} are larger than Planck values at higher redshifts. Since $\Omega_m$ and $h$ must be constant if the $\Lambda$CDM model is correct, this is a conflict between the model and observation. We have independently confirmed these observations in this paper in section \ref{sec:JWST}. On the other hand, the Risaliti-Lusso QSOs are discrepant with the Planck-$\Lambda$CDM model and this discrepancy is most transparent at higher redshifts, where the luminosity distance/distance modulus falls off relative to Planck \cite{Risaliti:2018reu}. Translated into the $\Lambda$CDM cosmology, as remarked in \cite{Yang:2019vgk, Velten:2019vwo}, this corresponds to a Universe with little or no dark energy, i. e. $\Omega_m \sim 1$. 

However, this may not be the full story. In \cite{Colgain:2022nlb} it was observed that $\Omega_m$ increases with effective redshift in the QSO sample. Importantly, as we have seen in Fig. \ref{fig:QSO_OM_ZEFF}, QSOs recover the Planck value $\Omega_m \sim 0.3$ at lower redshifts. In this paper we have revisited the observation with profile likelihoods and confirmed that the $68 \%$ confidence intervals are shifted to larger $\Omega_m$ values as one increases the effective redshift of the sample. {Moreover, profile likelihood peaks in $\Omega_m > 1$ parameter space are consistent with negative dark energy density.} Since $\Omega_m$ \textit{theoretically} must be a constant in the $\Lambda$CDM model if matter is pressureless, which incidentally is an assumption that Bayesian cosmologists need never question, \textit{observationally} $\Omega_m$ need not be a constant. In particular, a non-constant $\Omega_m$ is either a problem with the $\Lambda$CDM model or the data set, here the Risaliti-Lusso QSO data set \cite{Lusso:2020pdb}. If the problem is on the data set side, this adds credence to claims that QSOs are not standardisable \cite{Khadka:2020vlh, Khadka:2020tlm, Khadka:2021xcc, Khadka:2022aeg, Singal:2022nto, Petrosian:2022tlp, Zajacek:2023qjm}. 

In section \ref{sec:methods} we explained the limitations with the existing frequentist confidence interval literature  and highlighted new methodology \cite{Gomez-Valent:2022hkb, Colgain:2023bge} (see also \cite{Herold:2021ksg}). Where the profile likelihoods were better constrained, we observed that the new methodology resulted in larger, more conservative confidence intervals than more established methodology that approximates profile likelihoods as Gaussian distributions, most recently \cite{Holm:2023laa, holm2023prospect}. It should be obvious that given the assumptions in Wilks' theorem \cite{Wilks} (recall  (\ref{eq:PL})) that both methodologies show best agreement when profile likelihoods are closest to Gaussian.     

However, given the question marks over QSOs as standardisable candles \cite{Khadka:2020vlh, Khadka:2020tlm, Khadka:2021xcc, Khadka:2022aeg, Singal:2022nto, Petrosian:2022tlp, Zajacek:2023qjm}, we analysed independent GRB samples \cite{Cao:2024vmo, Khadka:2021vqa}. For both samples, we find larger values of $\Omega_m$ than Planck expectations. Observations that GRBs prefer larger $\Omega_m$ values than Planck are widespread in the GRB literature \cite{Demianski:2016zxi, Demianski:2016dsa, Luongo:2020hyk, Khadka:2021vqa, Alfano:2024ukk}. Moreover, we find that the larger 220 GRB sample \cite{Cao:2024vmo} exhibits an increasing $\Omega_m$ trend with effective redshift, which is in sync with our QSO observation, but there is no hint of evolution of $\Omega_m$ in the smaller 118 GRB sample \cite{Khadka:2021vqa}. In addition, we found that both the QSO sample and 220 GRB sample showed a strong $\gtrsim 4 \sigma$ tension with the Planck $\Omega_m$ value with standard frequentist methods (see appendix for Bayesian methods). Given the disagreement in GRB samples, it is imperative to revisit our methodology and results with other GRB samples, e. g. \cite{wang2024cosmological}. Note, as with Type Ia SNe, \textit{uncalibrated} QSOs and GRBs can constrain $\Omega_m$, so the goal is to identify GRB samples good enough to recover the Planck $\Omega_m$ value, \footnote{One can recover the Planck value from QSOs over extensive redshift ranges \cite{Dainotti:2022rfz}, but one does so by correcting or editing the raw data to promote (\ref{eq:lum}) to a relation intrinsic to QSO by removing by ansatz correlations between the UV and X-ray luminosities and redshift.} especially at lower redshifts. In contrast to QSOs, even at lower redshifts, neither of the GRB samples considered in this work succeed in convincingly recovering the Planck value without resorting to large errors. 

In the big picture, Planck data constrains the angular scale of the sound horizon at last scattering $\theta_* = r_*/D_{M}(z_*)$ almost model independently to high precision, where the scale $r_*$ depends on physics in the early Universe and $z_* \approx 1090$. A key point here is that $D_{M}(z_*) := c \int_0^{z_*} 1/H(z) \dd z$ is a weighted sum that attributes much greater weight to lower redshifts in the matter dominated regime where $H(z)$ is smaller ($1/H(z)$ is larger). In contrast, $r_{*} := \int_{z_*}^{\infty} c_s(z)/H(z) \dd z$, where $c_s(z)$ is the speed of sound in the plasma, is also a weighted sum, but it is dominated by assumptions in the radiation sector. What this means in practice is that deviations at higher redshifts in the matter dominated regime from Planck behaviour are poorly constrained by $\theta_*$. Admittedly, JWST observations, QSOs and GRBs, while they are relevent high redshift observables, they come with considerable uncertainties. For this reason, it is prudent to turn our attention to Type Ia SNe, where analogous hints of larger $\Omega_m$ values at higher redshifts exist in the literature \cite{Colgain:2022nlb, Colgain:2022rxy, Pasten:2023rpc, Malekjani:2023dky}. Interestingly, two recent SNe sample have favoured values of $\Omega_m$ larger than Planck \cite{Rubin:2023ovl, DES:2024jxu}, and it is evident that the DES SNe sample \cite{DES:2024jxu} has a high effective redshift. A tomographic analysis of the DES sample reveals that $\Omega_m$ increases with effective redshift \cite{Colgain:2024ksa}.

{As a final comment, we return to $H_0$ and $S_8$ tensions. $H_0$ is anti-correlated with $\Omega_m$ in the flat $\Lambda$CDM model, so if $\Omega_m$ is not a constant, then neither is $H_0$ (see Fig. 3 of \cite{Colgain:2024ksa} for visual confirmation in DES SNe). Moreover, $S_8 := \sigma_8 \sqrt{\Omega_m/0.3}$ depends on both the amplitude of matter fluctuations $\sigma_8$ and matter density $\Omega_m$. Recent DESI full-shape modelling results \cite{DESI:2024hhd, DESI:2024jis} show enough of an increase in $\sigma_8$ in tomographic redshift bins between the lowest and highest redshift bin at statistical significance $2.2 \sigma$ that one may question the constancy of $\sigma_8$. See \cite{Adil:2023jtu, Akarsu:2024hsu} for how these results translate to $S_8$ and may generalise to other observables. Nevertheless, DESI full-shape modelling \cite{DESI:2024jis} shows only a $0.8 \sigma$ increase in $\Omega_m$ between the lowest and highest redshift bin. The key point is that $S_8$ tension may in part be traced to the $\Omega_m$ discrepancies highlighted here. The increasing $\Omega_m$ shift in DESI full-shape modelling is consistent with observations in QSOs/GRBs/SNe, but we will need to see a reduction in errors to confirm if the signal is genuine or not.}

\begin{acknowledgments}
We would like to thank Stephen Appleby, Giacomo Galloni, Adri\`a G\'omez-Valent, Elisabeta Lusso and Saeed Pourojaghi for discussions on JWST anomalies, profile likelihoods, QSOs, GRBs, etc. We thank Orlando Luongo, Marco Muccino and Bharat Ratra for comments on a preliminary draft. We thank Dominik Schwarz, Yvonne Wong and an anonymous referee for inspiring section \ref{sec:methods} on frequentist confidence intervals. This article/publication is based upon work from COST Action CA21136 – “Addressing observational tensions in cosmology with systematics and fundamental physics (CosmoVerse)”, supported by COST (European Cooperation in Science and Technology). 
LY is supported by an appointment to the YST Program at the APCTP through the Science and Technology Promotion Fund and Lottery Fund of the Korean Government. 
\end{acknowledgments}

\appendix

\begin{figure}[htb]
   \centering
\includegraphics[width=80mm]{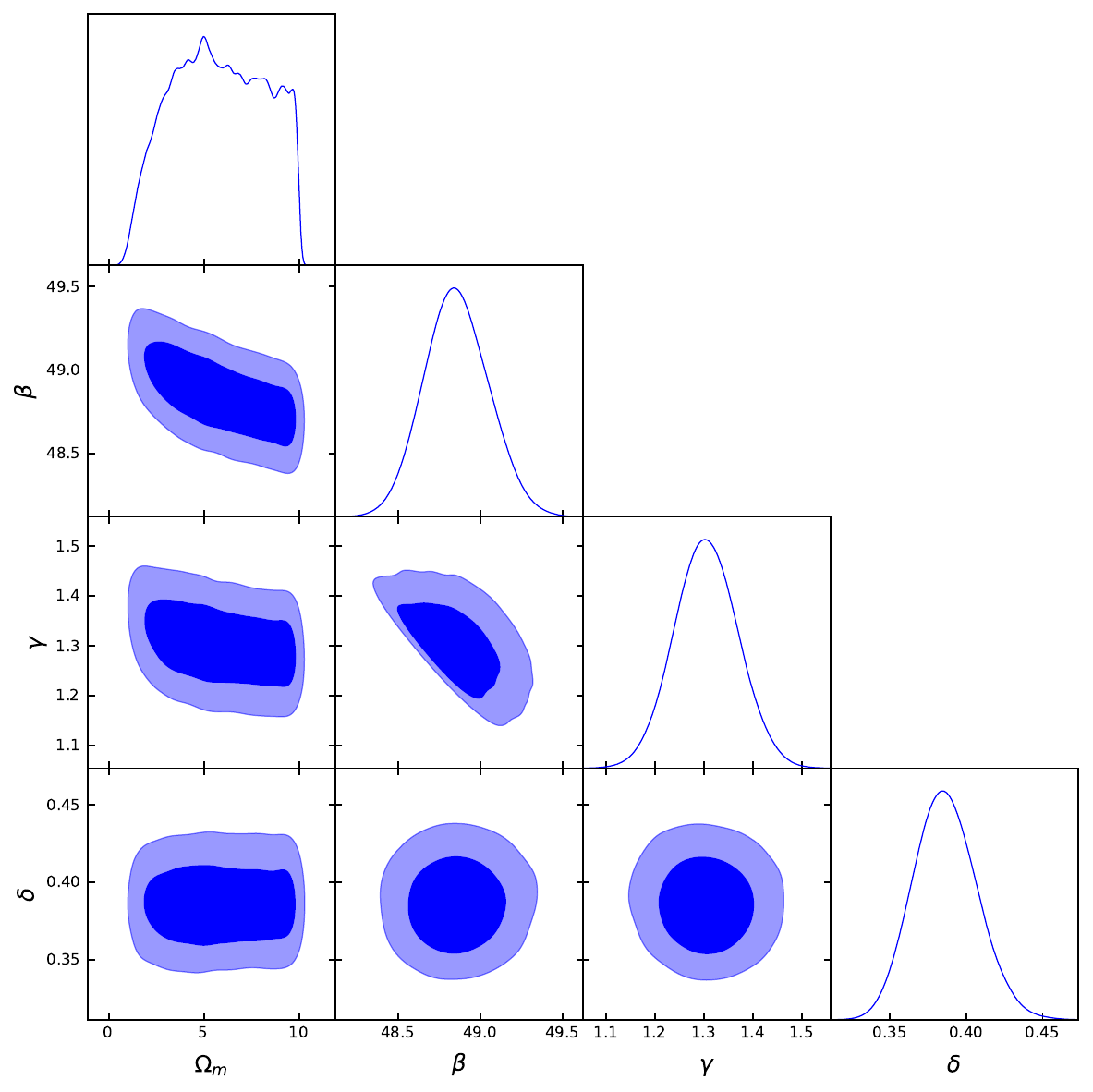} 
\caption{MCMC posteriors for the 220 GRB data set \cite{Cao:2024vmo} fitted to the $\Lambda$CDM model with $H_0$ fixed to $H_0 = 70$ km/s/Mpc to break a degeneracy.}
\label{fig:GRB_MCMC_4sigma}
\end{figure}

\section{MCMC confirmation of GRB tension}
The analysis in the text has led to strong $\gtrsim 4 \sigma$ tensions between GRB and QSO samples with Planck CMB data on the assumption that the $\Lambda$CDM model is correct. In this section, we confirm that one would arrive at the same conclusions with MCMC analysis. In Fig. \ref{fig:GRB_MCMC_4sigma} we show the MCMC posteriors for the data set of 220 GRBs \cite{Cao:2024vmo}, where we have made use of \textit{emcee} \cite{Foreman-Mackey:2012any} and \textit{GetDist} \cite{Lewis:2019xzd}. The $\Omega_m$ posterior should be compared with the cyan curve from Fig. \ref{fig:GRB_OM_ZEFF}. Both plots demonstrate a peak at $\Omega_m \sim 5$ with a sharp fall off towards smaller $\Omega_m$ values and a gradual fall off towards larger $\Omega_m$ values. Unfortunately, the tension with the Planck value is unclear from the corner plot. 

\begin{figure}[htb]
   \centering
\includegraphics[width=80mm]{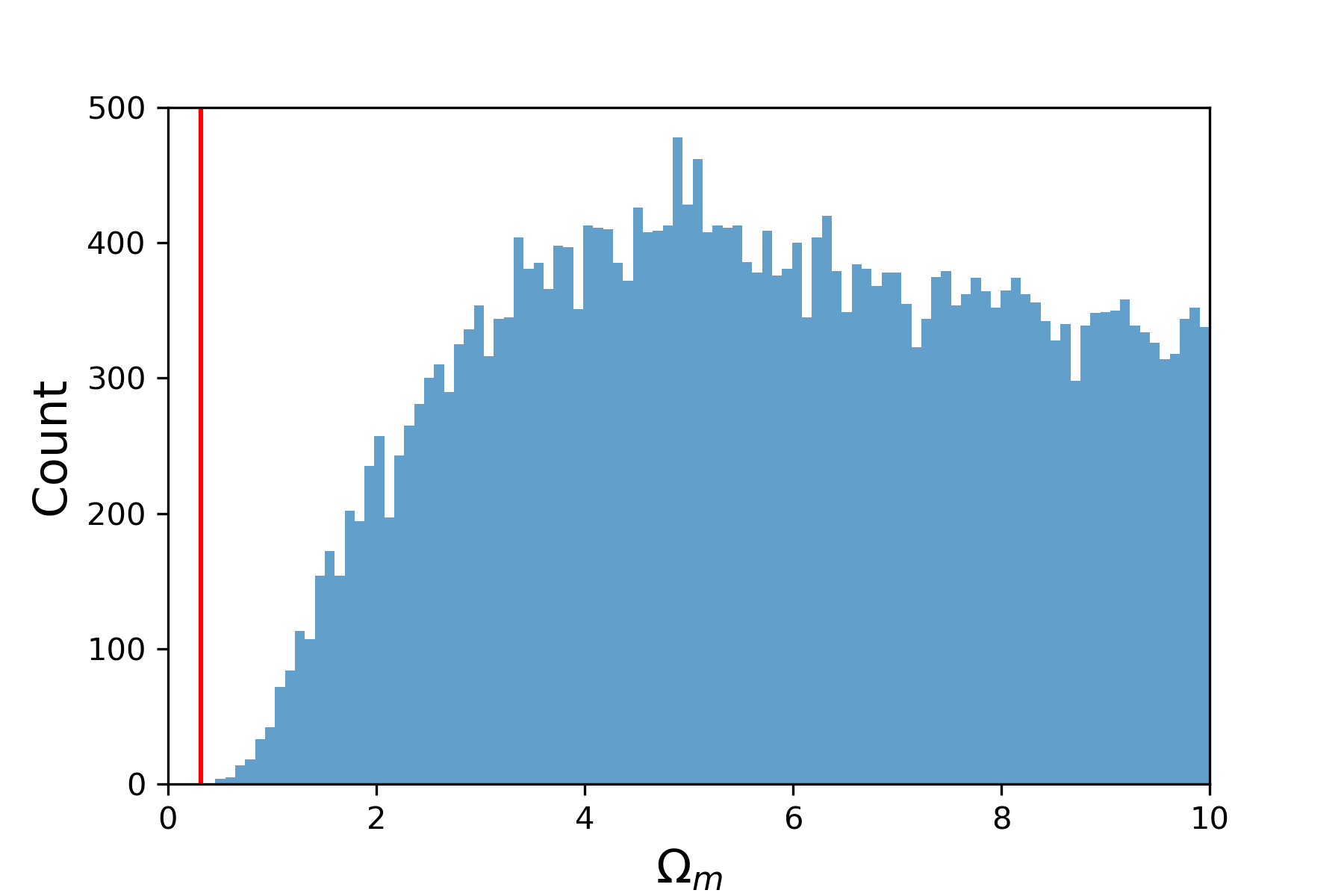} 
\caption{A histogram of $31,980$ $\Omega_m$ values from the MCMC hain for the 220 GRB data set \cite{Cao:2024vmo} fitted to the $\Lambda$CDM model with $H_0$ fixed to $H_0 = 70$ km/s/Mpc. The Planck $68 \%$ confidence interval is shown in red.}
\label{fig:GRB_MCMC_4sigma_hist}
\end{figure}

In Fig. \ref{fig:GRB_MCMC_4sigma_hist} we plot a histogram of the 31,980 $\Omega_m$ values from the MCMC chain, which confirms that we encounter no values of $\Omega_m$ smaller than the $1 \sigma$ upper bound on the Planck value $\Omega_m \leq 0.322$ highlighted in red. The lowest value we find is $\Omega_m \sim 0.455$. Removing non-unique points in ($\Omega_m, \beta, \gamma, \delta)$ parameter space from the MCMC chain, the number of unique configurations is $31,933$. Thus, the probability $p$ of getting a value of $\Omega_m$ consistent with Planck within $1 \sigma$ is conservatively less than $p < 1/31933$, which points to a tension between the GRB data set and Planck data that is bounded below at the $99.997 \%$ confidence level or $4 \sigma$. 

One could more accurately determine the tension from the MCMC chain, but it is clear that one needs to run a chain long enough to encounter the Planck value or its $1 \sigma$ upper bound. What is important here is that we arrived at $3.9 \sigma$ tension through a difference in the $\chi^2$ in section \ref{sec:GRB}, but it is clear from our MCMC analysis that this number is a lower bound. In principle, one could run a very long MCMC chain to see whether the tension as ascertained by MCMC is closer to the $4.5 \sigma$ we found with methodology based on equations (\ref{eq:w}) and (\ref{eq:conf}). This would require an MCMC chain approximately 10 times as long with $\sim 300,000$ entries. Given that the MCMC chain converges with $\sim 10,000$ entries, this represents a poor use of computation for relatively little gain. The main takeaway from this section is that the $3.9 \sigma$ tension between the GRB data set and Planck is recovered with Bayesian methodology based on MCMC, modulo the fact that long MCMC chains are required.   

\section{Derivation of chi-squared distribution PDF}
In this section we show that a chi-squared distribution PDF corresponding to one degree of freedom is easily derived from a standard normal distribution PDF. Let $Y$ be a random variable $Y=X^2$ where $X$ has a standard normal distribution, i. e. mean zero and variance one. Consider now the cumulative distribution function (CDF) for $X$: 
\begin{equation}
    F_{X}(x) = \int_{-\infty}^{x} \frac{1}{\sqrt{2 \pi}} e^{-\frac{1}{2}t^2 } \dd t, 
\end{equation}
which is the probability of $X \leq x$. For $y < 0$, the CDF for $Y$ is simply $F_{Y}(y) = P(Y \leq y) = 0$, where $P$ denotes probability, since $Y$ cannot be negative if $X \in \mathbb{R}$. We can then define $F_{Y}(y)$ for $y \geq 0$ as 
\begin{eqnarray}
 F_{Y}(y) &=& P(Y \leq y) = P(X^2 \leq y) = P(-\sqrt{y} \leq X \leq \sqrt{y}) \nn
 &=& F_{X}(\sqrt{y}) - F_{X}(-\sqrt{y}) = F_{X}(\sqrt{y}) - (1-F_{X}(\sqrt{y})) \nn 
 &=& 2 F_{X}(\sqrt{y}) - 1. 
\end{eqnarray}
Then using the fundamental theorem of calculus, it is easy to show that 
\begin{equation}
    f_{Y}(y) = \frac{\dd \phantom{x}}{\dd y} F_{Y}(y) = 2 \frac{\dd \phantom{x}}{\dd y} F_{X}(\sqrt{y}) = \frac{1}{\sqrt{2 \pi y}} e^{-\frac{1}{2}y}, 
\end{equation}
which is the integrand in equation (\ref{eq:alpha}) of the text.

\section{Feldman-Cousins versus Wilks' confidence intervals}

In this section we focus on Gaussian profile likelihoods and make a comparison between the confidence intervals inferred from Wilks' theorem \cite{Wilks} and the Feldman-Cousins prescription \cite{Feldman:1997qc}. We observe that in the large sample limit, Wilks' theorem concerns Gaussian profile likelihoods (see section \ref{sec:methods}). Here we do not have to do much work as we can simply import results from Table X of \cite{Feldman:1997qc}, where we suppress rows of the table by omitting measurements $x_0$ that do not contribute to the points being made. We focus on $68\%$ confidence intervals, corresponding to $\Delta \chi^2 = 1$ from Wilks' theorem, but the lessons learned apply more generally. The setting is a Gaussian profile likelihood with mean parameter $\mu$ and standard deviation $\sigma$ where the mean $\mu$ must be non-negative $\mu \geq 0$. In the absence of the boundary at $\mu = 0$, $\mu = x_0$ corresponds to the peak of the profile likelihood and the $68\%$ confidence intervals are $x_0 - \sigma \leq \mu \leq x_0 + \sigma$. In other words, without a boundary, the $\mu$ profile likelihood is peaked on the measurement $x_0$ and the confidence intervals terminate at $\pm 1 \sigma$. In Table \ref{tab:FC_vs_Wilks} we show the confidence intervals in units of $\sigma$.

\begin{table}[htb]
\centering 
\begin{tabular}{c|c|c}
 \rule{0pt}{3ex} $x_0$ & Feldman-Cousins &  Wilks $(\Delta \chi^2 =1)$ \\
 \hline
\rule{0pt}{3ex} $-3$ & $(0, 0.04)$ & $(0,0.16)$ \\
\rule{0pt}{3ex} $-2.5$ & $(0, 0.05)$ & $(0,0.19)$ \\
\rule{0pt}{3ex} $-2$ & $(0, 0.07)$ & $(0,0.24)$ \\
\rule{0pt}{3ex} $-1.5$ & $(0, 0.13)$ & $(0,0.30)$ \\
\rule{0pt}{3ex} $-1$ & $(0, 0.27)$ & $(0,0.41)$ \\
\rule{0pt}{3ex} $-0.5$ & $(0, 0.56)$ & $(0,0.62)$ \\
\rule{0pt}{3ex} $0$ & $(0, 1)$ & $(0,1)$ \\
\rule{0pt}{3ex} $0.5$ & $(0.02, 1.5)$ & $(0,1.5)$ \\ 
\rule{0pt}{3ex} $1$ & $(0.24, 2)$ & $(0,2)$ \\
\rule{0pt}{3ex} $1.5$ & $(0.56, 2.5)$ & $(0.5,2.5)$ \\
\rule{0pt}{3ex} $1.8$ & $(0.81, 2.8)$ & $(0.8,2.8)$ \\
\rule{0pt}{3ex} $2$ & $(1, 3)$ & $(1,3)$ \\
\rule{0pt}{3ex} $2.5$ & $(1.5, 3.5)$ & $(1.5,3.5)$ \\
\rule{0pt}{3ex} $3$ & $(2, 4)$ & $(2,4)$ \\
\end{tabular}
\caption{Comparison of $68\%$ confidence intervals from Feldman-Cousins and Wilks' theorem methods. The Feldman-Cousin entries are reproduced from Table X of \cite{Feldman:1997qc}.}
\label{tab:FC_vs_Wilks}
\end{table}

Starting from a measurement $x_0 \geq 2$, at least $2 \sigma$ removed from boundary at $\mu =0$, in line with expectations we see Feldman-Cousins and Wilks' methods agree. Furthermore, we see that the $68\%$ confidence intervals are $x_0 - 1 \leq \mu \leq x_0 + 1$ in units of $\sigma$. However, once the peak of the profile likelihood comes within $2 \sigma$ of the boundary, $x_0 = 1.8$, we start to see a difference between the two methods. This is interesting because the boundary is still outside of the naive $68\%$ confidence interval as defined by Wilks' $\Delta \chi^2 =1$. Only when the measurement becomes $x_0 = 1$ does the boundary at $\mu = 0$ start to impact the naive Wilks' confidence intervals. When this happens, we simply adopt $\mu = 0$ as the lower bound on the $68\%$ confidence interval as negative values of $\mu$ are not permitted. An interesting point here is that the Feldman-Cousins correction to the Wilks' $\Delta \chi^2 =1$ confidence intervals always makes the confidence intervals smaller in the range $0 < x_0 < 2$. This is also evident in the $1 \leq x_0 < 2$ range where despite the boundary being outside the $\Delta \chi^2 =1$ interval, it is clear that Feldman-Cousins narrows the confidence interval on the side of the profile likelihood facing the boundary.

Interestingly, the confidence interval on the opposite side of the profile likelihood away from the boundary is unaffected up to the point that the measurement coincides with the $\mu = 0$ boundary at $x_0 = 0$. As explained earlier, this corresponds to a Gaussian profile likelihood peaked at $\mu = 0$, where both the Feldman-Cousins and Wilks' methods agree that the upper bound on the $68\%$ confidence interval is at $1 \sigma$. Nevertheless, for measurements $x_0 < 0$, the peak of the Gaussian profile likelihood moves into the prohibited $\mu < 0$ regime. Once again, one has to make a choice with Wilks' theorem and the choice one could make is to simply interpret the boundary as the peak of the likelihood. This is not an unreasonable assumption as typically in the sum of the neutrino masses, a non-negative prior on the sum of the masses cuts off the peak, e. g. \cite{eBOSS:2020yzd, craig2024, naredotuero2024}. In Table \ref{tab:FC_vs_Wilks} the Wilks' confidence intervals for $x_0 < 0$ come from $\Delta \chi^2=1$ starting from the boundary at $\mu =0$. Since the profile likelihood is Gaussian by assumption, this is easily calculated by solving for $\mu$ in the following equation:
\begin{equation}
(\mu - x_0)^2 = (0-x_0)^2+1,  
\end{equation}
where on the right hand side we impose $\mu = 0$ and allow for $\Delta \chi^2 =1$. This equation gives the values for the Wilks' confidence interval upper bounds quoted in the table. Once again we witness the same feature, namely the Feldman-Cousins prescription leads to narrower confidence intervals.

Let us summarise the lessons learned for a Gaussian profile likelihood impacted by a boundary. First, the boundary leads to corrections to the naive Wilks' confidence intervals on the side of the profile likelihood facing the boundary, but does not affect the confidence intervals on the other side up to the point that the peak of the profile likelihood coincides with the boundary at $x_0 = 0$. The corrections become relevant when the boundary is less than $ 2 \sigma$ away. Once the likelihood profile peak passes through the boundary, the Feldman-Cousins prescription leads to corrections to the confidence intervals in the allowed parameter regime. Throughout, these corrections are such that they \textit{narrow} confidence intervals. The flip side of this is that if tensions are present, the Feldman-Cousins prescription will lead to larger tensions than methods based on Wilks' theorem. Our analysis here explains why the authors of ref. \cite{naredotuero2024} find that Wilks' theorem leads to larger errors than the Feldman-Cousins prescription. For Gaussian profile likelihoods this is a general result.

\bibliography{refs}

\begin{thebibliography}{139}%
\makeatletter
\providecommand \@ifxundefined [1]{%
 \@ifx{#1\undefined}
}%
\providecommand \@ifnum [1]{%
 \ifnum #1\expandafter \@firstoftwo
 \else \expandafter \@secondoftwo
 \fi
}%
\providecommand \@ifx [1]{%
 \ifx #1\expandafter \@firstoftwo
 \else \expandafter \@secondoftwo
 \fi
}%
\providecommand \natexlab [1]{#1}%
\providecommand \enquote  [1]{``#1''}%
\providecommand \bibnamefont  [1]{#1}%
\providecommand \bibfnamefont [1]{#1}%
\providecommand \citenamefont [1]{#1}%
\providecommand \href@noop [0]{\@secondoftwo}%
\providecommand \href [0]{\begingroup \@sanitize@url \@href}%
\providecommand \@href[1]{\@@startlink{#1}\@@href}%
\providecommand \@@href[1]{\endgroup#1\@@endlink}%
\providecommand \@sanitize@url [0]{\catcode `\\12\catcode `\$12\catcode `\&12\catcode `\#12\catcode `\^12\catcode `\_12\catcode `\%12\relax}%
\providecommand \@@startlink[1]{}%
\providecommand \@@endlink[0]{}%
\providecommand \url  [0]{\begingroup\@sanitize@url \@url }%
\providecommand \@url [1]{\endgroup\@href {#1}{\urlprefix }}%
\providecommand \urlprefix  [0]{URL }%
\providecommand \Eprint [0]{\href }%
\providecommand \doibase [0]{https://doi.org/}%
\providecommand \selectlanguage [0]{\@gobble}%
\providecommand \bibinfo  [0]{\@secondoftwo}%
\providecommand \bibfield  [0]{\@secondoftwo}%
\providecommand \translation [1]{[#1]}%
\providecommand \BibitemOpen [0]{}%
\providecommand \bibitemStop [0]{}%
\providecommand \bibitemNoStop [0]{.\EOS\space}%
\providecommand \EOS [0]{\spacefactor3000\relax}%
\providecommand \BibitemShut  [1]{\csname bibitem#1\endcsname}%
\let\auto@bib@innerbib\@empty
\bibitem [{\citenamefont {Adams}\ \emph {et~al.}(2022)\citenamefont {Adams}, \citenamefont {Conselice}, \citenamefont {Ferreira}, \citenamefont {Austin}, \citenamefont {Trussler}, \citenamefont {Juodžbalis}, \citenamefont {Wilkins}, \citenamefont {Caruana}, \citenamefont {Dayal}, \citenamefont {Verma},\ and\ \citenamefont {Vijayan}}]{10.1093/mnras/stac3347}%
  \BibitemOpen
  \bibfield  {author} {\bibinfo {author} {\bibfnamefont {N.~J.}\ \bibnamefont {Adams}}, \bibinfo {author} {\bibfnamefont {C.~J.}\ \bibnamefont {Conselice}}, \bibinfo {author} {\bibfnamefont {L.}~\bibnamefont {Ferreira}}, \bibinfo {author} {\bibfnamefont {D.}~\bibnamefont {Austin}}, \bibinfo {author} {\bibfnamefont {J.~A.~A.}\ \bibnamefont {Trussler}}, \bibinfo {author} {\bibfnamefont {I.}~\bibnamefont {Juodžbalis}}, \bibinfo {author} {\bibfnamefont {S.~M.}\ \bibnamefont {Wilkins}}, \bibinfo {author} {\bibfnamefont {J.}~\bibnamefont {Caruana}}, \bibinfo {author} {\bibfnamefont {P.}~\bibnamefont {Dayal}}, \bibinfo {author} {\bibfnamefont {A.}~\bibnamefont {Verma}},\ and\ \bibinfo {author} {\bibfnamefont {A.~P.}\ \bibnamefont {Vijayan}},\ }\bibfield  {title} {\bibinfo {title} {{Discovery and properties of ultra-high redshift galaxies ($9 \leq z \leq 12$) in the JWST ERO SMACS 0723 Field}},\ }\href {https://doi.org/10.1093/mnras/stac3347} {\bibfield  {journal} {\bibinfo  {journal} {Monthly Notices of the Royal
  Astronomical Society}\ }\textbf {\bibinfo {volume} {518}},\ \bibinfo {pages} {4755} (\bibinfo {year} {2022})},\ \Eprint {https://arxiv.org/abs/https://academic.oup.com/mnras/article-pdf/518/3/4755/47749914/stac3347.pdf} {https://academic.oup.com/mnras/article-pdf/518/3/4755/47749914/stac3347.pdf} \BibitemShut {NoStop}%
\bibitem [{\citenamefont {Labbe}\ \emph {et~al.}(2023)\citenamefont {Labbe} \emph {et~al.}}]{Labbe:2022ahb}%
  \BibitemOpen
  \bibfield  {author} {\bibinfo {author} {\bibfnamefont {I.}~\bibnamefont {Labbe}} \emph {et~al.},\ }\bibfield  {title} {\bibinfo {title} {{A population of red candidate massive galaxies \textasciitilde{}600 Myr after the Big Bang}},\ }\href {https://doi.org/10.1038/s41586-023-05786-2} {\bibfield  {journal} {\bibinfo  {journal} {Nature}\ }\textbf {\bibinfo {volume} {616}},\ \bibinfo {pages} {266} (\bibinfo {year} {2023})},\ \Eprint {https://arxiv.org/abs/2207.12446} {arXiv:2207.12446 [astro-ph.GA]} \BibitemShut {NoStop}%
\bibitem [{\citenamefont {Castellano}\ \emph {et~al.}(2023)\citenamefont {Castellano} \emph {et~al.}}]{Castellano:2022ikm}%
  \BibitemOpen
  \bibfield  {author} {\bibinfo {author} {\bibfnamefont {M.}~\bibnamefont {Castellano}} \emph {et~al.},\ }\bibfield  {title} {\bibinfo {title} {{Early Results from GLASS-JWST. XIX. A High Density of Bright Galaxies at z \ensuremath{\approx} 10 in the A2744 Region}},\ }\href {https://doi.org/10.3847/2041-8213/accea5} {\bibfield  {journal} {\bibinfo  {journal} {Astrophys. J. Lett.}\ }\textbf {\bibinfo {volume} {948}},\ \bibinfo {pages} {L14} (\bibinfo {year} {2023})},\ \Eprint {https://arxiv.org/abs/2212.06666} {arXiv:2212.06666 [astro-ph.GA]} \BibitemShut {NoStop}%
\bibitem [{\citenamefont {Finkelstein}\ \emph {et~al.}(2023)\citenamefont {Finkelstein}, \citenamefont {Bagley}, \citenamefont {Ferguson}, \citenamefont {Wilkins}, \citenamefont {Kartaltepe}, \citenamefont {Papovich}, \citenamefont {Yung}, \citenamefont {Arrabal~Haro}, \citenamefont {Behroozi}, \citenamefont {Dickinson}, \citenamefont {Kocevski}, \citenamefont {Koekemoer}, \citenamefont {Larson}, \citenamefont {Le~Bail}, \citenamefont {Morales}, \citenamefont {Pérez-González}, \citenamefont {Burgarella}, \citenamefont {Davé}, \citenamefont {Hirschmann}, \citenamefont {Somerville}, \citenamefont {Wuyts}, \citenamefont {Bromm}, \citenamefont {Casey}, \citenamefont {Fontana}, \citenamefont {Fujimoto}, \citenamefont {Gardner}, \citenamefont {Giavalisco}, \citenamefont {Grazian}, \citenamefont {Grogin}, \citenamefont {Hathi}, \citenamefont {Hutchison}, \citenamefont {Jha}, \citenamefont {Jogee}, \citenamefont {Kewley}, \citenamefont {Kirkpatrick}, \citenamefont {Long}, \citenamefont {Lotz}, \citenamefont
  {Pentericci}, \citenamefont {Pierel}, \citenamefont {Pirzkal}, \citenamefont {Ravindranath}, \citenamefont {Ryan}, \citenamefont {Trump}, \citenamefont {Yang}, \citenamefont {Bhatawdekar}, \citenamefont {Bisigello}, \citenamefont {Buat}, \citenamefont {Calabrò}, \citenamefont {Castellano}, \citenamefont {Cleri}, \citenamefont {Cooper}, \citenamefont {Croton}, \citenamefont {Daddi}, \citenamefont {Dekel}, \citenamefont {Elbaz}, \citenamefont {Franco}, \citenamefont {Gawiser}, \citenamefont {Holwerda}, \citenamefont {Huertas-Company}, \citenamefont {Jaskot}, \citenamefont {Leung}, \citenamefont {Lucas}, \citenamefont {Mobasher}, \citenamefont {Pandya}, \citenamefont {Tacchella}, \citenamefont {Weiner},\ and\ \citenamefont {Zavala}}]{Finkelstein:2023}%
  \BibitemOpen
  \bibfield  {author} {\bibinfo {author} {\bibfnamefont {S.~L.}\ \bibnamefont {Finkelstein}}, \bibinfo {author} {\bibfnamefont {M.~B.}\ \bibnamefont {Bagley}}, \bibinfo {author} {\bibfnamefont {H.~C.}\ \bibnamefont {Ferguson}}, \bibinfo {author} {\bibfnamefont {S.~M.}\ \bibnamefont {Wilkins}}, \bibinfo {author} {\bibfnamefont {J.~S.}\ \bibnamefont {Kartaltepe}}, \bibinfo {author} {\bibfnamefont {C.}~\bibnamefont {Papovich}}, \bibinfo {author} {\bibfnamefont {L.~Y.~A.}\ \bibnamefont {Yung}}, \bibinfo {author} {\bibfnamefont {P.}~\bibnamefont {Arrabal~Haro}}, \bibinfo {author} {\bibfnamefont {P.}~\bibnamefont {Behroozi}}, \bibinfo {author} {\bibfnamefont {M.}~\bibnamefont {Dickinson}}, \bibinfo {author} {\bibfnamefont {D.~D.}\ \bibnamefont {Kocevski}}, \bibinfo {author} {\bibfnamefont {A.~M.}\ \bibnamefont {Koekemoer}}, \bibinfo {author} {\bibfnamefont {R.~L.}\ \bibnamefont {Larson}}, \bibinfo {author} {\bibfnamefont {A.}~\bibnamefont {Le~Bail}}, \bibinfo {author} {\bibfnamefont {A.~M.}\ \bibnamefont
  {Morales}}, \bibinfo {author} {\bibfnamefont {P.~G.}\ \bibnamefont {Pérez-González}}, \bibinfo {author} {\bibfnamefont {D.}~\bibnamefont {Burgarella}}, \bibinfo {author} {\bibfnamefont {R.}~\bibnamefont {Davé}}, \bibinfo {author} {\bibfnamefont {M.}~\bibnamefont {Hirschmann}}, \bibinfo {author} {\bibfnamefont {R.~S.}\ \bibnamefont {Somerville}}, \bibinfo {author} {\bibfnamefont {S.}~\bibnamefont {Wuyts}}, \bibinfo {author} {\bibfnamefont {V.}~\bibnamefont {Bromm}}, \bibinfo {author} {\bibfnamefont {C.~M.}\ \bibnamefont {Casey}}, \bibinfo {author} {\bibfnamefont {A.}~\bibnamefont {Fontana}}, \bibinfo {author} {\bibfnamefont {S.}~\bibnamefont {Fujimoto}}, \bibinfo {author} {\bibfnamefont {J.~P.}\ \bibnamefont {Gardner}}, \bibinfo {author} {\bibfnamefont {M.}~\bibnamefont {Giavalisco}}, \bibinfo {author} {\bibfnamefont {A.}~\bibnamefont {Grazian}}, \bibinfo {author} {\bibfnamefont {N.~A.}\ \bibnamefont {Grogin}}, \bibinfo {author} {\bibfnamefont {N.~P.}\ \bibnamefont {Hathi}}, \bibinfo {author}
  {\bibfnamefont {T.~A.}\ \bibnamefont {Hutchison}}, \bibinfo {author} {\bibfnamefont {S.~W.}\ \bibnamefont {Jha}}, \bibinfo {author} {\bibfnamefont {S.}~\bibnamefont {Jogee}}, \bibinfo {author} {\bibfnamefont {L.~J.}\ \bibnamefont {Kewley}}, \bibinfo {author} {\bibfnamefont {A.}~\bibnamefont {Kirkpatrick}}, \bibinfo {author} {\bibfnamefont {A.~S.}\ \bibnamefont {Long}}, \bibinfo {author} {\bibfnamefont {J.~M.}\ \bibnamefont {Lotz}}, \bibinfo {author} {\bibfnamefont {L.}~\bibnamefont {Pentericci}}, \bibinfo {author} {\bibfnamefont {J.~D.~R.}\ \bibnamefont {Pierel}}, \bibinfo {author} {\bibfnamefont {N.}~\bibnamefont {Pirzkal}}, \bibinfo {author} {\bibfnamefont {S.}~\bibnamefont {Ravindranath}}, \bibinfo {author} {\bibfnamefont {R.~E.}\ \bibnamefont {Ryan}}, \bibinfo {author} {\bibfnamefont {J.~R.}\ \bibnamefont {Trump}}, \bibinfo {author} {\bibfnamefont {G.}~\bibnamefont {Yang}}, \bibinfo {author} {\bibfnamefont {R.}~\bibnamefont {Bhatawdekar}}, \bibinfo {author} {\bibfnamefont {L.}~\bibnamefont {Bisigello}},
  \bibinfo {author} {\bibfnamefont {V.}~\bibnamefont {Buat}}, \bibinfo {author} {\bibfnamefont {A.}~\bibnamefont {Calabrò}}, \bibinfo {author} {\bibfnamefont {M.}~\bibnamefont {Castellano}}, \bibinfo {author} {\bibfnamefont {N.~J.}\ \bibnamefont {Cleri}}, \bibinfo {author} {\bibfnamefont {M.~C.}\ \bibnamefont {Cooper}}, \bibinfo {author} {\bibfnamefont {D.}~\bibnamefont {Croton}}, \bibinfo {author} {\bibfnamefont {E.}~\bibnamefont {Daddi}}, \bibinfo {author} {\bibfnamefont {A.}~\bibnamefont {Dekel}}, \bibinfo {author} {\bibfnamefont {D.}~\bibnamefont {Elbaz}}, \bibinfo {author} {\bibfnamefont {M.}~\bibnamefont {Franco}}, \bibinfo {author} {\bibfnamefont {E.}~\bibnamefont {Gawiser}}, \bibinfo {author} {\bibfnamefont {B.~W.}\ \bibnamefont {Holwerda}}, \bibinfo {author} {\bibfnamefont {M.}~\bibnamefont {Huertas-Company}}, \bibinfo {author} {\bibfnamefont {A.~E.}\ \bibnamefont {Jaskot}}, \bibinfo {author} {\bibfnamefont {G.~C.~K.}\ \bibnamefont {Leung}}, \bibinfo {author} {\bibfnamefont {R.~A.}\ \bibnamefont
  {Lucas}}, \bibinfo {author} {\bibfnamefont {B.}~\bibnamefont {Mobasher}}, \bibinfo {author} {\bibfnamefont {V.}~\bibnamefont {Pandya}}, \bibinfo {author} {\bibfnamefont {S.}~\bibnamefont {Tacchella}}, \bibinfo {author} {\bibfnamefont {B.~J.}\ \bibnamefont {Weiner}},\ and\ \bibinfo {author} {\bibfnamefont {J.~A.}\ \bibnamefont {Zavala}},\ }\bibfield  {title} {\bibinfo {title} {Ceers key paper. i. an early look into the first 500 myr of galaxy formation with jwst},\ }\href {https://doi.org/10.3847/2041-8213/acade4} {\bibfield  {journal} {\bibinfo  {journal} {The Astrophysical Journal Letters}\ }\textbf {\bibinfo {volume} {946}},\ \bibinfo {pages} {L13} (\bibinfo {year} {2023})}\BibitemShut {NoStop}%
\bibitem [{\citenamefont {{Atek}}\ \emph {et~al.}(2023)\citenamefont {{Atek}}, \citenamefont {{Shuntov}}, \citenamefont {{Furtak}}, \citenamefont {{Richard}}, \citenamefont {{Kneib}}, \citenamefont {{Mahler}}, \citenamefont {{Zitrin}}, \citenamefont {{McCracken}}, \citenamefont {{Charlot}}, \citenamefont {{Chevallard}},\ and\ \citenamefont {{Chemerynska}}}]{2023MNRAS.519.1201A}%
  \BibitemOpen
  \bibfield  {author} {\bibinfo {author} {\bibfnamefont {H.}~\bibnamefont {{Atek}}}, \bibinfo {author} {\bibfnamefont {M.}~\bibnamefont {{Shuntov}}}, \bibinfo {author} {\bibfnamefont {L.~J.}\ \bibnamefont {{Furtak}}}, \bibinfo {author} {\bibfnamefont {J.}~\bibnamefont {{Richard}}}, \bibinfo {author} {\bibfnamefont {J.-P.}\ \bibnamefont {{Kneib}}}, \bibinfo {author} {\bibfnamefont {G.}~\bibnamefont {{Mahler}}}, \bibinfo {author} {\bibfnamefont {A.}~\bibnamefont {{Zitrin}}}, \bibinfo {author} {\bibfnamefont {H.~J.}\ \bibnamefont {{McCracken}}}, \bibinfo {author} {\bibfnamefont {S.}~\bibnamefont {{Charlot}}}, \bibinfo {author} {\bibfnamefont {J.}~\bibnamefont {{Chevallard}}},\ and\ \bibinfo {author} {\bibfnamefont {I.}~\bibnamefont {{Chemerynska}}},\ }\bibfield  {title} {\bibinfo {title} {{Revealing galaxy candidates out to z 16 with JWST observations of the lensing cluster SMACS0723}},\ }\href {https://doi.org/10.1093/mnras/stac3144} {\bibfield  {journal} {\bibinfo  {journal} {mnras}\ }\textbf {\bibinfo
  {volume} {519}},\ \bibinfo {pages} {1201} (\bibinfo {year} {2023})},\ \Eprint {https://arxiv.org/abs/2207.12338} {arXiv:2207.12338 [astro-ph.GA]} \BibitemShut {NoStop}%
\bibitem [{\citenamefont {{Donnan}}\ \emph {et~al.}(2023)\citenamefont {{Donnan}}, \citenamefont {{McLeod}}, \citenamefont {{Dunlop}}, \citenamefont {{McLure}}, \citenamefont {{Carnall}}, \citenamefont {{Begley}}, \citenamefont {{Cullen}}, \citenamefont {{Hamadouche}}, \citenamefont {{Bowler}}, \citenamefont {{Magee}}, \citenamefont {{McCracken}}, \citenamefont {{Milvang-Jensen}}, \citenamefont {{Moneti}},\ and\ \citenamefont {{Targett}}}]{2023MNRAS.518.6011D}%
  \BibitemOpen
  \bibfield  {author} {\bibinfo {author} {\bibfnamefont {C.~T.}\ \bibnamefont {{Donnan}}}, \bibinfo {author} {\bibfnamefont {D.~J.}\ \bibnamefont {{McLeod}}}, \bibinfo {author} {\bibfnamefont {J.~S.}\ \bibnamefont {{Dunlop}}}, \bibinfo {author} {\bibfnamefont {R.~J.}\ \bibnamefont {{McLure}}}, \bibinfo {author} {\bibfnamefont {A.~C.}\ \bibnamefont {{Carnall}}}, \bibinfo {author} {\bibfnamefont {R.}~\bibnamefont {{Begley}}}, \bibinfo {author} {\bibfnamefont {F.}~\bibnamefont {{Cullen}}}, \bibinfo {author} {\bibfnamefont {M.~L.}\ \bibnamefont {{Hamadouche}}}, \bibinfo {author} {\bibfnamefont {R.~A.~A.}\ \bibnamefont {{Bowler}}}, \bibinfo {author} {\bibfnamefont {D.}~\bibnamefont {{Magee}}}, \bibinfo {author} {\bibfnamefont {H.~J.}\ \bibnamefont {{McCracken}}}, \bibinfo {author} {\bibfnamefont {B.}~\bibnamefont {{Milvang-Jensen}}}, \bibinfo {author} {\bibfnamefont {A.}~\bibnamefont {{Moneti}}},\ and\ \bibinfo {author} {\bibfnamefont {T.}~\bibnamefont {{Targett}}},\ }\bibfield  {title} {\bibinfo {title} {{The
  evolution of the galaxy UV luminosity function at redshifts $z \lesssim 8 - 15$ from deep JWST and ground-based near-infrared imaging}},\ }\href {https://doi.org/10.1093/mnras/stac3472} {\bibfield  {journal} {\bibinfo  {journal} {mnras}\ }\textbf {\bibinfo {volume} {518}},\ \bibinfo {pages} {6011} (\bibinfo {year} {2023})},\ \Eprint {https://arxiv.org/abs/2207.12356} {arXiv:2207.12356 [astro-ph.GA]} \BibitemShut {NoStop}%
\bibitem [{\citenamefont {{Naidu}}\ \emph {et~al.}(2022)\citenamefont {{Naidu}}, \citenamefont {{Oesch}}, \citenamefont {{van Dokkum}}, \citenamefont {{Nelson}}, \citenamefont {{Suess}}, \citenamefont {{Brammer}}, \citenamefont {{Whitaker}}, \citenamefont {{Illingworth}}, \citenamefont {{Bouwens}}, \citenamefont {{Tacchella}}, \citenamefont {{Matthee}}, \citenamefont {{Allen}}, \citenamefont {{Bezanson}}, \citenamefont {{Conroy}}, \citenamefont {{Labbe}}, \citenamefont {{Leja}}, \citenamefont {{Leonova}}, \citenamefont {{Magee}}, \citenamefont {{Price}}, \citenamefont {{Setton}}, \citenamefont {{Strait}}, \citenamefont {{Stefanon}}, \citenamefont {{Toft}}, \citenamefont {{Weaver}},\ and\ \citenamefont {{Weibel}}}]{2022ApJ...940L..14N}%
  \BibitemOpen
  \bibfield  {author} {\bibinfo {author} {\bibfnamefont {R.~P.}\ \bibnamefont {{Naidu}}}, \bibinfo {author} {\bibfnamefont {P.~A.}\ \bibnamefont {{Oesch}}}, \bibinfo {author} {\bibfnamefont {P.}~\bibnamefont {{van Dokkum}}}, \bibinfo {author} {\bibfnamefont {E.~J.}\ \bibnamefont {{Nelson}}}, \bibinfo {author} {\bibfnamefont {K.~A.}\ \bibnamefont {{Suess}}}, \bibinfo {author} {\bibfnamefont {G.}~\bibnamefont {{Brammer}}}, \bibinfo {author} {\bibfnamefont {K.~E.}\ \bibnamefont {{Whitaker}}}, \bibinfo {author} {\bibfnamefont {G.}~\bibnamefont {{Illingworth}}}, \bibinfo {author} {\bibfnamefont {R.}~\bibnamefont {{Bouwens}}}, \bibinfo {author} {\bibfnamefont {S.}~\bibnamefont {{Tacchella}}}, \bibinfo {author} {\bibfnamefont {J.}~\bibnamefont {{Matthee}}}, \bibinfo {author} {\bibfnamefont {N.}~\bibnamefont {{Allen}}}, \bibinfo {author} {\bibfnamefont {R.}~\bibnamefont {{Bezanson}}}, \bibinfo {author} {\bibfnamefont {C.}~\bibnamefont {{Conroy}}}, \bibinfo {author} {\bibfnamefont {I.}~\bibnamefont {{Labbe}}},
  \bibinfo {author} {\bibfnamefont {J.}~\bibnamefont {{Leja}}}, \bibinfo {author} {\bibfnamefont {E.}~\bibnamefont {{Leonova}}}, \bibinfo {author} {\bibfnamefont {D.}~\bibnamefont {{Magee}}}, \bibinfo {author} {\bibfnamefont {S.~H.}\ \bibnamefont {{Price}}}, \bibinfo {author} {\bibfnamefont {D.~J.}\ \bibnamefont {{Setton}}}, \bibinfo {author} {\bibfnamefont {V.}~\bibnamefont {{Strait}}}, \bibinfo {author} {\bibfnamefont {M.}~\bibnamefont {{Stefanon}}}, \bibinfo {author} {\bibfnamefont {S.}~\bibnamefont {{Toft}}}, \bibinfo {author} {\bibfnamefont {J.~R.}\ \bibnamefont {{Weaver}}},\ and\ \bibinfo {author} {\bibfnamefont {A.}~\bibnamefont {{Weibel}}},\ }\bibfield  {title} {\bibinfo {title} {{Two Remarkably Luminous Galaxy Candidates at z {\ensuremath{\approx}} 10-12 Revealed by JWST}},\ }\href {https://doi.org/10.3847/2041-8213/ac9b22} {\bibfield  {journal} {\bibinfo  {journal} {apjl}\ }\textbf {\bibinfo {volume} {940}},\ \bibinfo {eid} {L14} (\bibinfo {year} {2022})},\ \Eprint {https://arxiv.org/abs/2207.09434}
  {arXiv:2207.09434 [astro-ph.GA]} \BibitemShut {NoStop}%
\bibitem [{\citenamefont {{Yan}}\ \emph {et~al.}(2023)\citenamefont {{Yan}}, \citenamefont {{Ma}}, \citenamefont {{Ling}}, \citenamefont {{Cheng}},\ and\ \citenamefont {{Huang}}}]{2023ApJ...942L...9Y}%
  \BibitemOpen
  \bibfield  {author} {\bibinfo {author} {\bibfnamefont {H.}~\bibnamefont {{Yan}}}, \bibinfo {author} {\bibfnamefont {Z.}~\bibnamefont {{Ma}}}, \bibinfo {author} {\bibfnamefont {C.}~\bibnamefont {{Ling}}}, \bibinfo {author} {\bibfnamefont {C.}~\bibnamefont {{Cheng}}},\ and\ \bibinfo {author} {\bibfnamefont {J.-S.}\ \bibnamefont {{Huang}}},\ }\bibfield  {title} {\bibinfo {title} {{First Batch of z {\ensuremath{\approx}} 11-20 Candidate Objects Revealed by the James Webb Space Telescope Early Release Observations on SMACS 0723-73}},\ }\href {https://doi.org/10.3847/2041-8213/aca80c} {\bibfield  {journal} {\bibinfo  {journal} {apjl}\ }\textbf {\bibinfo {volume} {942}},\ \bibinfo {eid} {L9} (\bibinfo {year} {2023})},\ \Eprint {https://arxiv.org/abs/2207.11558} {arXiv:2207.11558 [astro-ph.GA]} \BibitemShut {NoStop}%
\bibitem [{\citenamefont {Lovell}\ \emph {et~al.}(2022)\citenamefont {Lovell}, \citenamefont {Harrison}, \citenamefont {Harikane}, \citenamefont {Tacchella},\ and\ \citenamefont {Wilkins}}]{Lovell:2022bhx}%
  \BibitemOpen
  \bibfield  {author} {\bibinfo {author} {\bibfnamefont {C.~C.}\ \bibnamefont {Lovell}}, \bibinfo {author} {\bibfnamefont {I.}~\bibnamefont {Harrison}}, \bibinfo {author} {\bibfnamefont {Y.}~\bibnamefont {Harikane}}, \bibinfo {author} {\bibfnamefont {S.}~\bibnamefont {Tacchella}},\ and\ \bibinfo {author} {\bibfnamefont {S.~M.}\ \bibnamefont {Wilkins}},\ }\bibfield  {title} {\bibinfo {title} {{Extreme value statistics of the halo and stellar mass distributions at high redshift: are JWST results in tension with \ensuremath{\Lambda}CDM?}},\ }\href {https://doi.org/10.1093/mnras/stac3224} {\bibfield  {journal} {\bibinfo  {journal} {Mon. Not. Roy. Astron. Soc.}\ }\textbf {\bibinfo {volume} {518}},\ \bibinfo {pages} {2511} (\bibinfo {year} {2022})},\ \Eprint {https://arxiv.org/abs/2208.10479} {arXiv:2208.10479 [astro-ph.GA]} \BibitemShut {NoStop}%
\bibitem [{\citenamefont {Fujimoto}\ \emph {et~al.}(2023)\citenamefont {Fujimoto} \emph {et~al.}}]{Fujimoto:2022aco}%
  \BibitemOpen
  \bibfield  {author} {\bibinfo {author} {\bibfnamefont {S.}~\bibnamefont {Fujimoto}} \emph {et~al.},\ }\bibfield  {title} {\bibinfo {title} {{ALMA FIR View of Ultra-high-redshift Galaxy Candidates at z \ensuremath{\sim} 11\textendash{}17: Blue Monsters or Low-z Red Interlopers?}},\ }\href {https://doi.org/10.3847/1538-4357/aceb67} {\bibfield  {journal} {\bibinfo  {journal} {Astrophys. J.}\ }\textbf {\bibinfo {volume} {955}},\ \bibinfo {pages} {130} (\bibinfo {year} {2023})},\ \Eprint {https://arxiv.org/abs/2211.03896} {arXiv:2211.03896 [astro-ph.GA]} \BibitemShut {NoStop}%
\bibitem [{\citenamefont {Prada}\ \emph {et~al.}(2023)\citenamefont {Prada}, \citenamefont {Behroozi}, \citenamefont {Ishiyama}, \citenamefont {Klypin},\ and\ \citenamefont {Pérez}}]{Prada:2023dix}%
  \BibitemOpen
  \bibfield  {author} {\bibinfo {author} {\bibfnamefont {F.}~\bibnamefont {Prada}}, \bibinfo {author} {\bibfnamefont {P.}~\bibnamefont {Behroozi}}, \bibinfo {author} {\bibfnamefont {T.}~\bibnamefont {Ishiyama}}, \bibinfo {author} {\bibfnamefont {A.}~\bibnamefont {Klypin}},\ and\ \bibinfo {author} {\bibfnamefont {E.}~\bibnamefont {Pérez}},\ }\href@noop {} {\bibinfo {title} {Confirmation of the standard cosmological model from red massive galaxies $\sim600$ myr after the big bang}} (\bibinfo {year} {2023}),\ \Eprint {https://arxiv.org/abs/2304.11911} {arXiv:2304.11911 [astro-ph.GA]} \BibitemShut {NoStop}%
\bibitem [{\citenamefont {Chen}\ \emph {et~al.}(2023)\citenamefont {Chen}, \citenamefont {Mo},\ and\ \citenamefont {Wang}}]{Chen:2023ugq}%
  \BibitemOpen
  \bibfield  {author} {\bibinfo {author} {\bibfnamefont {Y.}~\bibnamefont {Chen}}, \bibinfo {author} {\bibfnamefont {H.~J.}\ \bibnamefont {Mo}},\ and\ \bibinfo {author} {\bibfnamefont {K.}~\bibnamefont {Wang}},\ }\bibfield  {title} {\bibinfo {title} {{Massive dark matter haloes at high redshift: implications for observations in the JWST era}},\ }\href {https://doi.org/10.1093/mnras/stad2866} {\bibfield  {journal} {\bibinfo  {journal} {Mon. Not. Roy. Astron. Soc.}\ }\textbf {\bibinfo {volume} {526}},\ \bibinfo {pages} {2542} (\bibinfo {year} {2023})},\ \Eprint {https://arxiv.org/abs/2304.13890} {arXiv:2304.13890 [astro-ph.GA]} \BibitemShut {NoStop}%
\bibitem [{\citenamefont {Forconi}\ \emph {et~al.}(2023{\natexlab{a}})\citenamefont {Forconi}, \citenamefont {Ruchika}, \citenamefont {Melchiorri}, \citenamefont {Mena},\ and\ \citenamefont {Menci}}]{Forconi:2023izg}%
  \BibitemOpen
  \bibfield  {author} {\bibinfo {author} {\bibfnamefont {M.}~\bibnamefont {Forconi}}, \bibinfo {author} {\bibnamefont {Ruchika}}, \bibinfo {author} {\bibfnamefont {A.}~\bibnamefont {Melchiorri}}, \bibinfo {author} {\bibfnamefont {O.}~\bibnamefont {Mena}},\ and\ \bibinfo {author} {\bibfnamefont {N.}~\bibnamefont {Menci}},\ }\bibfield  {title} {\bibinfo {title} {{Do the early galaxies observed by JWST disagree with Planck's CMB polarization measurements?}},\ }\href {https://doi.org/10.1088/1475-7516/2023/10/012} {\bibfield  {journal} {\bibinfo  {journal} {JCAP}\ }\textbf {\bibinfo {volume} {10}},\ \bibinfo {pages} {012}},\ \Eprint {https://arxiv.org/abs/2306.07781} {arXiv:2306.07781 [astro-ph.CO]} \BibitemShut {NoStop}%
\bibitem [{\citenamefont {Steinhardt}\ \emph {et~al.}(2023)\citenamefont {Steinhardt}, \citenamefont {Sneppen}, \citenamefont {Clausen}, \citenamefont {Katz}, \citenamefont {Rey},\ and\ \citenamefont {Stahlschmidt}}]{Steinhardt:2023oow}%
  \BibitemOpen
  \bibfield  {author} {\bibinfo {author} {\bibfnamefont {C.~L.}\ \bibnamefont {Steinhardt}}, \bibinfo {author} {\bibfnamefont {A.}~\bibnamefont {Sneppen}}, \bibinfo {author} {\bibfnamefont {T.}~\bibnamefont {Clausen}}, \bibinfo {author} {\bibfnamefont {H.}~\bibnamefont {Katz}}, \bibinfo {author} {\bibfnamefont {M.~P.}\ \bibnamefont {Rey}},\ and\ \bibinfo {author} {\bibfnamefont {J.}~\bibnamefont {Stahlschmidt}},\ }\href@noop {} {\bibinfo {title} {The highest-redshift balmer breaks as a test of $\lambda$cdm}} (\bibinfo {year} {2023}),\ \Eprint {https://arxiv.org/abs/2305.15459} {arXiv:2305.15459 [astro-ph.GA]} \BibitemShut {NoStop}%
\bibitem [{\citenamefont {Vikaeus}\ \emph {et~al.}(2023)\citenamefont {Vikaeus}, \citenamefont {Zackrisson}, \citenamefont {Wilkins}, \citenamefont {Nabizadeh}, \citenamefont {Kokorev}, \citenamefont {Abdurrouf}, \citenamefont {Bradley}, \citenamefont {Coe}, \citenamefont {Dayal},\ and\ \citenamefont {Ricotti}}]{Vikaeus:2023cyi}%
  \BibitemOpen
  \bibfield  {author} {\bibinfo {author} {\bibfnamefont {A.}~\bibnamefont {Vikaeus}}, \bibinfo {author} {\bibfnamefont {E.}~\bibnamefont {Zackrisson}}, \bibinfo {author} {\bibfnamefont {S.}~\bibnamefont {Wilkins}}, \bibinfo {author} {\bibfnamefont {A.}~\bibnamefont {Nabizadeh}}, \bibinfo {author} {\bibfnamefont {V.}~\bibnamefont {Kokorev}}, \bibinfo {author} {\bibnamefont {Abdurrouf}}, \bibinfo {author} {\bibfnamefont {L.~D.}\ \bibnamefont {Bradley}}, \bibinfo {author} {\bibfnamefont {D.}~\bibnamefont {Coe}}, \bibinfo {author} {\bibfnamefont {P.}~\bibnamefont {Dayal}},\ and\ \bibinfo {author} {\bibfnamefont {M.}~\bibnamefont {Ricotti}},\ }\href@noop {} {\bibinfo {title} {To be, or not to be: Balmer breaks in high-z galaxies with jwst}} (\bibinfo {year} {2023}),\ \Eprint {https://arxiv.org/abs/2309.02504} {arXiv:2309.02504 [astro-ph.GA]} \BibitemShut {NoStop}%
\bibitem [{\citenamefont {Robertson}\ \emph {et~al.}(2023)\citenamefont {Robertson} \emph {et~al.}}]{Robertson:2022gdk}%
  \BibitemOpen
  \bibfield  {author} {\bibinfo {author} {\bibfnamefont {B.~E.}\ \bibnamefont {Robertson}} \emph {et~al.},\ }\bibfield  {title} {\bibinfo {title} {{Identification and properties of intense star-forming galaxies at redshifts z\,\ensuremath{>}\,10}},\ }\href {https://doi.org/10.1038/s41550-023-01921-1} {\bibfield  {journal} {\bibinfo  {journal} {Nature Astron.}\ }\textbf {\bibinfo {volume} {7}},\ \bibinfo {pages} {611} (\bibinfo {year} {2023})},\ \Eprint {https://arxiv.org/abs/2212.04480} {arXiv:2212.04480 [astro-ph.GA]} \BibitemShut {NoStop}%
\bibitem [{\citenamefont {Parashari}\ and\ \citenamefont {Laha}(2023)}]{Parashari:2023cui}%
  \BibitemOpen
  \bibfield  {author} {\bibinfo {author} {\bibfnamefont {P.}~\bibnamefont {Parashari}}\ and\ \bibinfo {author} {\bibfnamefont {R.}~\bibnamefont {Laha}},\ }\bibfield  {title} {\bibinfo {title} {{Primordial power spectrum in light of JWST observations of high redshift galaxies}},\ }\href {https://doi.org/10.1093/mnrasl/slad107} {\bibfield  {journal} {\bibinfo  {journal} {Mon. Not. Roy. Astron. Soc.}\ }\textbf {\bibinfo {volume} {526}},\ \bibinfo {pages} {L63} (\bibinfo {year} {2023})},\ \Eprint {https://arxiv.org/abs/2305.00999} {arXiv:2305.00999 [astro-ph.CO]} \BibitemShut {NoStop}%
\bibitem [{\citenamefont {Sabti}\ \emph {et~al.}(2024)\citenamefont {Sabti}, \citenamefont {Mu\~noz},\ and\ \citenamefont {Kamionkowski}}]{Sabti:2023xwo}%
  \BibitemOpen
  \bibfield  {author} {\bibinfo {author} {\bibfnamefont {N.}~\bibnamefont {Sabti}}, \bibinfo {author} {\bibfnamefont {J.~B.}\ \bibnamefont {Mu\~noz}},\ and\ \bibinfo {author} {\bibfnamefont {M.}~\bibnamefont {Kamionkowski}},\ }\bibfield  {title} {\bibinfo {title} {{Insights from HST into Ultramassive Galaxies and Early-Universe Cosmology}},\ }\href {https://doi.org/10.1103/PhysRevLett.132.061002} {\bibfield  {journal} {\bibinfo  {journal} {Phys. Rev. Lett.}\ }\textbf {\bibinfo {volume} {132}},\ \bibinfo {pages} {061002} (\bibinfo {year} {2024})},\ \Eprint {https://arxiv.org/abs/2305.07049} {arXiv:2305.07049 [astro-ph.CO]} \BibitemShut {NoStop}%
\bibitem [{\citenamefont {Pallottini}\ and\ \citenamefont {Ferrara}(2023)}]{Pallottini:2023yqg}%
  \BibitemOpen
  \bibfield  {author} {\bibinfo {author} {\bibfnamefont {A.}~\bibnamefont {Pallottini}}\ and\ \bibinfo {author} {\bibfnamefont {A.}~\bibnamefont {Ferrara}},\ }\bibfield  {title} {\bibinfo {title} {{Stochastic star formation in early galaxies: Implications for the James Webb Space Telescope}},\ }\href {https://doi.org/10.1051/0004-6361/202347384} {\bibfield  {journal} {\bibinfo  {journal} {Astron. Astrophys.}\ }\textbf {\bibinfo {volume} {677}},\ \bibinfo {pages} {L4} (\bibinfo {year} {2023})},\ \Eprint {https://arxiv.org/abs/2307.03219} {arXiv:2307.03219 [astro-ph.GA]} \BibitemShut {NoStop}%
\bibitem [{\citenamefont {Tkachev}\ \emph {et~al.}(2023)\citenamefont {Tkachev}, \citenamefont {Pilipenko}, \citenamefont {Mikheeva},\ and\ \citenamefont {Lukash}}]{Tkachev:2023acf}%
  \BibitemOpen
  \bibfield  {author} {\bibinfo {author} {\bibfnamefont {M.~V.}\ \bibnamefont {Tkachev}}, \bibinfo {author} {\bibfnamefont {S.~V.}\ \bibnamefont {Pilipenko}}, \bibinfo {author} {\bibfnamefont {E.~V.}\ \bibnamefont {Mikheeva}},\ and\ \bibinfo {author} {\bibfnamefont {V.~N.}\ \bibnamefont {Lukash}},\ }\bibfield  {title} {\bibinfo {title} {{Excess of high-z galaxies as a test for bumpy power spectrum of density perturbations}},\ }\href {https://doi.org/10.1093/mnras/stad3279} {\bibfield  {journal} {\bibinfo  {journal} {Mon. Not. Roy. Astron. Soc.}\ }\textbf {\bibinfo {volume} {527}},\ \bibinfo {pages} {1381} (\bibinfo {year} {2023})},\ \Eprint {https://arxiv.org/abs/2307.13774} {arXiv:2307.13774 [astro-ph.CO]} \BibitemShut {NoStop}%
\bibitem [{\citenamefont {Wang}\ \emph {et~al.}(2023{\natexlab{a}})\citenamefont {Wang}, \citenamefont {Lei}, \citenamefont {Yuan},\ and\ \citenamefont {Fan}}]{Wang:2023xmm}%
  \BibitemOpen
  \bibfield  {author} {\bibinfo {author} {\bibfnamefont {Y.-Y.}\ \bibnamefont {Wang}}, \bibinfo {author} {\bibfnamefont {L.}~\bibnamefont {Lei}}, \bibinfo {author} {\bibfnamefont {G.-W.}\ \bibnamefont {Yuan}},\ and\ \bibinfo {author} {\bibfnamefont {Y.-Z.}\ \bibnamefont {Fan}},\ }\bibfield  {title} {\bibinfo {title} {{Modeling the JWST High-redshift Galaxies with a General Formation Scenario and the Consistency with the \ensuremath{\Lambda}CDM Model}},\ }\href {https://doi.org/10.3847/2041-8213/acf46c} {\bibfield  {journal} {\bibinfo  {journal} {Astrophys. J. Lett.}\ }\textbf {\bibinfo {volume} {954}},\ \bibinfo {pages} {L48} (\bibinfo {year} {2023}{\natexlab{a}})},\ \Eprint {https://arxiv.org/abs/2307.12487} {arXiv:2307.12487 [astro-ph.GA]} \BibitemShut {NoStop}%
\bibitem [{\citenamefont {Wang}\ \emph {et~al.}(2023{\natexlab{b}})\citenamefont {Wang}, \citenamefont {Huang}, \citenamefont {Huang},\ and\ \citenamefont {Liu}}]{Wang:2023gla}%
  \BibitemOpen
  \bibfield  {author} {\bibinfo {author} {\bibfnamefont {J.}~\bibnamefont {Wang}}, \bibinfo {author} {\bibfnamefont {Z.}~\bibnamefont {Huang}}, \bibinfo {author} {\bibfnamefont {L.}~\bibnamefont {Huang}},\ and\ \bibinfo {author} {\bibfnamefont {J.}~\bibnamefont {Liu}},\ }\href@noop {} {\bibinfo {title} {Quantifying the tension between cosmological models and jwst red candidate massive galaxies}} (\bibinfo {year} {2023}{\natexlab{b}}),\ \Eprint {https://arxiv.org/abs/2311.02866} {arXiv:2311.02866 [astro-ph.CO]} \BibitemShut {NoStop}%
\bibitem [{\citenamefont {Inayoshi}\ and\ \citenamefont {Ichikawa}(2024)}]{Inayoshi:2024xwv}%
  \BibitemOpen
  \bibfield  {author} {\bibinfo {author} {\bibfnamefont {K.}~\bibnamefont {Inayoshi}}\ and\ \bibinfo {author} {\bibfnamefont {K.}~\bibnamefont {Ichikawa}},\ }\href@noop {} {\bibinfo {title} {Birth of rapidly spinning, overmassive black holes in the early universe}} (\bibinfo {year} {2024}),\ \Eprint {https://arxiv.org/abs/2402.14706} {arXiv:2402.14706 [astro-ph.GA]} \BibitemShut {NoStop}%
\bibitem [{\citenamefont {Jeon}\ \emph {et~al.}(2024)\citenamefont {Jeon}, \citenamefont {Bromm}, \citenamefont {Liu},\ and\ \citenamefont {Finkelstein}}]{Jeon:2024iml}%
  \BibitemOpen
  \bibfield  {author} {\bibinfo {author} {\bibfnamefont {J.}~\bibnamefont {Jeon}}, \bibinfo {author} {\bibfnamefont {V.}~\bibnamefont {Bromm}}, \bibinfo {author} {\bibfnamefont {B.}~\bibnamefont {Liu}},\ and\ \bibinfo {author} {\bibfnamefont {S.~L.}\ \bibnamefont {Finkelstein}},\ }\href@noop {} {\bibinfo {title} {Physical pathways for jwst-observed supermassive black holes in the early universe}} (\bibinfo {year} {2024}),\ \Eprint {https://arxiv.org/abs/2402.18773} {arXiv:2402.18773 [astro-ph.GA]} \BibitemShut {NoStop}%
\bibitem [{\citenamefont {Liu}\ and\ \citenamefont {Bromm}(2022)}]{Liu:2022bvr}%
  \BibitemOpen
  \bibfield  {author} {\bibinfo {author} {\bibfnamefont {B.}~\bibnamefont {Liu}}\ and\ \bibinfo {author} {\bibfnamefont {V.}~\bibnamefont {Bromm}},\ }\bibfield  {title} {\bibinfo {title} {{Accelerating Early Massive Galaxy Formation with Primordial Black Holes}},\ }\href {https://doi.org/10.3847/2041-8213/ac927f} {\bibfield  {journal} {\bibinfo  {journal} {Astrophys. J. Lett.}\ }\textbf {\bibinfo {volume} {937}},\ \bibinfo {pages} {L30} (\bibinfo {year} {2022})},\ \Eprint {https://arxiv.org/abs/2208.13178} {arXiv:2208.13178 [astro-ph.CO]} \BibitemShut {NoStop}%
\bibitem [{\citenamefont {Huang}\ \emph {et~al.}(2023)\citenamefont {Huang}, \citenamefont {Cai}, \citenamefont {Jiang}, \citenamefont {Zhang},\ and\ \citenamefont {Piao}}]{Huang:2023chx}%
  \BibitemOpen
  \bibfield  {author} {\bibinfo {author} {\bibfnamefont {H.-L.}\ \bibnamefont {Huang}}, \bibinfo {author} {\bibfnamefont {Y.}~\bibnamefont {Cai}}, \bibinfo {author} {\bibfnamefont {J.-Q.}\ \bibnamefont {Jiang}}, \bibinfo {author} {\bibfnamefont {J.}~\bibnamefont {Zhang}},\ and\ \bibinfo {author} {\bibfnamefont {Y.-S.}\ \bibnamefont {Piao}},\ }\href@noop {} {\bibinfo {title} {Supermassive primordial black holes in multiverse: for nano-hertz gravitational wave and high-redshift jwst galaxies}} (\bibinfo {year} {2023}),\ \Eprint {https://arxiv.org/abs/2306.17577} {arXiv:2306.17577 [gr-qc]} \BibitemShut {NoStop}%
\bibitem [{\citenamefont {Gouttenoire}\ \emph {et~al.}(2023)\citenamefont {Gouttenoire}, \citenamefont {Trifinopoulos}, \citenamefont {Valogiannis},\ and\ \citenamefont {Vanvlasselaer}}]{Gouttenoire:2023nzr}%
  \BibitemOpen
  \bibfield  {author} {\bibinfo {author} {\bibfnamefont {Y.}~\bibnamefont {Gouttenoire}}, \bibinfo {author} {\bibfnamefont {S.}~\bibnamefont {Trifinopoulos}}, \bibinfo {author} {\bibfnamefont {G.}~\bibnamefont {Valogiannis}},\ and\ \bibinfo {author} {\bibfnamefont {M.}~\bibnamefont {Vanvlasselaer}},\ }\href@noop {} {\bibinfo {title} {Scrutinizing the primordial black holes interpretation of pta gravitational waves and jwst early galaxies}} (\bibinfo {year} {2023}),\ \Eprint {https://arxiv.org/abs/2307.01457} {arXiv:2307.01457 [astro-ph.CO]} \BibitemShut {NoStop}%
\bibitem [{\citenamefont {Ralegankar}\ \emph {et~al.}(2024)\citenamefont {Ralegankar}, \citenamefont {Pavičević},\ and\ \citenamefont {Viel}}]{Ralegankar:2024ekl}%
  \BibitemOpen
  \bibfield  {author} {\bibinfo {author} {\bibfnamefont {P.}~\bibnamefont {Ralegankar}}, \bibinfo {author} {\bibfnamefont {M.}~\bibnamefont {Pavičević}},\ and\ \bibinfo {author} {\bibfnamefont {M.}~\bibnamefont {Viel}},\ }\href@noop {} {\bibinfo {title} {Primordial magnetic fields: consistent initial conditions and impact on high-z structures}} (\bibinfo {year} {2024}),\ \Eprint {https://arxiv.org/abs/2402.14079} {arXiv:2402.14079 [astro-ph.CO]} \BibitemShut {NoStop}%
\bibitem [{\citenamefont {Hirano}\ and\ \citenamefont {Yoshida}(2024)}]{Hirano:2023auh}%
  \BibitemOpen
  \bibfield  {author} {\bibinfo {author} {\bibfnamefont {S.}~\bibnamefont {Hirano}}\ and\ \bibinfo {author} {\bibfnamefont {N.}~\bibnamefont {Yoshida}},\ }\bibfield  {title} {\bibinfo {title} {{Early Structure Formation from Primordial Density Fluctuations with a Blue, Tilted Power Spectrum: High-redshift Galaxies}},\ }\href {https://doi.org/10.3847/1538-4357/ad22e0} {\bibfield  {journal} {\bibinfo  {journal} {Astrophys. J.}\ }\textbf {\bibinfo {volume} {963}},\ \bibinfo {pages} {2} (\bibinfo {year} {2024})},\ \Eprint {https://arxiv.org/abs/2306.11993} {arXiv:2306.11993 [astro-ph.GA]} \BibitemShut {NoStop}%
\bibitem [{\citenamefont {Jiao}\ \emph {et~al.}(2023)\citenamefont {Jiao}, \citenamefont {Brandenberger},\ and\ \citenamefont {Refregier}}]{Jiao:2023wcn}%
  \BibitemOpen
  \bibfield  {author} {\bibinfo {author} {\bibfnamefont {H.}~\bibnamefont {Jiao}}, \bibinfo {author} {\bibfnamefont {R.}~\bibnamefont {Brandenberger}},\ and\ \bibinfo {author} {\bibfnamefont {A.}~\bibnamefont {Refregier}},\ }\bibfield  {title} {\bibinfo {title} {{Early structure formation from cosmic string loops in light of early JWST observations}},\ }\href {https://doi.org/10.1103/PhysRevD.108.043510} {\bibfield  {journal} {\bibinfo  {journal} {Phys. Rev. D}\ }\textbf {\bibinfo {volume} {108}},\ \bibinfo {pages} {043510} (\bibinfo {year} {2023})},\ \Eprint {https://arxiv.org/abs/2304.06429} {arXiv:2304.06429 [astro-ph.CO]} \BibitemShut {NoStop}%
\bibitem [{\citenamefont {Wang}\ \emph {et~al.}(2023{\natexlab{c}})\citenamefont {Wang}, \citenamefont {Lei}, \citenamefont {Jiao}, \citenamefont {Feng},\ and\ \citenamefont {Fan}}]{Wang:2023len}%
  \BibitemOpen
  \bibfield  {author} {\bibinfo {author} {\bibfnamefont {Z.}~\bibnamefont {Wang}}, \bibinfo {author} {\bibfnamefont {L.}~\bibnamefont {Lei}}, \bibinfo {author} {\bibfnamefont {H.}~\bibnamefont {Jiao}}, \bibinfo {author} {\bibfnamefont {L.}~\bibnamefont {Feng}},\ and\ \bibinfo {author} {\bibfnamefont {Y.-Z.}\ \bibnamefont {Fan}},\ }\bibfield  {title} {\bibinfo {title} {{The nanohertz stochastic gravitational wave background from cosmic string loops and the abundant high redshift massive galaxies}},\ }\href {https://doi.org/10.1007/s11433-023-2262-0} {\bibfield  {journal} {\bibinfo  {journal} {Sci. China Phys. Mech. Astron.}\ }\textbf {\bibinfo {volume} {66}},\ \bibinfo {pages} {120403} (\bibinfo {year} {2023}{\natexlab{c}})},\ \Eprint {https://arxiv.org/abs/2306.17150} {arXiv:2306.17150 [astro-ph.HE]} \BibitemShut {NoStop}%
\bibitem [{\citenamefont {Jiao}\ \emph {et~al.}(2024)\citenamefont {Jiao}, \citenamefont {Brandenberger},\ and\ \citenamefont {Refregier}}]{Jiao:2024rcr}%
  \BibitemOpen
  \bibfield  {author} {\bibinfo {author} {\bibfnamefont {H.}~\bibnamefont {Jiao}}, \bibinfo {author} {\bibfnamefont {R.}~\bibnamefont {Brandenberger}},\ and\ \bibinfo {author} {\bibfnamefont {A.}~\bibnamefont {Refregier}},\ }\href@noop {} {\bibinfo {title} {N-body simulation of early structure formation from cosmic string loops}} (\bibinfo {year} {2024}),\ \Eprint {https://arxiv.org/abs/2402.06235} {arXiv:2402.06235 [astro-ph.CO]} \BibitemShut {NoStop}%
\bibitem [{\citenamefont {Maio}\ and\ \citenamefont {Viel}(2023)}]{Maio:2022lzg}%
  \BibitemOpen
  \bibfield  {author} {\bibinfo {author} {\bibfnamefont {U.}~\bibnamefont {Maio}}\ and\ \bibinfo {author} {\bibfnamefont {M.}~\bibnamefont {Viel}},\ }\bibfield  {title} {\bibinfo {title} {{JWST high-redshift galaxy constraints on warm and cold dark matter models}},\ }\href {https://doi.org/10.1051/0004-6361/202345851} {\bibfield  {journal} {\bibinfo  {journal} {Astron. Astrophys.}\ }\textbf {\bibinfo {volume} {672}},\ \bibinfo {pages} {A71} (\bibinfo {year} {2023})},\ \Eprint {https://arxiv.org/abs/2211.03620} {arXiv:2211.03620 [astro-ph.CO]} \BibitemShut {NoStop}%
\bibitem [{\citenamefont {Gong}\ \emph {et~al.}(2023)\citenamefont {Gong}, \citenamefont {Yue}, \citenamefont {Cao},\ and\ \citenamefont {Chen}}]{Gong:2022qjx}%
  \BibitemOpen
  \bibfield  {author} {\bibinfo {author} {\bibfnamefont {Y.}~\bibnamefont {Gong}}, \bibinfo {author} {\bibfnamefont {B.}~\bibnamefont {Yue}}, \bibinfo {author} {\bibfnamefont {Y.}~\bibnamefont {Cao}},\ and\ \bibinfo {author} {\bibfnamefont {X.}~\bibnamefont {Chen}},\ }\bibfield  {title} {\bibinfo {title} {{Fuzzy Dark Matter as a Solution to Reconcile the Stellar Mass Density of High-z Massive Galaxies and Reionization History}},\ }\href {https://doi.org/10.3847/1538-4357/acc109} {\bibfield  {journal} {\bibinfo  {journal} {Astrophys. J.}\ }\textbf {\bibinfo {volume} {947}},\ \bibinfo {pages} {28} (\bibinfo {year} {2023})},\ \Eprint {https://arxiv.org/abs/2209.13757} {arXiv:2209.13757 [astro-ph.CO]} \BibitemShut {NoStop}%
\bibitem [{\citenamefont {H\"utsi}\ \emph {et~al.}(2023)\citenamefont {H\"utsi}, \citenamefont {Raidal}, \citenamefont {Urrutia}, \citenamefont {Vaskonen},\ and\ \citenamefont {Veerm\"ae}}]{Hutsi:2022fzw}%
  \BibitemOpen
  \bibfield  {author} {\bibinfo {author} {\bibfnamefont {G.}~\bibnamefont {H\"utsi}}, \bibinfo {author} {\bibfnamefont {M.}~\bibnamefont {Raidal}}, \bibinfo {author} {\bibfnamefont {J.}~\bibnamefont {Urrutia}}, \bibinfo {author} {\bibfnamefont {V.}~\bibnamefont {Vaskonen}},\ and\ \bibinfo {author} {\bibfnamefont {H.}~\bibnamefont {Veerm\"ae}},\ }\bibfield  {title} {\bibinfo {title} {{Did JWST observe imprints of axion miniclusters or primordial black holes?}},\ }\href {https://doi.org/10.1103/PhysRevD.107.043502} {\bibfield  {journal} {\bibinfo  {journal} {Phys. Rev. D}\ }\textbf {\bibinfo {volume} {107}},\ \bibinfo {pages} {043502} (\bibinfo {year} {2023})},\ \Eprint {https://arxiv.org/abs/2211.02651} {arXiv:2211.02651 [astro-ph.CO]} \BibitemShut {NoStop}%
\bibitem [{\citenamefont {Haslbauer}\ \emph {et~al.}(2022)\citenamefont {Haslbauer}, \citenamefont {Kroupa}, \citenamefont {Zonoozi},\ and\ \citenamefont {Haghi}}]{Haslbauer:2022vnq}%
  \BibitemOpen
  \bibfield  {author} {\bibinfo {author} {\bibfnamefont {M.}~\bibnamefont {Haslbauer}}, \bibinfo {author} {\bibfnamefont {P.}~\bibnamefont {Kroupa}}, \bibinfo {author} {\bibfnamefont {A.~H.}\ \bibnamefont {Zonoozi}},\ and\ \bibinfo {author} {\bibfnamefont {H.}~\bibnamefont {Haghi}},\ }\bibfield  {title} {\bibinfo {title} {{Has JWST Already Falsified Dark-matter-driven Galaxy Formation?}},\ }\href {https://doi.org/10.3847/2041-8213/ac9a50} {\bibfield  {journal} {\bibinfo  {journal} {Astrophys. J. Lett.}\ }\textbf {\bibinfo {volume} {939}},\ \bibinfo {pages} {L31} (\bibinfo {year} {2022})},\ \Eprint {https://arxiv.org/abs/2210.14915} {arXiv:2210.14915 [astro-ph.GA]} \BibitemShut {NoStop}%
\bibitem [{\citenamefont {Dom\`enech}\ \emph {et~al.}(2023)\citenamefont {Dom\`enech}, \citenamefont {Inman}, \citenamefont {Kusenko},\ and\ \citenamefont {Sasaki}}]{Domenech:2023afs}%
  \BibitemOpen
  \bibfield  {author} {\bibinfo {author} {\bibfnamefont {G.}~\bibnamefont {Dom\`enech}}, \bibinfo {author} {\bibfnamefont {D.}~\bibnamefont {Inman}}, \bibinfo {author} {\bibfnamefont {A.}~\bibnamefont {Kusenko}},\ and\ \bibinfo {author} {\bibfnamefont {M.}~\bibnamefont {Sasaki}},\ }\bibfield  {title} {\bibinfo {title} {{Halo formation from Yukawa forces in the very early Universe}},\ }\href {https://doi.org/10.1103/PhysRevD.108.103543} {\bibfield  {journal} {\bibinfo  {journal} {Phys. Rev. D}\ }\textbf {\bibinfo {volume} {108}},\ \bibinfo {pages} {103543} (\bibinfo {year} {2023})},\ \Eprint {https://arxiv.org/abs/2304.13053} {arXiv:2304.13053 [astro-ph.CO]} \BibitemShut {NoStop}%
\bibitem [{\citenamefont {Lin}\ \emph {et~al.}(2024)\citenamefont {Lin}, \citenamefont {Gong}, \citenamefont {Yue},\ and\ \citenamefont {Chen}}]{Lin:2023ewc}%
  \BibitemOpen
  \bibfield  {author} {\bibinfo {author} {\bibfnamefont {H.}~\bibnamefont {Lin}}, \bibinfo {author} {\bibfnamefont {Y.}~\bibnamefont {Gong}}, \bibinfo {author} {\bibfnamefont {B.}~\bibnamefont {Yue}},\ and\ \bibinfo {author} {\bibfnamefont {X.}~\bibnamefont {Chen}},\ }\bibfield  {title} {\bibinfo {title} {{Implications of the Stellar Mass Density of High-z Massive Galaxies from JWST on Warm Dark Matter}},\ }\href {https://doi.org/10.1088/1674-4527/ad0864} {\bibfield  {journal} {\bibinfo  {journal} {Res. Astron. Astrophys.}\ }\textbf {\bibinfo {volume} {24}},\ \bibinfo {pages} {015009} (\bibinfo {year} {2024})},\ \Eprint {https://arxiv.org/abs/2306.05648} {arXiv:2306.05648 [astro-ph.CO]} \BibitemShut {NoStop}%
\bibitem [{\citenamefont {Bird}\ \emph {et~al.}(2023)\citenamefont {Bird}, \citenamefont {Chang}, \citenamefont {Cui},\ and\ \citenamefont {Yang}}]{Bird:2023pkr}%
  \BibitemOpen
  \bibfield  {author} {\bibinfo {author} {\bibfnamefont {S.}~\bibnamefont {Bird}}, \bibinfo {author} {\bibfnamefont {C.-F.}\ \bibnamefont {Chang}}, \bibinfo {author} {\bibfnamefont {Y.}~\bibnamefont {Cui}},\ and\ \bibinfo {author} {\bibfnamefont {D.}~\bibnamefont {Yang}},\ }\href@noop {} {\bibinfo {title} {Enhanced early galaxy formation in jwst from axion dark matter?}} (\bibinfo {year} {2023}),\ \Eprint {https://arxiv.org/abs/2307.10302} {arXiv:2307.10302 [hep-ph]} \BibitemShut {NoStop}%
\bibitem [{\citenamefont {Davari}\ \emph {et~al.}(2023)\citenamefont {Davari}, \citenamefont {Ashoorioon},\ and\ \citenamefont {Rezazadeh}}]{Davari:2023tam}%
  \BibitemOpen
  \bibfield  {author} {\bibinfo {author} {\bibfnamefont {Z.}~\bibnamefont {Davari}}, \bibinfo {author} {\bibfnamefont {A.}~\bibnamefont {Ashoorioon}},\ and\ \bibinfo {author} {\bibfnamefont {K.}~\bibnamefont {Rezazadeh}},\ }\href@noop {} {\bibinfo {title} {Spherical collapse approach for non-standard cold dark matter models and enhanced early galaxy formation in jwst}} (\bibinfo {year} {2023}),\ \Eprint {https://arxiv.org/abs/2311.15083} {arXiv:2311.15083 [astro-ph.CO]} \BibitemShut {NoStop}%
\bibitem [{\citenamefont {Santini}\ \emph {et~al.}(2022)\citenamefont {Santini}, \citenamefont {Menci},\ and\ \citenamefont {Castellano}}]{Santini:2022bib}%
  \BibitemOpen
  \bibfield  {author} {\bibinfo {author} {\bibfnamefont {P.}~\bibnamefont {Santini}}, \bibinfo {author} {\bibfnamefont {N.}~\bibnamefont {Menci}},\ and\ \bibinfo {author} {\bibfnamefont {M.}~\bibnamefont {Castellano}},\ }\bibfield  {title} {\bibinfo {title} {{CONSTRAINTS ON DARK ENERGY FROM THE ABUNDANCE OF MASSIVE GALAXIES}},\ }\href@noop {} {\bibfield  {journal} {\bibinfo  {journal} {Frascati Phys. Ser.}\ }\textbf {\bibinfo {volume} {74}},\ \bibinfo {pages} {239} (\bibinfo {year} {2022})},\ \Eprint {https://arxiv.org/abs/2301.03892} {arXiv:2301.03892 [astro-ph.CO]} \BibitemShut {NoStop}%
\bibitem [{\citenamefont {Menci}\ \emph {et~al.}(2022)\citenamefont {Menci}, \citenamefont {Castellano}, \citenamefont {Santini}, \citenamefont {Merlin}, \citenamefont {Fontana},\ and\ \citenamefont {Shankar}}]{Menci:2022wia}%
  \BibitemOpen
  \bibfield  {author} {\bibinfo {author} {\bibfnamefont {N.}~\bibnamefont {Menci}}, \bibinfo {author} {\bibfnamefont {M.}~\bibnamefont {Castellano}}, \bibinfo {author} {\bibfnamefont {P.}~\bibnamefont {Santini}}, \bibinfo {author} {\bibfnamefont {E.}~\bibnamefont {Merlin}}, \bibinfo {author} {\bibfnamefont {A.}~\bibnamefont {Fontana}},\ and\ \bibinfo {author} {\bibfnamefont {F.}~\bibnamefont {Shankar}},\ }\bibfield  {title} {\bibinfo {title} {{High-redshift Galaxies from Early JWST Observations: Constraints on Dark Energy Models}},\ }\href {https://doi.org/10.3847/2041-8213/ac96e9} {\bibfield  {journal} {\bibinfo  {journal} {Astrophys. J. Lett.}\ }\textbf {\bibinfo {volume} {938}},\ \bibinfo {pages} {L5} (\bibinfo {year} {2022})},\ \Eprint {https://arxiv.org/abs/2208.11471} {arXiv:2208.11471 [astro-ph.CO]} \BibitemShut {NoStop}%
\bibitem [{\citenamefont {Wang}\ and\ \citenamefont {Liu}(2023)}]{Wang:2022jvx}%
  \BibitemOpen
  \bibfield  {author} {\bibinfo {author} {\bibfnamefont {D.}~\bibnamefont {Wang}}\ and\ \bibinfo {author} {\bibfnamefont {Y.}~\bibnamefont {Liu}},\ }\href@noop {} {\bibinfo {title} {Jwst high redshift galaxy observations have a strong tension with planck cmb measurements}} (\bibinfo {year} {2023}),\ \Eprint {https://arxiv.org/abs/2301.00347} {arXiv:2301.00347 [astro-ph.CO]} \BibitemShut {NoStop}%
\bibitem [{\citenamefont {Wang}\ \emph {et~al.}(2023{\natexlab{d}})\citenamefont {Wang}, \citenamefont {Su}, \citenamefont {Zu}, \citenamefont {Yang},\ and\ \citenamefont {Feng}}]{Wang:2023ros}%
  \BibitemOpen
  \bibfield  {author} {\bibinfo {author} {\bibfnamefont {P.}~\bibnamefont {Wang}}, \bibinfo {author} {\bibfnamefont {B.-Y.}\ \bibnamefont {Su}}, \bibinfo {author} {\bibfnamefont {L.}~\bibnamefont {Zu}}, \bibinfo {author} {\bibfnamefont {Y.}~\bibnamefont {Yang}},\ and\ \bibinfo {author} {\bibfnamefont {L.}~\bibnamefont {Feng}},\ }\href@noop {} {\bibinfo {title} {Exploring the dark energy equation of state with jwst}} (\bibinfo {year} {2023}{\natexlab{d}}),\ \Eprint {https://arxiv.org/abs/2307.11374} {arXiv:2307.11374 [astro-ph.CO]} \BibitemShut {NoStop}%
\bibitem [{\citenamefont {Adil}\ \emph {et~al.}(2023{\natexlab{a}})\citenamefont {Adil}, \citenamefont {Mukhopadhyay}, \citenamefont {Sen},\ and\ \citenamefont {Vagnozzi}}]{Adil:2023ara}%
  \BibitemOpen
  \bibfield  {author} {\bibinfo {author} {\bibfnamefont {S.~A.}\ \bibnamefont {Adil}}, \bibinfo {author} {\bibfnamefont {U.}~\bibnamefont {Mukhopadhyay}}, \bibinfo {author} {\bibfnamefont {A.~A.}\ \bibnamefont {Sen}},\ and\ \bibinfo {author} {\bibfnamefont {S.}~\bibnamefont {Vagnozzi}},\ }\bibfield  {title} {\bibinfo {title} {{Dark energy in light of the early JWST observations: case for a negative cosmological constant?}},\ }\href {https://doi.org/10.1088/1475-7516/2023/10/072} {\bibfield  {journal} {\bibinfo  {journal} {JCAP}\ }\textbf {\bibinfo {volume} {10}},\ \bibinfo {pages} {072}},\ \Eprint {https://arxiv.org/abs/2307.12763} {arXiv:2307.12763 [astro-ph.CO]} \BibitemShut {NoStop}%
\bibitem [{\citenamefont {Menci}\ \emph {et~al.}(2024)\citenamefont {Menci}, \citenamefont {Adil}, \citenamefont {Mukhopadhyay}, \citenamefont {Sen},\ and\ \citenamefont {Vagnozzi}}]{Menci:2024rbq}%
  \BibitemOpen
  \bibfield  {author} {\bibinfo {author} {\bibfnamefont {N.}~\bibnamefont {Menci}}, \bibinfo {author} {\bibfnamefont {S.~A.}\ \bibnamefont {Adil}}, \bibinfo {author} {\bibfnamefont {U.}~\bibnamefont {Mukhopadhyay}}, \bibinfo {author} {\bibfnamefont {A.~A.}\ \bibnamefont {Sen}},\ and\ \bibinfo {author} {\bibfnamefont {S.}~\bibnamefont {Vagnozzi}},\ }\href@noop {} {\bibinfo {title} {Negative cosmological constant in the dark energy sector: tests from jwst photometric and spectroscopic observations of high-redshift galaxies}} (\bibinfo {year} {2024}),\ \Eprint {https://arxiv.org/abs/2401.12659} {arXiv:2401.12659 [astro-ph.CO]} \BibitemShut {NoStop}%
\bibitem [{\citenamefont {Melia}(2023)}]{Melia:2023dsy}%
  \BibitemOpen
  \bibfield  {author} {\bibinfo {author} {\bibfnamefont {F.}~\bibnamefont {Melia}},\ }\bibfield  {title} {\bibinfo {title} {{The cosmic timeline implied by the JWST high-redshift galaxies}},\ }\href {https://doi.org/10.1093/mnrasl/slad025} {\bibfield  {journal} {\bibinfo  {journal} {Mon. Not. Roy. Astron. Soc.}\ }\textbf {\bibinfo {volume} {521}},\ \bibinfo {pages} {L85} (\bibinfo {year} {2023})},\ \Eprint {https://arxiv.org/abs/2302.10103} {arXiv:2302.10103 [astro-ph.CO]} \BibitemShut {NoStop}%
\bibitem [{\citenamefont {Binici}\ \emph {et~al.}(2024)\citenamefont {Binici}, \citenamefont {Deliduman},\ and\ \citenamefont {Şakir Dilsiz}}]{Binici:2024smk}%
  \BibitemOpen
  \bibfield  {author} {\bibinfo {author} {\bibfnamefont {S.~S.}\ \bibnamefont {Binici}}, \bibinfo {author} {\bibfnamefont {C.}~\bibnamefont {Deliduman}},\ and\ \bibinfo {author} {\bibfnamefont {F.}~\bibnamefont {Şakir Dilsiz}},\ }\href@noop {} {\bibinfo {title} {The ages of the oldest astrophysical objects in an ellipsoidal universe}} (\bibinfo {year} {2024}),\ \Eprint {https://arxiv.org/abs/2402.16646} {arXiv:2402.16646 [astro-ph.CO]} \BibitemShut {NoStop}%
\bibitem [{\citenamefont {Lopez-Corredoira}\ \emph {et~al.}(2024)\citenamefont {Lopez-Corredoira}, \citenamefont {Melia}, \citenamefont {Wei},\ and\ \citenamefont {Gao}}]{lopezcorredoira2024age}%
  \BibitemOpen
  \bibfield  {author} {\bibinfo {author} {\bibfnamefont {M.}~\bibnamefont {Lopez-Corredoira}}, \bibinfo {author} {\bibfnamefont {F.}~\bibnamefont {Melia}}, \bibinfo {author} {\bibfnamefont {J.~J.}\ \bibnamefont {Wei}},\ and\ \bibinfo {author} {\bibfnamefont {C.~Y.}\ \bibnamefont {Gao}},\ }\href@noop {} {\bibinfo {title} {Age of massive galaxies at redshift 8}} (\bibinfo {year} {2024}),\ \Eprint {https://arxiv.org/abs/2405.12665} {arXiv:2405.12665 [astro-ph.CO]} \BibitemShut {NoStop}%
\bibitem [{\citenamefont {Keller}\ \emph {et~al.}(2023)\citenamefont {Keller}, \citenamefont {Munshi}, \citenamefont {Trebitsch},\ and\ \citenamefont {Tremmel}}]{Keller:2022mnb}%
  \BibitemOpen
  \bibfield  {author} {\bibinfo {author} {\bibfnamefont {B.~W.}\ \bibnamefont {Keller}}, \bibinfo {author} {\bibfnamefont {F.}~\bibnamefont {Munshi}}, \bibinfo {author} {\bibfnamefont {M.}~\bibnamefont {Trebitsch}},\ and\ \bibinfo {author} {\bibfnamefont {M.}~\bibnamefont {Tremmel}},\ }\bibfield  {title} {\bibinfo {title} {{Can Cosmological Simulations Reproduce the Spectroscopically Confirmed Galaxies Seen at z \ensuremath{\geq} 10?}},\ }\href {https://doi.org/10.3847/2041-8213/acb148} {\bibfield  {journal} {\bibinfo  {journal} {Astrophys. J. Lett.}\ }\textbf {\bibinfo {volume} {943}},\ \bibinfo {pages} {L28} (\bibinfo {year} {2023})},\ \Eprint {https://arxiv.org/abs/2212.12804} {arXiv:2212.12804 [astro-ph.GA]} \BibitemShut {NoStop}%
\bibitem [{\citenamefont {{Desprez}}\ \emph {et~al.}(2023)\citenamefont {{Desprez}}, \citenamefont {{Martis}}, \citenamefont {{Asada}}, \citenamefont {{Sawicki}}, \citenamefont {{Willott}}, \citenamefont {{Muzzin}}, \citenamefont {{Abraham}}, \citenamefont {{Brada{\v{c}}}}, \citenamefont {{Brammer}}, \citenamefont {{Estrada-Carpenter}}, \citenamefont {{Iyer}}, \citenamefont {{Matharu}}, \citenamefont {{Mowla}}, \citenamefont {{Noirot}}, \citenamefont {{Sarrouh}}, \citenamefont {{Strait}}, \citenamefont {{Gledhill}},\ and\ \citenamefont {{Rihtar{\v{s}}i{\v{c}}}}}]{Desprez}%
  \BibitemOpen
  \bibfield  {author} {\bibinfo {author} {\bibfnamefont {G.}~\bibnamefont {{Desprez}}}, \bibinfo {author} {\bibfnamefont {N.~S.}\ \bibnamefont {{Martis}}}, \bibinfo {author} {\bibfnamefont {Y.}~\bibnamefont {{Asada}}}, \bibinfo {author} {\bibfnamefont {M.}~\bibnamefont {{Sawicki}}}, \bibinfo {author} {\bibfnamefont {C.~J.}\ \bibnamefont {{Willott}}}, \bibinfo {author} {\bibfnamefont {A.}~\bibnamefont {{Muzzin}}}, \bibinfo {author} {\bibfnamefont {R.~G.}\ \bibnamefont {{Abraham}}}, \bibinfo {author} {\bibfnamefont {M.}~\bibnamefont {{Brada{\v{c}}}}}, \bibinfo {author} {\bibfnamefont {G.}~\bibnamefont {{Brammer}}}, \bibinfo {author} {\bibfnamefont {V.}~\bibnamefont {{Estrada-Carpenter}}}, \bibinfo {author} {\bibfnamefont {K.~G.}\ \bibnamefont {{Iyer}}}, \bibinfo {author} {\bibfnamefont {J.}~\bibnamefont {{Matharu}}}, \bibinfo {author} {\bibfnamefont {L.}~\bibnamefont {{Mowla}}}, \bibinfo {author} {\bibfnamefont {G.}~\bibnamefont {{Noirot}}}, \bibinfo {author} {\bibfnamefont {G.~T.~E.}\ \bibnamefont
  {{Sarrouh}}}, \bibinfo {author} {\bibfnamefont {V.}~\bibnamefont {{Strait}}}, \bibinfo {author} {\bibfnamefont {R.}~\bibnamefont {{Gledhill}}},\ and\ \bibinfo {author} {\bibfnamefont {G.}~\bibnamefont {{Rihtar{\v{s}}i{\v{c}}}}},\ }\bibfield  {title} {\bibinfo {title} {{$\Lambda$CDM not dead yet: massive high-z Balmer break galaxies are less common than previously reported}},\ }\href {https://doi.org/10.48550/arXiv.2310.03063} {\bibfield  {journal} {\bibinfo  {journal} {arXiv e-prints}\ ,\ \bibinfo {eid} {arXiv:2310.03063}} (\bibinfo {year} {2023})},\ \Eprint {https://arxiv.org/abs/2310.03063} {arXiv:2310.03063 [astro-ph.GA]} \BibitemShut {NoStop}%
\bibitem [{\citenamefont {Sun}\ \emph {et~al.}(2023)\citenamefont {Sun}, \citenamefont {Faucher-Gigu\`ere}, \citenamefont {Hayward}, \citenamefont {Shen}, \citenamefont {Wetzel},\ and\ \citenamefont {Cochrane}}]{Sun:2023ocn}%
  \BibitemOpen
  \bibfield  {author} {\bibinfo {author} {\bibfnamefont {G.}~\bibnamefont {Sun}}, \bibinfo {author} {\bibfnamefont {C.-A.}\ \bibnamefont {Faucher-Gigu\`ere}}, \bibinfo {author} {\bibfnamefont {C.~C.}\ \bibnamefont {Hayward}}, \bibinfo {author} {\bibfnamefont {X.}~\bibnamefont {Shen}}, \bibinfo {author} {\bibfnamefont {A.}~\bibnamefont {Wetzel}},\ and\ \bibinfo {author} {\bibfnamefont {R.~K.}\ \bibnamefont {Cochrane}},\ }\bibfield  {title} {\bibinfo {title} {{Bursty Star Formation Naturally Explains the Abundance of Bright Galaxies at Cosmic Dawn}},\ }\href {https://doi.org/10.3847/2041-8213/acf85a} {\bibfield  {journal} {\bibinfo  {journal} {Astrophys. J. Lett.}\ }\textbf {\bibinfo {volume} {955}},\ \bibinfo {pages} {L35} (\bibinfo {year} {2023})},\ \Eprint {https://arxiv.org/abs/2307.15305} {arXiv:2307.15305 [astro-ph.GA]} \BibitemShut {NoStop}%
\bibitem [{\citenamefont {McCaffrey}\ \emph {et~al.}(2023)\citenamefont {McCaffrey}, \citenamefont {Hardin}, \citenamefont {Wise},\ and\ \citenamefont {Regan}}]{McCaffrey_2023}%
  \BibitemOpen
  \bibfield  {author} {\bibinfo {author} {\bibfnamefont {J.}~\bibnamefont {McCaffrey}}, \bibinfo {author} {\bibfnamefont {S.}~\bibnamefont {Hardin}}, \bibinfo {author} {\bibfnamefont {J.~H.}\ \bibnamefont {Wise}},\ and\ \bibinfo {author} {\bibfnamefont {J.~A.}\ \bibnamefont {Regan}},\ }\bibfield  {title} {\bibinfo {title} {No tension: Jwst galaxies at $z>10$ consistent with cosmological simulations},\ }\bibfield  {journal} {\bibinfo  {journal} {The Open Journal of Astrophysics}\ }\textbf {\bibinfo {volume} {6}},\ \href {https://doi.org/10.21105/astro.2304.13755} {10.21105/astro.2304.13755} (\bibinfo {year} {2023})\BibitemShut {NoStop}%
\bibitem [{\citenamefont {Di~Valentino}\ \emph {et~al.}(2021)\citenamefont {Di~Valentino}, \citenamefont {Mena}, \citenamefont {Pan}, \citenamefont {Visinelli}, \citenamefont {Yang}, \citenamefont {Melchiorri}, \citenamefont {Mota}, \citenamefont {Riess},\ and\ \citenamefont {Silk}}]{DiValentino:2021izs}%
  \BibitemOpen
  \bibfield  {author} {\bibinfo {author} {\bibfnamefont {E.}~\bibnamefont {Di~Valentino}}, \bibinfo {author} {\bibfnamefont {O.}~\bibnamefont {Mena}}, \bibinfo {author} {\bibfnamefont {S.}~\bibnamefont {Pan}}, \bibinfo {author} {\bibfnamefont {L.}~\bibnamefont {Visinelli}}, \bibinfo {author} {\bibfnamefont {W.}~\bibnamefont {Yang}}, \bibinfo {author} {\bibfnamefont {A.}~\bibnamefont {Melchiorri}}, \bibinfo {author} {\bibfnamefont {D.~F.}\ \bibnamefont {Mota}}, \bibinfo {author} {\bibfnamefont {A.~G.}\ \bibnamefont {Riess}},\ and\ \bibinfo {author} {\bibfnamefont {J.}~\bibnamefont {Silk}},\ }\bibfield  {title} {\bibinfo {title} {{In the realm of the Hubble tension\textemdash{}a review of solutions}},\ }\href {https://doi.org/10.1088/1361-6382/ac086d} {\bibfield  {journal} {\bibinfo  {journal} {Class. Quant. Grav.}\ }\textbf {\bibinfo {volume} {38}},\ \bibinfo {pages} {153001} (\bibinfo {year} {2021})},\ \Eprint {https://arxiv.org/abs/2103.01183} {arXiv:2103.01183 [astro-ph.CO]} \BibitemShut {NoStop}%
\bibitem [{\citenamefont {Perivolaropoulos}\ and\ \citenamefont {Skara}(2022)}]{Perivolaropoulos:2021jda}%
  \BibitemOpen
  \bibfield  {author} {\bibinfo {author} {\bibfnamefont {L.}~\bibnamefont {Perivolaropoulos}}\ and\ \bibinfo {author} {\bibfnamefont {F.}~\bibnamefont {Skara}},\ }\bibfield  {title} {\bibinfo {title} {{Challenges for \ensuremath{\Lambda}CDM: An update}},\ }\href {https://doi.org/10.1016/j.newar.2022.101659} {\bibfield  {journal} {\bibinfo  {journal} {New Astron. Rev.}\ }\textbf {\bibinfo {volume} {95}},\ \bibinfo {pages} {101659} (\bibinfo {year} {2022})},\ \Eprint {https://arxiv.org/abs/2105.05208} {arXiv:2105.05208 [astro-ph.CO]} \BibitemShut {NoStop}%
\bibitem [{\citenamefont {Abdalla}\ \emph {et~al.}(2022)\citenamefont {Abdalla} \emph {et~al.}}]{Abdalla:2022yfr}%
  \BibitemOpen
  \bibfield  {author} {\bibinfo {author} {\bibfnamefont {E.}~\bibnamefont {Abdalla}} \emph {et~al.},\ }\bibfield  {title} {\bibinfo {title} {{Cosmology intertwined: A review of the particle physics, astrophysics, and cosmology associated with the cosmological tensions and anomalies}},\ }\href {https://doi.org/10.1016/j.jheap.2022.04.002} {\bibfield  {journal} {\bibinfo  {journal} {JHEAp}\ }\textbf {\bibinfo {volume} {34}},\ \bibinfo {pages} {49} (\bibinfo {year} {2022})},\ \Eprint {https://arxiv.org/abs/2203.06142} {arXiv:2203.06142 [astro-ph.CO]} \BibitemShut {NoStop}%
\bibitem [{\citenamefont {Aghanim}\ \emph {et~al.}(2020)\citenamefont {Aghanim} \emph {et~al.}}]{Planck:2018vyg}%
  \BibitemOpen
  \bibfield  {author} {\bibinfo {author} {\bibfnamefont {N.}~\bibnamefont {Aghanim}} \emph {et~al.} (\bibinfo {collaboration} {Planck}),\ }\bibfield  {title} {\bibinfo {title} {{Planck 2018 results. VI. Cosmological parameters}},\ }\href {https://doi.org/10.1051/0004-6361/201833910} {\bibfield  {journal} {\bibinfo  {journal} {Astron. Astrophys.}\ }\textbf {\bibinfo {volume} {641}},\ \bibinfo {pages} {A6} (\bibinfo {year} {2020})},\ \bibinfo {note} {[Erratum: Astron.Astrophys. 652, C4 (2021)]},\ \Eprint {https://arxiv.org/abs/1807.06209} {arXiv:1807.06209 [astro-ph.CO]} \BibitemShut {NoStop}%
\bibitem [{\citenamefont {Knox}\ and\ \citenamefont {Millea}(2020)}]{Knox:2019rjx}%
  \BibitemOpen
  \bibfield  {author} {\bibinfo {author} {\bibfnamefont {L.}~\bibnamefont {Knox}}\ and\ \bibinfo {author} {\bibfnamefont {M.}~\bibnamefont {Millea}},\ }\bibfield  {title} {\bibinfo {title} {{Hubble constant hunter\textquoteright{}s guide}},\ }\href {https://doi.org/10.1103/PhysRevD.101.043533} {\bibfield  {journal} {\bibinfo  {journal} {Phys. Rev. D}\ }\textbf {\bibinfo {volume} {101}},\ \bibinfo {pages} {043533} (\bibinfo {year} {2020})},\ \Eprint {https://arxiv.org/abs/1908.03663} {arXiv:1908.03663 [astro-ph.CO]} \BibitemShut {NoStop}%
\bibitem [{\citenamefont {Poulin}\ \emph {et~al.}(2019)\citenamefont {Poulin}, \citenamefont {Smith}, \citenamefont {Karwal},\ and\ \citenamefont {Kamionkowski}}]{Poulin:2018cxd}%
  \BibitemOpen
  \bibfield  {author} {\bibinfo {author} {\bibfnamefont {V.}~\bibnamefont {Poulin}}, \bibinfo {author} {\bibfnamefont {T.~L.}\ \bibnamefont {Smith}}, \bibinfo {author} {\bibfnamefont {T.}~\bibnamefont {Karwal}},\ and\ \bibinfo {author} {\bibfnamefont {M.}~\bibnamefont {Kamionkowski}},\ }\bibfield  {title} {\bibinfo {title} {{Early Dark Energy Can Resolve The Hubble Tension}},\ }\href {https://doi.org/10.1103/PhysRevLett.122.221301} {\bibfield  {journal} {\bibinfo  {journal} {Phys. Rev. Lett.}\ }\textbf {\bibinfo {volume} {122}},\ \bibinfo {pages} {221301} (\bibinfo {year} {2019})},\ \Eprint {https://arxiv.org/abs/1811.04083} {arXiv:1811.04083 [astro-ph.CO]} \BibitemShut {NoStop}%
\bibitem [{\citenamefont {Agrawal}\ \emph {et~al.}(2023)\citenamefont {Agrawal}, \citenamefont {Cyr-Racine}, \citenamefont {Pinner},\ and\ \citenamefont {Randall}}]{Agrawal:2019lmo}%
  \BibitemOpen
  \bibfield  {author} {\bibinfo {author} {\bibfnamefont {P.}~\bibnamefont {Agrawal}}, \bibinfo {author} {\bibfnamefont {F.-Y.}\ \bibnamefont {Cyr-Racine}}, \bibinfo {author} {\bibfnamefont {D.}~\bibnamefont {Pinner}},\ and\ \bibinfo {author} {\bibfnamefont {L.}~\bibnamefont {Randall}},\ }\bibfield  {title} {\bibinfo {title} {{Rock \textquoteleft{}n\textquoteright{} roll solutions to the Hubble tension}},\ }\href {https://doi.org/10.1016/j.dark.2023.101347} {\bibfield  {journal} {\bibinfo  {journal} {Phys. Dark Univ.}\ }\textbf {\bibinfo {volume} {42}},\ \bibinfo {pages} {101347} (\bibinfo {year} {2023})},\ \Eprint {https://arxiv.org/abs/1904.01016} {arXiv:1904.01016 [astro-ph.CO]} \BibitemShut {NoStop}%
\bibitem [{\citenamefont {Lin}\ \emph {et~al.}(2019)\citenamefont {Lin}, \citenamefont {Benevento}, \citenamefont {Hu},\ and\ \citenamefont {Raveri}}]{Lin:2019qug}%
  \BibitemOpen
  \bibfield  {author} {\bibinfo {author} {\bibfnamefont {M.-X.}\ \bibnamefont {Lin}}, \bibinfo {author} {\bibfnamefont {G.}~\bibnamefont {Benevento}}, \bibinfo {author} {\bibfnamefont {W.}~\bibnamefont {Hu}},\ and\ \bibinfo {author} {\bibfnamefont {M.}~\bibnamefont {Raveri}},\ }\bibfield  {title} {\bibinfo {title} {{Acoustic Dark Energy: Potential Conversion of the Hubble Tension}},\ }\href {https://doi.org/10.1103/PhysRevD.100.063542} {\bibfield  {journal} {\bibinfo  {journal} {Phys. Rev. D}\ }\textbf {\bibinfo {volume} {100}},\ \bibinfo {pages} {063542} (\bibinfo {year} {2019})},\ \Eprint {https://arxiv.org/abs/1905.12618} {arXiv:1905.12618 [astro-ph.CO]} \BibitemShut {NoStop}%
\bibitem [{\citenamefont {Niedermann}\ and\ \citenamefont {Sloth}(2021)}]{Niedermann:2019olb}%
  \BibitemOpen
  \bibfield  {author} {\bibinfo {author} {\bibfnamefont {F.}~\bibnamefont {Niedermann}}\ and\ \bibinfo {author} {\bibfnamefont {M.~S.}\ \bibnamefont {Sloth}},\ }\bibfield  {title} {\bibinfo {title} {{New early dark energy}},\ }\href {https://doi.org/10.1103/PhysRevD.103.L041303} {\bibfield  {journal} {\bibinfo  {journal} {Phys. Rev. D}\ }\textbf {\bibinfo {volume} {103}},\ \bibinfo {pages} {L041303} (\bibinfo {year} {2021})},\ \Eprint {https://arxiv.org/abs/1910.10739} {arXiv:1910.10739 [astro-ph.CO]} \BibitemShut {NoStop}%
\bibitem [{\citenamefont {Ye}\ and\ \citenamefont {Piao}(2020)}]{Ye:2020btb}%
  \BibitemOpen
  \bibfield  {author} {\bibinfo {author} {\bibfnamefont {G.}~\bibnamefont {Ye}}\ and\ \bibinfo {author} {\bibfnamefont {Y.-S.}\ \bibnamefont {Piao}},\ }\bibfield  {title} {\bibinfo {title} {{Is the Hubble tension a hint of AdS phase around recombination?}},\ }\href {https://doi.org/10.1103/PhysRevD.101.083507} {\bibfield  {journal} {\bibinfo  {journal} {Phys. Rev. D}\ }\textbf {\bibinfo {volume} {101}},\ \bibinfo {pages} {083507} (\bibinfo {year} {2020})},\ \Eprint {https://arxiv.org/abs/2001.02451} {arXiv:2001.02451 [astro-ph.CO]} \BibitemShut {NoStop}%
\bibitem [{\citenamefont {Poulin}\ \emph {et~al.}(2023)\citenamefont {Poulin}, \citenamefont {Smith},\ and\ \citenamefont {Karwal}}]{Poulin:2023lkg}%
  \BibitemOpen
  \bibfield  {author} {\bibinfo {author} {\bibfnamefont {V.}~\bibnamefont {Poulin}}, \bibinfo {author} {\bibfnamefont {T.~L.}\ \bibnamefont {Smith}},\ and\ \bibinfo {author} {\bibfnamefont {T.}~\bibnamefont {Karwal}},\ }\bibfield  {title} {\bibinfo {title} {{The Ups and Downs of Early Dark Energy solutions to the Hubble tension: A review of models, hints and constraints circa 2023}},\ }\href {https://doi.org/10.1016/j.dark.2023.101348} {\bibfield  {journal} {\bibinfo  {journal} {Phys. Dark Univ.}\ }\textbf {\bibinfo {volume} {42}},\ \bibinfo {pages} {101348} (\bibinfo {year} {2023})},\ \Eprint {https://arxiv.org/abs/2302.09032} {arXiv:2302.09032 [astro-ph.CO]} \BibitemShut {NoStop}%
\bibitem [{\citenamefont {Boylan-Kolchin}(2023)}]{Boylan-Kolchin:2022kae}%
  \BibitemOpen
  \bibfield  {author} {\bibinfo {author} {\bibfnamefont {M.}~\bibnamefont {Boylan-Kolchin}},\ }\bibfield  {title} {\bibinfo {title} {{Stress testing \ensuremath{\Lambda}CDM with high-redshift galaxy candidates}},\ }\href {https://doi.org/10.1038/s41550-023-01937-7} {\bibfield  {journal} {\bibinfo  {journal} {Nature Astron.}\ }\textbf {\bibinfo {volume} {7}},\ \bibinfo {pages} {731} (\bibinfo {year} {2023})},\ \Eprint {https://arxiv.org/abs/2208.01611} {arXiv:2208.01611 [astro-ph.CO]} \BibitemShut {NoStop}%
\bibitem [{\citenamefont {McGaugh}(2024)}]{McGaugh:2023nkc}%
  \BibitemOpen
  \bibfield  {author} {\bibinfo {author} {\bibfnamefont {S.~S.}\ \bibnamefont {McGaugh}},\ }\bibfield  {title} {\bibinfo {title} {{Discord in Concordance Cosmology and Anomalously Massive Early Galaxies}},\ }\href {https://doi.org/10.3390/universe10010048} {\bibfield  {journal} {\bibinfo  {journal} {Universe}\ }\textbf {\bibinfo {volume} {10}},\ \bibinfo {pages} {48} (\bibinfo {year} {2024})},\ \Eprint {https://arxiv.org/abs/2312.03127} {arXiv:2312.03127 [astro-ph.CO]} \BibitemShut {NoStop}%
\bibitem [{\citenamefont {Liu}\ \emph {et~al.}(2024)\citenamefont {Liu}, \citenamefont {Zhan}, \citenamefont {Gong},\ and\ \citenamefont {Wang}}]{Liu:2024yan}%
  \BibitemOpen
  \bibfield  {author} {\bibinfo {author} {\bibfnamefont {W.}~\bibnamefont {Liu}}, \bibinfo {author} {\bibfnamefont {H.}~\bibnamefont {Zhan}}, \bibinfo {author} {\bibfnamefont {Y.}~\bibnamefont {Gong}},\ and\ \bibinfo {author} {\bibfnamefont {X.}~\bibnamefont {Wang}},\ }\href@noop {} {\bibinfo {title} {Can early dark energy be probed by the high-redshift galaxy abundance?}} (\bibinfo {year} {2024}),\ \Eprint {https://arxiv.org/abs/2402.14339} {arXiv:2402.14339 [astro-ph.CO]} \BibitemShut {NoStop}%
\bibitem [{\citenamefont {Forconi}\ \emph {et~al.}(2023{\natexlab{b}})\citenamefont {Forconi}, \citenamefont {Giarè}, \citenamefont {Mena}, \citenamefont {Ruchika}, \citenamefont {Valentino}, \citenamefont {Melchiorri},\ and\ \citenamefont {Nunes}}]{Forconi:2023hsj}%
  \BibitemOpen
  \bibfield  {author} {\bibinfo {author} {\bibfnamefont {M.}~\bibnamefont {Forconi}}, \bibinfo {author} {\bibfnamefont {W.}~\bibnamefont {Giarè}}, \bibinfo {author} {\bibfnamefont {O.}~\bibnamefont {Mena}}, \bibinfo {author} {\bibnamefont {Ruchika}}, \bibinfo {author} {\bibfnamefont {E.~D.}\ \bibnamefont {Valentino}}, \bibinfo {author} {\bibfnamefont {A.}~\bibnamefont {Melchiorri}},\ and\ \bibinfo {author} {\bibfnamefont {R.~C.}\ \bibnamefont {Nunes}},\ }\href@noop {} {\bibinfo {title} {A double take on early and interacting dark energy from jwst}} (\bibinfo {year} {2023}{\natexlab{b}}),\ \Eprint {https://arxiv.org/abs/2312.11074} {arXiv:2312.11074 [astro-ph.CO]} \BibitemShut {NoStop}%
\bibitem [{\citenamefont {Jedamzik}\ \emph {et~al.}(2021)\citenamefont {Jedamzik}, \citenamefont {Pogosian},\ and\ \citenamefont {Zhao}}]{Jedamzik:2020zmd}%
  \BibitemOpen
  \bibfield  {author} {\bibinfo {author} {\bibfnamefont {K.}~\bibnamefont {Jedamzik}}, \bibinfo {author} {\bibfnamefont {L.}~\bibnamefont {Pogosian}},\ and\ \bibinfo {author} {\bibfnamefont {G.-B.}\ \bibnamefont {Zhao}},\ }\bibfield  {title} {\bibinfo {title} {{Why reducing the cosmic sound horizon alone can not fully resolve the Hubble tension}},\ }\href {https://doi.org/10.1038/s42005-021-00628-x} {\bibfield  {journal} {\bibinfo  {journal} {Commun. in Phys.}\ }\textbf {\bibinfo {volume} {4}},\ \bibinfo {pages} {123} (\bibinfo {year} {2021})},\ \Eprint {https://arxiv.org/abs/2010.04158} {arXiv:2010.04158 [astro-ph.CO]} \BibitemShut {NoStop}%
\bibitem [{\citenamefont {van Putten}(2024)}]{vanPutten:2023ths}%
  \BibitemOpen
  \bibfield  {author} {\bibinfo {author} {\bibfnamefont {M.~H. P.~M.}\ \bibnamefont {van Putten}},\ }\bibfield  {title} {\bibinfo {title} {{The Fast and Furious in JWST high-z galaxies}},\ }\href {https://doi.org/10.1016/j.dark.2023.101417} {\bibfield  {journal} {\bibinfo  {journal} {Phys. Dark Univ.}\ }\textbf {\bibinfo {volume} {43}},\ \bibinfo {pages} {101417} (\bibinfo {year} {2024})},\ \Eprint {https://arxiv.org/abs/2312.16692} {arXiv:2312.16692 [astro-ph.CO]} \BibitemShut {NoStop}%
\bibitem [{\citenamefont {Risaliti}\ and\ \citenamefont {Lusso}(2015)}]{Risaliti:2015zla}%
  \BibitemOpen
  \bibfield  {author} {\bibinfo {author} {\bibfnamefont {G.}~\bibnamefont {Risaliti}}\ and\ \bibinfo {author} {\bibfnamefont {E.}~\bibnamefont {Lusso}},\ }\bibfield  {title} {\bibinfo {title} {{A Hubble Diagram for Quasars}},\ }\href {https://doi.org/10.1088/0004-637X/815/1/33} {\bibfield  {journal} {\bibinfo  {journal} {Astrophys. J.}\ }\textbf {\bibinfo {volume} {815}},\ \bibinfo {pages} {33} (\bibinfo {year} {2015})},\ \Eprint {https://arxiv.org/abs/1505.07118} {arXiv:1505.07118 [astro-ph.CO]} \BibitemShut {NoStop}%
\bibitem [{\citenamefont {Risaliti}\ and\ \citenamefont {Lusso}(2019)}]{Risaliti:2018reu}%
  \BibitemOpen
  \bibfield  {author} {\bibinfo {author} {\bibfnamefont {G.}~\bibnamefont {Risaliti}}\ and\ \bibinfo {author} {\bibfnamefont {E.}~\bibnamefont {Lusso}},\ }\bibfield  {title} {\bibinfo {title} {{Cosmological constraints from the Hubble diagram of quasars at high redshifts}},\ }\href {https://doi.org/10.1038/s41550-018-0657-z} {\bibfield  {journal} {\bibinfo  {journal} {Nature Astron.}\ }\textbf {\bibinfo {volume} {3}},\ \bibinfo {pages} {272} (\bibinfo {year} {2019})},\ \Eprint {https://arxiv.org/abs/1811.02590} {arXiv:1811.02590 [astro-ph.CO]} \BibitemShut {NoStop}%
\bibitem [{\citenamefont {Lusso}\ \emph {et~al.}(2020)\citenamefont {Lusso} \emph {et~al.}}]{Lusso:2020pdb}%
  \BibitemOpen
  \bibfield  {author} {\bibinfo {author} {\bibfnamefont {E.}~\bibnamefont {Lusso}} \emph {et~al.},\ }\bibfield  {title} {\bibinfo {title} {{Quasars as standard candles III. Validation of a new sample for cosmological studies}},\ }\href {https://doi.org/10.1051/0004-6361/202038899} {\bibfield  {journal} {\bibinfo  {journal} {Astron. Astrophys.}\ }\textbf {\bibinfo {volume} {642}},\ \bibinfo {pages} {A150} (\bibinfo {year} {2020})},\ \Eprint {https://arxiv.org/abs/2008.08586} {arXiv:2008.08586 [astro-ph.GA]} \BibitemShut {NoStop}%
\bibitem [{\citenamefont {Yang}\ \emph {et~al.}(2020)\citenamefont {Yang}, \citenamefont {Banerjee},\ and\ \citenamefont {\'O~Colg\'ain}}]{Yang:2019vgk}%
  \BibitemOpen
  \bibfield  {author} {\bibinfo {author} {\bibfnamefont {T.}~\bibnamefont {Yang}}, \bibinfo {author} {\bibfnamefont {A.}~\bibnamefont {Banerjee}},\ and\ \bibinfo {author} {\bibfnamefont {E.}~\bibnamefont {\'O~Colg\'ain}},\ }\bibfield  {title} {\bibinfo {title} {{Cosmography and flat $\Lambda$CDM tensions at high redshift}},\ }\href {https://doi.org/10.1103/PhysRevD.102.123532} {\bibfield  {journal} {\bibinfo  {journal} {Phys. Rev. D}\ }\textbf {\bibinfo {volume} {102}},\ \bibinfo {pages} {123532} (\bibinfo {year} {2020})},\ \Eprint {https://arxiv.org/abs/1911.01681} {arXiv:1911.01681 [astro-ph.CO]} \BibitemShut {NoStop}%
\bibitem [{\citenamefont {Khadka}\ and\ \citenamefont {Ratra}(2020)}]{Khadka:2020vlh}%
  \BibitemOpen
  \bibfield  {author} {\bibinfo {author} {\bibfnamefont {N.}~\bibnamefont {Khadka}}\ and\ \bibinfo {author} {\bibfnamefont {B.}~\bibnamefont {Ratra}},\ }\bibfield  {title} {\bibinfo {title} {{Using quasar X-ray and UV flux measurements to constrain cosmological model parameters}},\ }\href {https://doi.org/10.1093/mnras/staa1855} {\bibfield  {journal} {\bibinfo  {journal} {Mon. Not. Roy. Astron. Soc.}\ }\textbf {\bibinfo {volume} {497}},\ \bibinfo {pages} {263} (\bibinfo {year} {2020})},\ \Eprint {https://arxiv.org/abs/2004.09979} {arXiv:2004.09979 [astro-ph.CO]} \BibitemShut {NoStop}%
\bibitem [{\citenamefont {Khadka}\ and\ \citenamefont {Ratra}(2021)}]{Khadka:2020tlm}%
  \BibitemOpen
  \bibfield  {author} {\bibinfo {author} {\bibfnamefont {N.}~\bibnamefont {Khadka}}\ and\ \bibinfo {author} {\bibfnamefont {B.}~\bibnamefont {Ratra}},\ }\bibfield  {title} {\bibinfo {title} {{Determining the range of validity of quasar X-ray and UV flux measurements for constraining cosmological model parameters}},\ }\href {https://doi.org/10.1093/mnras/stab486} {\bibfield  {journal} {\bibinfo  {journal} {Mon. Not. Roy. Astron. Soc.}\ }\textbf {\bibinfo {volume} {502}},\ \bibinfo {pages} {6140} (\bibinfo {year} {2021})},\ \Eprint {https://arxiv.org/abs/2012.09291} {arXiv:2012.09291 [astro-ph.CO]} \BibitemShut {NoStop}%
\bibitem [{\citenamefont {Khadka}\ and\ \citenamefont {Ratra}(2022)}]{Khadka:2021xcc}%
  \BibitemOpen
  \bibfield  {author} {\bibinfo {author} {\bibfnamefont {N.}~\bibnamefont {Khadka}}\ and\ \bibinfo {author} {\bibfnamefont {B.}~\bibnamefont {Ratra}},\ }\bibfield  {title} {\bibinfo {title} {{Do quasar X-ray and UV flux measurements provide a useful test of cosmological models?}},\ }\href {https://doi.org/10.1093/mnras/stab3678} {\bibfield  {journal} {\bibinfo  {journal} {Mon. Not. Roy. Astron. Soc.}\ }\textbf {\bibinfo {volume} {510}},\ \bibinfo {pages} {2753} (\bibinfo {year} {2022})},\ \Eprint {https://arxiv.org/abs/2107.07600} {arXiv:2107.07600 [astro-ph.CO]} \BibitemShut {NoStop}%
\bibitem [{\citenamefont {Khadka}\ \emph {et~al.}(2023)\citenamefont {Khadka}, \citenamefont {Zaja\v{c}ek}, \citenamefont {Prince}, \citenamefont {Panda}, \citenamefont {Czerny}, \citenamefont {Mart\'\i{}nez-Aldama}, \citenamefont {Jaiswal},\ and\ \citenamefont {Ratra}}]{Khadka:2022aeg}%
  \BibitemOpen
  \bibfield  {author} {\bibinfo {author} {\bibfnamefont {N.}~\bibnamefont {Khadka}}, \bibinfo {author} {\bibfnamefont {M.}~\bibnamefont {Zaja\v{c}ek}}, \bibinfo {author} {\bibfnamefont {R.}~\bibnamefont {Prince}}, \bibinfo {author} {\bibfnamefont {S.}~\bibnamefont {Panda}}, \bibinfo {author} {\bibfnamefont {B.}~\bibnamefont {Czerny}}, \bibinfo {author} {\bibfnamefont {M.~L.}\ \bibnamefont {Mart\'\i{}nez-Aldama}}, \bibinfo {author} {\bibfnamefont {V.~K.}\ \bibnamefont {Jaiswal}},\ and\ \bibinfo {author} {\bibfnamefont {B.}~\bibnamefont {Ratra}},\ }\bibfield  {title} {\bibinfo {title} {{Quasar UV/X-ray relation luminosity distances are shorter than reverberation-measured radius\textendash{}luminosity relation luminosity distances}},\ }\href {https://doi.org/10.1093/mnras/stad1040} {\bibfield  {journal} {\bibinfo  {journal} {Mon. Not. Roy. Astron. Soc.}\ }\textbf {\bibinfo {volume} {522}},\ \bibinfo {pages} {1247} (\bibinfo {year} {2023})},\ \Eprint {https://arxiv.org/abs/2212.10483} {arXiv:2212.10483
  [astro-ph.CO]} \BibitemShut {NoStop}%
\bibitem [{\citenamefont {Singal}\ \emph {et~al.}(2022)\citenamefont {Singal}, \citenamefont {Mutchnick},\ and\ \citenamefont {Petrosian}}]{Singal:2022nto}%
  \BibitemOpen
  \bibfield  {author} {\bibinfo {author} {\bibfnamefont {J.}~\bibnamefont {Singal}}, \bibinfo {author} {\bibfnamefont {S.}~\bibnamefont {Mutchnick}},\ and\ \bibinfo {author} {\bibfnamefont {V.}~\bibnamefont {Petrosian}},\ }\bibfield  {title} {\bibinfo {title} {{The X-Ray Luminosity Function Evolution of Quasars and the Correlation between the X-Ray and Ultraviolet Luminosities}},\ }\href {https://doi.org/10.3847/1538-4357/ac6f06} {\bibfield  {journal} {\bibinfo  {journal} {Astrophys. J.}\ }\textbf {\bibinfo {volume} {932}},\ \bibinfo {pages} {111} (\bibinfo {year} {2022})},\ \Eprint {https://arxiv.org/abs/2203.13374} {arXiv:2203.13374 [astro-ph.CO]} \BibitemShut {NoStop}%
\bibitem [{\citenamefont {Petrosian}\ \emph {et~al.}(2022)\citenamefont {Petrosian}, \citenamefont {Singal},\ and\ \citenamefont {Mutchnick}}]{Petrosian:2022tlp}%
  \BibitemOpen
  \bibfield  {author} {\bibinfo {author} {\bibfnamefont {V.}~\bibnamefont {Petrosian}}, \bibinfo {author} {\bibfnamefont {J.}~\bibnamefont {Singal}},\ and\ \bibinfo {author} {\bibfnamefont {S.}~\bibnamefont {Mutchnick}},\ }\bibfield  {title} {\bibinfo {title} {{Can the Distance-Redshift Relation be Determined from Correlations between Luminosities?}},\ }\href {https://doi.org/10.3847/2041-8213/ac85ac} {\bibfield  {journal} {\bibinfo  {journal} {Astrophys. J. Lett.}\ }\textbf {\bibinfo {volume} {935}},\ \bibinfo {pages} {L19} (\bibinfo {year} {2022})},\ \Eprint {https://arxiv.org/abs/2205.07981} {arXiv:2205.07981 [astro-ph.CO]} \BibitemShut {NoStop}%
\bibitem [{\citenamefont {Zaja\v{c}ek}\ \emph {et~al.}(2024)\citenamefont {Zaja\v{c}ek}, \citenamefont {Czerny}, \citenamefont {Khadka}, \citenamefont {Mart\'\i{}nez-Aldama}, \citenamefont {Prince}, \citenamefont {Panda},\ and\ \citenamefont {Ratra}}]{Zajacek:2023qjm}%
  \BibitemOpen
  \bibfield  {author} {\bibinfo {author} {\bibfnamefont {M.}~\bibnamefont {Zaja\v{c}ek}}, \bibinfo {author} {\bibfnamefont {B.}~\bibnamefont {Czerny}}, \bibinfo {author} {\bibfnamefont {N.}~\bibnamefont {Khadka}}, \bibinfo {author} {\bibfnamefont {M.~L.}\ \bibnamefont {Mart\'\i{}nez-Aldama}}, \bibinfo {author} {\bibfnamefont {R.}~\bibnamefont {Prince}}, \bibinfo {author} {\bibfnamefont {S.}~\bibnamefont {Panda}},\ and\ \bibinfo {author} {\bibfnamefont {B.}~\bibnamefont {Ratra}},\ }\bibfield  {title} {\bibinfo {title} {{Effect of Extinction on Quasar Luminosity Distances Determined from UV and X-Ray Flux Measurements}},\ }\href {https://doi.org/10.3847/1538-4357/ad11dc} {\bibfield  {journal} {\bibinfo  {journal} {Astrophys. J.}\ }\textbf {\bibinfo {volume} {961}},\ \bibinfo {pages} {229} (\bibinfo {year} {2024})},\ \Eprint {https://arxiv.org/abs/2305.08179} {arXiv:2305.08179 [astro-ph.GA]} \BibitemShut {NoStop}%
\bibitem [{\citenamefont {Cao}\ and\ \citenamefont {Ratra}(2024)}]{Cao:2024vmo}%
  \BibitemOpen
  \bibfield  {author} {\bibinfo {author} {\bibfnamefont {S.}~\bibnamefont {Cao}}\ and\ \bibinfo {author} {\bibfnamefont {B.}~\bibnamefont {Ratra}},\ }\href@noop {} {\bibinfo {title} {Testing the standardizability of, and deriving cosmological constraints from, a new amati-correlated gamma-ray burst data compilation}} (\bibinfo {year} {2024}),\ \Eprint {https://arxiv.org/abs/2404.08697} {arXiv:2404.08697 [astro-ph.CO]} \BibitemShut {NoStop}%
\bibitem [{\citenamefont {G\'omez-Valent}(2022)}]{Gomez-Valent:2022hkb}%
  \BibitemOpen
  \bibfield  {author} {\bibinfo {author} {\bibfnamefont {A.}~\bibnamefont {G\'omez-Valent}},\ }\bibfield  {title} {\bibinfo {title} {{Fast test to assess the impact of marginalization in Monte~Carlo analyses and its application to cosmology}},\ }\href {https://doi.org/10.1103/PhysRevD.106.063506} {\bibfield  {journal} {\bibinfo  {journal} {Phys. Rev. D}\ }\textbf {\bibinfo {volume} {106}},\ \bibinfo {pages} {063506} (\bibinfo {year} {2022})},\ \Eprint {https://arxiv.org/abs/2203.16285} {arXiv:2203.16285 [astro-ph.CO]} \BibitemShut {NoStop}%
\bibitem [{\citenamefont {Herold}\ \emph {et~al.}(2022)\citenamefont {Herold}, \citenamefont {Ferreira},\ and\ \citenamefont {Komatsu}}]{Herold:2021ksg}%
  \BibitemOpen
  \bibfield  {author} {\bibinfo {author} {\bibfnamefont {L.}~\bibnamefont {Herold}}, \bibinfo {author} {\bibfnamefont {E.~G.~M.}\ \bibnamefont {Ferreira}},\ and\ \bibinfo {author} {\bibfnamefont {E.}~\bibnamefont {Komatsu}},\ }\bibfield  {title} {\bibinfo {title} {{New Constraint on Early Dark Energy from Planck and BOSS Data Using the Profile Likelihood}},\ }\href {https://doi.org/10.3847/2041-8213/ac63a3} {\bibfield  {journal} {\bibinfo  {journal} {Astrophys. J. Lett.}\ }\textbf {\bibinfo {volume} {929}},\ \bibinfo {pages} {L16} (\bibinfo {year} {2022})},\ \Eprint {https://arxiv.org/abs/2112.12140} {arXiv:2112.12140 [astro-ph.CO]} \BibitemShut {NoStop}%
\bibitem [{\citenamefont {Colg\'ain}\ \emph {et~al.}(2023)\citenamefont {Colg\'ain}, \citenamefont {Pourojaghi}, \citenamefont {Sheikh-Jabbari},\ and\ \citenamefont {Sherwin}}]{Colgain:2023bge}%
  \BibitemOpen
  \bibfield  {author} {\bibinfo {author} {\bibfnamefont {E.~O.}\ \bibnamefont {Colg\'ain}}, \bibinfo {author} {\bibfnamefont {S.}~\bibnamefont {Pourojaghi}}, \bibinfo {author} {\bibfnamefont {M.~M.}\ \bibnamefont {Sheikh-Jabbari}},\ and\ \bibinfo {author} {\bibfnamefont {D.}~\bibnamefont {Sherwin}},\ }\href@noop {} {\bibinfo {title} {{MCMC Marginalisation Bias and $\Lambda$CDM tensions}}} (\bibinfo {year} {2023}),\ \Eprint {https://arxiv.org/abs/2307.16349} {arXiv:2307.16349 [astro-ph.CO]} \BibitemShut {NoStop}%
\bibitem [{\citenamefont {Sheth}\ and\ \citenamefont {Tormen}(1999)}]{Sheth:1999mn}%
  \BibitemOpen
  \bibfield  {author} {\bibinfo {author} {\bibfnamefont {R.~K.}\ \bibnamefont {Sheth}}\ and\ \bibinfo {author} {\bibfnamefont {G.}~\bibnamefont {Tormen}},\ }\bibfield  {title} {\bibinfo {title} {{Large scale bias and the peak background split}},\ }\href {https://doi.org/10.1046/j.1365-8711.1999.02692.x} {\bibfield  {journal} {\bibinfo  {journal} {Mon. Not. Roy. Astron. Soc.}\ }\textbf {\bibinfo {volume} {308}},\ \bibinfo {pages} {119} (\bibinfo {year} {1999})},\ \Eprint {https://arxiv.org/abs/astro-ph/9901122} {arXiv:astro-ph/9901122} \BibitemShut {NoStop}%
\bibitem [{\citenamefont {Akarsu}\ \emph {et~al.}(2024{\natexlab{a}})\citenamefont {Akarsu}, \citenamefont {{\'O Colg\'ain}}, \citenamefont {Sen},\ and\ \citenamefont {Sheikh-Jabbari}}]{akarsu2024lambdacdm}%
  \BibitemOpen
  \bibfield  {author} {\bibinfo {author} {\bibfnamefont {O.}~\bibnamefont {Akarsu}}, \bibinfo {author} {\bibfnamefont {E.}~\bibnamefont {{\'O Colg\'ain}}}, \bibinfo {author} {\bibfnamefont {A.~A.}\ \bibnamefont {Sen}},\ and\ \bibinfo {author} {\bibfnamefont {M.~M.}\ \bibnamefont {Sheikh-Jabbari}},\ }\href@noop {} {\bibinfo {title} {$\lambda$cdm tensions: Localising missing physics through consistency checks}} (\bibinfo {year} {2024}{\natexlab{a}}),\ \Eprint {https://arxiv.org/abs/2402.04767} {arXiv:2402.04767 [astro-ph.CO]} \BibitemShut {NoStop}%
\bibitem [{\citenamefont {Trotta}(2017)}]{Trotta:2017wnx}%
  \BibitemOpen
  \bibfield  {author} {\bibinfo {author} {\bibfnamefont {R.}~\bibnamefont {Trotta}},\ }\bibfield  {title} {\bibinfo {title} {{Bayesian Methods in Cosmology}}\ }(\bibinfo {year} {2017})\ \Eprint {https://arxiv.org/abs/1701.01467} {arXiv:1701.01467 [astro-ph.CO]} \BibitemShut {NoStop}%
\bibitem [{\citenamefont {Wilks}(1938)}]{Wilks}%
  \BibitemOpen
  \bibfield  {author} {\bibinfo {author} {\bibfnamefont {S.~S.}\ \bibnamefont {Wilks}},\ }\bibfield  {title} {\bibinfo {title} {{The Large-Sample Distribution of the Likelihood Ratio for Testing Composite Hypotheses}},\ }\href {https://doi.org/10.1214/aoms/1177732360} {\bibfield  {journal} {\bibinfo  {journal} {The Annals of Mathematical Statistics}\ }\textbf {\bibinfo {volume} {9}},\ \bibinfo {pages} {60 } (\bibinfo {year} {1938})}\BibitemShut {NoStop}%
\bibitem [{\citenamefont {Feldman}\ and\ \citenamefont {Cousins}(1998)}]{Feldman:1997qc}%
  \BibitemOpen
  \bibfield  {author} {\bibinfo {author} {\bibfnamefont {G.~J.}\ \bibnamefont {Feldman}}\ and\ \bibinfo {author} {\bibfnamefont {R.~D.}\ \bibnamefont {Cousins}},\ }\bibfield  {title} {\bibinfo {title} {{A Unified approach to the classical statistical analysis of small signals}},\ }\href {https://doi.org/10.1103/PhysRevD.57.3873} {\bibfield  {journal} {\bibinfo  {journal} {Phys. Rev. D}\ }\textbf {\bibinfo {volume} {57}},\ \bibinfo {pages} {3873} (\bibinfo {year} {1998})},\ \Eprint {https://arxiv.org/abs/physics/9711021} {arXiv:physics/9711021} \BibitemShut {NoStop}%
\bibitem [{\citenamefont {{Galloni}}\ \emph {et~al.}(2024)\citenamefont {{Galloni}}, \citenamefont {{Henrot-Versill{\'e}}},\ and\ \citenamefont {{Tristram}}}]{2024arXiv240504455G}%
  \BibitemOpen
  \bibfield  {author} {\bibinfo {author} {\bibfnamefont {G.}~\bibnamefont {{Galloni}}}, \bibinfo {author} {\bibfnamefont {S.}~\bibnamefont {{Henrot-Versill{\'e}}}},\ and\ \bibinfo {author} {\bibfnamefont {M.}~\bibnamefont {{Tristram}}},\ }\bibfield  {title} {\bibinfo {title} {{Robust constraints on tensor perturbations from cosmological data: a comparative analysis from Bayesian and frequentist perspectives}},\ }\href {https://doi.org/10.48550/arXiv.2405.04455} {\bibfield  {journal} {\bibinfo  {journal} {arXiv e-prints}\ ,\ \bibinfo {eid} {arXiv:2405.04455}} (\bibinfo {year} {2024})},\ \Eprint {https://arxiv.org/abs/2405.04455} {arXiv:2405.04455 [astro-ph.CO]} \BibitemShut {NoStop}%
\bibitem [{\citenamefont {Karwal}\ \emph {et~al.}(2024)\citenamefont {Karwal}, \citenamefont {Patel}, \citenamefont {Bartlett}, \citenamefont {Poulin}, \citenamefont {Smith},\ and\ \citenamefont {Pfeffer}}]{Karwal:2024qpt}%
  \BibitemOpen
  \bibfield  {author} {\bibinfo {author} {\bibfnamefont {T.}~\bibnamefont {Karwal}}, \bibinfo {author} {\bibfnamefont {Y.}~\bibnamefont {Patel}}, \bibinfo {author} {\bibfnamefont {A.}~\bibnamefont {Bartlett}}, \bibinfo {author} {\bibfnamefont {V.}~\bibnamefont {Poulin}}, \bibinfo {author} {\bibfnamefont {T.~L.}\ \bibnamefont {Smith}},\ and\ \bibinfo {author} {\bibfnamefont {D.~N.}\ \bibnamefont {Pfeffer}},\ }\href@noop {} {\bibinfo {title} {Procoli: Profiles of cosmological likelihoods}} (\bibinfo {year} {2024}),\ \Eprint {https://arxiv.org/abs/2401.14225} {arXiv:2401.14225 [astro-ph.CO]} \BibitemShut {NoStop}%
\bibitem [{\citenamefont {\'O~Colg\'ain}\ \emph {et~al.}(2022)\citenamefont {\'O~Colg\'ain}, \citenamefont {Sheikh-Jabbari}, \citenamefont {Solomon}, \citenamefont {Bargiacchi}, \citenamefont {Capozziello}, \citenamefont {Dainotti},\ and\ \citenamefont {Stojkovic}}]{Colgain:2022nlb}%
  \BibitemOpen
  \bibfield  {author} {\bibinfo {author} {\bibfnamefont {E.}~\bibnamefont {\'O~Colg\'ain}}, \bibinfo {author} {\bibfnamefont {M.~M.}\ \bibnamefont {Sheikh-Jabbari}}, \bibinfo {author} {\bibfnamefont {R.}~\bibnamefont {Solomon}}, \bibinfo {author} {\bibfnamefont {G.}~\bibnamefont {Bargiacchi}}, \bibinfo {author} {\bibfnamefont {S.}~\bibnamefont {Capozziello}}, \bibinfo {author} {\bibfnamefont {M.~G.}\ \bibnamefont {Dainotti}},\ and\ \bibinfo {author} {\bibfnamefont {D.}~\bibnamefont {Stojkovic}},\ }\bibfield  {title} {\bibinfo {title} {{Revealing intrinsic flat \ensuremath{\Lambda}CDM biases with standardizable candles}},\ }\href {https://doi.org/10.1103/PhysRevD.106.L041301} {\bibfield  {journal} {\bibinfo  {journal} {Phys. Rev. D}\ }\textbf {\bibinfo {volume} {106}},\ \bibinfo {pages} {L041301} (\bibinfo {year} {2022})},\ \Eprint {https://arxiv.org/abs/2203.10558} {arXiv:2203.10558 [astro-ph.CO]} \BibitemShut {NoStop}%
\bibitem [{\citenamefont {\'O~Colg\'ain}\ \emph {et~al.}(2024)\citenamefont {\'O~Colg\'ain}, \citenamefont {Sheikh-Jabbari}, \citenamefont {Solomon}, \citenamefont {Dainotti},\ and\ \citenamefont {Stojkovic}}]{Colgain:2022rxy}%
  \BibitemOpen
  \bibfield  {author} {\bibinfo {author} {\bibfnamefont {E.}~\bibnamefont {\'O~Colg\'ain}}, \bibinfo {author} {\bibfnamefont {M.~M.}\ \bibnamefont {Sheikh-Jabbari}}, \bibinfo {author} {\bibfnamefont {R.}~\bibnamefont {Solomon}}, \bibinfo {author} {\bibfnamefont {M.~G.}\ \bibnamefont {Dainotti}},\ and\ \bibinfo {author} {\bibfnamefont {D.}~\bibnamefont {Stojkovic}},\ }\bibfield  {title} {\bibinfo {title} {{Putting flat \ensuremath{\Lambda}CDM in the (Redshift) bin}},\ }\href {https://doi.org/10.1016/j.dark.2024.101464} {\bibfield  {journal} {\bibinfo  {journal} {Phys. Dark Univ.}\ }\textbf {\bibinfo {volume} {44}},\ \bibinfo {pages} {101464} (\bibinfo {year} {2024})},\ \Eprint {https://arxiv.org/abs/2206.11447} {arXiv:2206.11447 [astro-ph.CO]} \BibitemShut {NoStop}%
\bibitem [{\citenamefont {Past\'en}\ and\ \citenamefont {C\'ardenas}(2023)}]{Pasten:2023rpc}%
  \BibitemOpen
  \bibfield  {author} {\bibinfo {author} {\bibfnamefont {E.}~\bibnamefont {Past\'en}}\ and\ \bibinfo {author} {\bibfnamefont {V.~H.}\ \bibnamefont {C\'ardenas}},\ }\bibfield  {title} {\bibinfo {title} {{Testing \ensuremath{\Lambda}CDM cosmology in a binned universe: Anomalies in the deceleration parameter}},\ }\href {https://doi.org/10.1016/j.dark.2023.101224} {\bibfield  {journal} {\bibinfo  {journal} {Phys. Dark Univ.}\ }\textbf {\bibinfo {volume} {40}},\ \bibinfo {pages} {101224} (\bibinfo {year} {2023})},\ \Eprint {https://arxiv.org/abs/2301.10740} {arXiv:2301.10740 [astro-ph.CO]} \BibitemShut {NoStop}%
\bibitem [{\citenamefont {Malekjani}\ \emph {et~al.}(2024)\citenamefont {Malekjani}, \citenamefont {Conville}, \citenamefont {Colg\'ain}, \citenamefont {Pourojaghi},\ and\ \citenamefont {Sheikh-Jabbari}}]{Malekjani:2023dky}%
  \BibitemOpen
  \bibfield  {author} {\bibinfo {author} {\bibfnamefont {M.}~\bibnamefont {Malekjani}}, \bibinfo {author} {\bibfnamefont {R.~M.}\ \bibnamefont {Conville}}, \bibinfo {author} {\bibfnamefont {E.~O.}\ \bibnamefont {Colg\'ain}}, \bibinfo {author} {\bibfnamefont {S.}~\bibnamefont {Pourojaghi}},\ and\ \bibinfo {author} {\bibfnamefont {M.~M.}\ \bibnamefont {Sheikh-Jabbari}},\ }\bibfield  {title} {\bibinfo {title} {{On redshift evolution and negative dark energy density in Pantheon + Supernovae}},\ }\href {https://doi.org/10.1140/epjc/s10052-024-12667-z} {\bibfield  {journal} {\bibinfo  {journal} {Eur. Phys. J. C}\ }\textbf {\bibinfo {volume} {84}},\ \bibinfo {pages} {317} (\bibinfo {year} {2024})},\ \Eprint {https://arxiv.org/abs/2301.12725} {arXiv:2301.12725 [astro-ph.CO]} \BibitemShut {NoStop}%
\bibitem [{\citenamefont {Weinberg}(1989)}]{Weinberg:1988cp}%
  \BibitemOpen
  \bibfield  {author} {\bibinfo {author} {\bibfnamefont {S.}~\bibnamefont {Weinberg}},\ }\bibfield  {title} {\bibinfo {title} {{The Cosmological Constant Problem}},\ }\href {https://doi.org/10.1103/RevModPhys.61.1} {\bibfield  {journal} {\bibinfo  {journal} {Rev. Mod. Phys.}\ }\textbf {\bibinfo {volume} {61}},\ \bibinfo {pages} {1} (\bibinfo {year} {1989})}\BibitemShut {NoStop}%
\bibitem [{\citenamefont {Dvali}\ and\ \citenamefont {Gomez}(2016)}]{Dvali:2014gua}%
  \BibitemOpen
  \bibfield  {author} {\bibinfo {author} {\bibfnamefont {G.}~\bibnamefont {Dvali}}\ and\ \bibinfo {author} {\bibfnamefont {C.}~\bibnamefont {Gomez}},\ }\bibfield  {title} {\bibinfo {title} {{Quantum Exclusion of Positive Cosmological Constant?}},\ }\href {https://doi.org/10.1002/andp.201500216} {\bibfield  {journal} {\bibinfo  {journal} {Annalen Phys.}\ }\textbf {\bibinfo {volume} {528}},\ \bibinfo {pages} {68} (\bibinfo {year} {2016})},\ \Eprint {https://arxiv.org/abs/1412.8077} {arXiv:1412.8077 [hep-th]} \BibitemShut {NoStop}%
\bibitem [{\citenamefont {Dvali}\ and\ \citenamefont {Gomez}(2019)}]{Dvali:2018fqu}%
  \BibitemOpen
  \bibfield  {author} {\bibinfo {author} {\bibfnamefont {G.}~\bibnamefont {Dvali}}\ and\ \bibinfo {author} {\bibfnamefont {C.}~\bibnamefont {Gomez}},\ }\bibfield  {title} {\bibinfo {title} {{On Exclusion of Positive Cosmological Constant}},\ }\href {https://doi.org/10.1002/prop.201800092} {\bibfield  {journal} {\bibinfo  {journal} {Fortsch. Phys.}\ }\textbf {\bibinfo {volume} {67}},\ \bibinfo {pages} {1800092} (\bibinfo {year} {2019})},\ \Eprint {https://arxiv.org/abs/1806.10877} {arXiv:1806.10877 [hep-th]} \BibitemShut {NoStop}%
\bibitem [{\citenamefont {{Obied}}\ \emph {et~al.}(2018)\citenamefont {{Obied}}, \citenamefont {{Ooguri}}, \citenamefont {{Spodyneiko}},\ and\ \citenamefont {{Vafa}}}]{Obied:2018sgi}%
  \BibitemOpen
  \bibfield  {author} {\bibinfo {author} {\bibfnamefont {G.}~\bibnamefont {{Obied}}}, \bibinfo {author} {\bibfnamefont {H.}~\bibnamefont {{Ooguri}}}, \bibinfo {author} {\bibfnamefont {L.}~\bibnamefont {{Spodyneiko}}},\ and\ \bibinfo {author} {\bibfnamefont {C.}~\bibnamefont {{Vafa}}},\ }\bibfield  {title} {\bibinfo {title} {{De Sitter Space and the Swampland}},\ }\href {https://doi.org/10.48550/arXiv.1806.08362} {\bibfield  {journal} {\bibinfo  {journal} {arXiv e-prints}\ ,\ \bibinfo {eid} {arXiv:1806.08362}} (\bibinfo {year} {2018})},\ \Eprint {https://arxiv.org/abs/1806.08362} {arXiv:1806.08362 [hep-th]} \BibitemShut {NoStop}%
\bibitem [{\citenamefont {Brout}\ \emph {et~al.}(2022)\citenamefont {Brout} \emph {et~al.}}]{Brout:2022vxf}%
  \BibitemOpen
  \bibfield  {author} {\bibinfo {author} {\bibfnamefont {D.}~\bibnamefont {Brout}} \emph {et~al.},\ }\bibfield  {title} {\bibinfo {title} {{The Pantheon+ Analysis: Cosmological Constraints}},\ }\href {https://doi.org/10.3847/1538-4357/ac8e04} {\bibfield  {journal} {\bibinfo  {journal} {Astrophys. J.}\ }\textbf {\bibinfo {volume} {938}},\ \bibinfo {pages} {110} (\bibinfo {year} {2022})},\ \Eprint {https://arxiv.org/abs/2202.04077} {arXiv:2202.04077 [astro-ph.CO]} \BibitemShut {NoStop}%
\bibitem [{\citenamefont {Lusso}\ and\ \citenamefont {Risaliti}(2017)}]{Lusso:2017hgz}%
  \BibitemOpen
  \bibfield  {author} {\bibinfo {author} {\bibfnamefont {E.}~\bibnamefont {Lusso}}\ and\ \bibinfo {author} {\bibfnamefont {G.}~\bibnamefont {Risaliti}},\ }\bibfield  {title} {\bibinfo {title} {{Quasars as standard candles I: The physical relation between disc and coronal emission}},\ }\href {https://doi.org/10.1051/0004-6361/201630079} {\bibfield  {journal} {\bibinfo  {journal} {Astron. Astrophys.}\ }\textbf {\bibinfo {volume} {602}},\ \bibinfo {pages} {A79} (\bibinfo {year} {2017})},\ \Eprint {https://arxiv.org/abs/1703.05299} {arXiv:1703.05299 [astro-ph.HE]} \BibitemShut {NoStop}%
\bibitem [{\citenamefont {Kubota}\ and\ \citenamefont {Done}(2018)}]{Kubota:2018cuj}%
  \BibitemOpen
  \bibfield  {author} {\bibinfo {author} {\bibfnamefont {A.}~\bibnamefont {Kubota}}\ and\ \bibinfo {author} {\bibfnamefont {C.}~\bibnamefont {Done}},\ }\bibfield  {title} {\bibinfo {title} {{A physical model of the broad-band continuum of AGN and its implications for the UV/X relation and optical variability}},\ }\href {https://doi.org/10.1093/mnras/sty1890} {\bibfield  {journal} {\bibinfo  {journal} {Mon. Not. Roy. Astron. Soc.}\ }\textbf {\bibinfo {volume} {480}},\ \bibinfo {pages} {1247} (\bibinfo {year} {2018})},\ \Eprint {https://arxiv.org/abs/1804.00171} {arXiv:1804.00171 [astro-ph.HE]} \BibitemShut {NoStop}%
\bibitem [{\citenamefont {Arcodia}\ \emph {et~al.}(2019)\citenamefont {Arcodia}, \citenamefont {Merloni}, \citenamefont {Nandra},\ and\ \citenamefont {Ponti}}]{Arcodia:2019oyq}%
  \BibitemOpen
  \bibfield  {author} {\bibinfo {author} {\bibfnamefont {R.}~\bibnamefont {Arcodia}}, \bibinfo {author} {\bibfnamefont {A.}~\bibnamefont {Merloni}}, \bibinfo {author} {\bibfnamefont {K.}~\bibnamefont {Nandra}},\ and\ \bibinfo {author} {\bibfnamefont {G.}~\bibnamefont {Ponti}},\ }\bibfield  {title} {\bibinfo {title} {{Testing the disk-corona interplay in radiatively-efficient broad-line AGN}},\ }\href {https://doi.org/10.1051/0004-6361/201935874} {\bibfield  {journal} {\bibinfo  {journal} {Astron. Astrophys.}\ }\textbf {\bibinfo {volume} {628}},\ \bibinfo {pages} {A135} (\bibinfo {year} {2019})},\ \Eprint {https://arxiv.org/abs/1907.10069} {arXiv:1907.10069 [astro-ph.HE]} \BibitemShut {NoStop}%
\bibitem [{\citenamefont {Lusso}\ \emph {et~al.}(2019)\citenamefont {Lusso}, \citenamefont {Piedipalumbo}, \citenamefont {Risaliti}, \citenamefont {Paolillo}, \citenamefont {Bisogni}, \citenamefont {Nardini},\ and\ \citenamefont {Amati}}]{Lusso:2019akb}%
  \BibitemOpen
  \bibfield  {author} {\bibinfo {author} {\bibfnamefont {E.}~\bibnamefont {Lusso}}, \bibinfo {author} {\bibfnamefont {E.}~\bibnamefont {Piedipalumbo}}, \bibinfo {author} {\bibfnamefont {G.}~\bibnamefont {Risaliti}}, \bibinfo {author} {\bibfnamefont {M.}~\bibnamefont {Paolillo}}, \bibinfo {author} {\bibfnamefont {S.}~\bibnamefont {Bisogni}}, \bibinfo {author} {\bibfnamefont {E.}~\bibnamefont {Nardini}},\ and\ \bibinfo {author} {\bibfnamefont {L.}~\bibnamefont {Amati}},\ }\bibfield  {title} {\bibinfo {title} {{Tension with the flat $\Lambda$CDM model from a high-redshift Hubble diagram of supernovae, quasars, and gamma-ray bursts}},\ }\href {https://doi.org/10.1051/0004-6361/201936223} {\bibfield  {journal} {\bibinfo  {journal} {Astron. Astrophys.}\ }\textbf {\bibinfo {volume} {628}},\ \bibinfo {pages} {L4} (\bibinfo {year} {2019})},\ \Eprint {https://arxiv.org/abs/1907.07692} {arXiv:1907.07692 [astro-ph.CO]} \BibitemShut {NoStop}%
\bibitem [{\citenamefont {Dainotti}\ \emph {et~al.}(2022)\citenamefont {Dainotti}, \citenamefont {Bardiacchi}, \citenamefont {Lenart}, \citenamefont {Capozziello}, \citenamefont {\'O~Colg\'ain}, \citenamefont {Solomon}, \citenamefont {Stojkovic},\ and\ \citenamefont {Sheikh-Jabbari}}]{Dainotti:2022rfz}%
  \BibitemOpen
  \bibfield  {author} {\bibinfo {author} {\bibfnamefont {M.~G.}\ \bibnamefont {Dainotti}}, \bibinfo {author} {\bibfnamefont {G.}~\bibnamefont {Bardiacchi}}, \bibinfo {author} {\bibfnamefont {A.~L.}\ \bibnamefont {Lenart}}, \bibinfo {author} {\bibfnamefont {S.}~\bibnamefont {Capozziello}}, \bibinfo {author} {\bibfnamefont {E.}~\bibnamefont {\'O~Colg\'ain}}, \bibinfo {author} {\bibfnamefont {R.}~\bibnamefont {Solomon}}, \bibinfo {author} {\bibfnamefont {D.}~\bibnamefont {Stojkovic}},\ and\ \bibinfo {author} {\bibfnamefont {M.~M.}\ \bibnamefont {Sheikh-Jabbari}},\ }\bibfield  {title} {\bibinfo {title} {{Quasar Standardization: Overcoming Selection Biases and Redshift Evolution}},\ }\href {https://doi.org/10.3847/1538-4357/ac6593} {\bibfield  {journal} {\bibinfo  {journal} {Astrophys. J.}\ }\textbf {\bibinfo {volume} {931}},\ \bibinfo {pages} {106} (\bibinfo {year} {2022})},\ \Eprint {https://arxiv.org/abs/2203.12914} {arXiv:2203.12914 [astro-ph.HE]} \BibitemShut {NoStop}%
\bibitem [{\citenamefont {Dainotti}\ \emph {et~al.}(2024)\citenamefont {Dainotti}, \citenamefont {Lenart}, \citenamefont {Yengejeh}, \citenamefont {Chakraborty}, \citenamefont {Fraija}, \citenamefont {Di~Valentino},\ and\ \citenamefont {Montani}}]{Dainotti:2024aha}%
  \BibitemOpen
  \bibfield  {author} {\bibinfo {author} {\bibfnamefont {M.~G.}\ \bibnamefont {Dainotti}}, \bibinfo {author} {\bibfnamefont {A.~L.}\ \bibnamefont {Lenart}}, \bibinfo {author} {\bibfnamefont {M.~G.}\ \bibnamefont {Yengejeh}}, \bibinfo {author} {\bibfnamefont {S.}~\bibnamefont {Chakraborty}}, \bibinfo {author} {\bibfnamefont {N.}~\bibnamefont {Fraija}}, \bibinfo {author} {\bibfnamefont {E.}~\bibnamefont {Di~Valentino}},\ and\ \bibinfo {author} {\bibfnamefont {G.}~\bibnamefont {Montani}},\ }\bibfield  {title} {\bibinfo {title} {{A new binning method to choose a standard set of Quasars}},\ }\href {https://doi.org/10.1016/j.dark.2024.101428} {\bibfield  {journal} {\bibinfo  {journal} {Phys. Dark Univ.}\ }\textbf {\bibinfo {volume} {44}},\ \bibinfo {pages} {101428} (\bibinfo {year} {2024})},\ \Eprint {https://arxiv.org/abs/2401.12847} {arXiv:2401.12847 [astro-ph.HE]} \BibitemShut {NoStop}%
\bibitem [{\citenamefont {Trefoloni}\ \emph {et~al.}(2024)\citenamefont {Trefoloni}, \citenamefont {Lusso}, \citenamefont {Nardini}, \citenamefont {Risaliti}, \citenamefont {Marconi}, \citenamefont {Bargiacchi}, \citenamefont {Sacchi},\ and\ \citenamefont {Signorini}}]{Trefoloni:2024dei}%
  \BibitemOpen
  \bibfield  {author} {\bibinfo {author} {\bibfnamefont {B.}~\bibnamefont {Trefoloni}}, \bibinfo {author} {\bibfnamefont {E.}~\bibnamefont {Lusso}}, \bibinfo {author} {\bibfnamefont {E.}~\bibnamefont {Nardini}}, \bibinfo {author} {\bibfnamefont {G.}~\bibnamefont {Risaliti}}, \bibinfo {author} {\bibfnamefont {A.}~\bibnamefont {Marconi}}, \bibinfo {author} {\bibfnamefont {G.}~\bibnamefont {Bargiacchi}}, \bibinfo {author} {\bibfnamefont {A.}~\bibnamefont {Sacchi}},\ and\ \bibinfo {author} {\bibfnamefont {M.}~\bibnamefont {Signorini}},\ }\href@noop {} {\bibinfo {title} {Quasars as standard candles vi: spectroscopic validation of the cosmological sample}} (\bibinfo {year} {2024}),\ \Eprint {https://arxiv.org/abs/2404.07205} {arXiv:2404.07205 [astro-ph.GA]} \BibitemShut {NoStop}%
\bibitem [{\citenamefont {Moresco}\ \emph {et~al.}(2022)\citenamefont {Moresco} \emph {et~al.}}]{Moresco:2022phi}%
  \BibitemOpen
  \bibfield  {author} {\bibinfo {author} {\bibfnamefont {M.}~\bibnamefont {Moresco}} \emph {et~al.},\ }\bibfield  {title} {\bibinfo {title} {{Unveiling the Universe with emerging cosmological probes}},\ }\href {https://doi.org/10.1007/s41114-022-00040-z} {\bibfield  {journal} {\bibinfo  {journal} {Living Rev. Rel.}\ }\textbf {\bibinfo {volume} {25}},\ \bibinfo {pages} {6} (\bibinfo {year} {2022})},\ \Eprint {https://arxiv.org/abs/2201.07241} {arXiv:2201.07241 [astro-ph.CO]} \BibitemShut {NoStop}%
\bibitem [{\citenamefont {Czerny}\ \emph {et~al.}(2023)\citenamefont {Czerny} \emph {et~al.}}]{Czerny:2022xfj}%
  \BibitemOpen
  \bibfield  {author} {\bibinfo {author} {\bibfnamefont {B.}~\bibnamefont {Czerny}} \emph {et~al.},\ }\bibfield  {title} {\bibinfo {title} {{Accretion disks, quasars and cosmology: meandering towards understanding}},\ }\href {https://doi.org/10.1007/s10509-023-04165-7} {\bibfield  {journal} {\bibinfo  {journal} {Astrophys. Space Sci.}\ }\textbf {\bibinfo {volume} {368}},\ \bibinfo {pages} {8} (\bibinfo {year} {2023})},\ \Eprint {https://arxiv.org/abs/2209.06563} {arXiv:2209.06563 [astro-ph.GA]} \BibitemShut {NoStop}%
\bibitem [{\citenamefont {Vignali}\ \emph {et~al.}(2003)\citenamefont {Vignali}, \citenamefont {Brandt},\ and\ \citenamefont {Schneider}}]{Vignali:2002ct}%
  \BibitemOpen
  \bibfield  {author} {\bibinfo {author} {\bibfnamefont {C.}~\bibnamefont {Vignali}}, \bibinfo {author} {\bibfnamefont {W.~N.}\ \bibnamefont {Brandt}},\ and\ \bibinfo {author} {\bibfnamefont {D.~P.}\ \bibnamefont {Schneider}},\ }\bibfield  {title} {\bibinfo {title} {{X-ray emission from radio - quiet quasars in the SDSS Early Data Release. The Alpha(ox) dependence upon UV luminosity}},\ }\href {https://doi.org/10.1086/345973} {\bibfield  {journal} {\bibinfo  {journal} {Astron. J.}\ }\textbf {\bibinfo {volume} {125}},\ \bibinfo {pages} {433} (\bibinfo {year} {2003})},\ \Eprint {https://arxiv.org/abs/astro-ph/0211125} {arXiv:astro-ph/0211125} \BibitemShut {NoStop}%
\bibitem [{\citenamefont {Just}\ \emph {et~al.}(2007)\citenamefont {Just}, \citenamefont {Brandt}, \citenamefont {Shemmer}, \citenamefont {Steffen}, \citenamefont {Schneider}, \citenamefont {Chartas},\ and\ \citenamefont {Garmire}}]{Just:2007se}%
  \BibitemOpen
  \bibfield  {author} {\bibinfo {author} {\bibfnamefont {D.~W.}\ \bibnamefont {Just}}, \bibinfo {author} {\bibfnamefont {W.~N.}\ \bibnamefont {Brandt}}, \bibinfo {author} {\bibfnamefont {O.}~\bibnamefont {Shemmer}}, \bibinfo {author} {\bibfnamefont {A.~T.}\ \bibnamefont {Steffen}}, \bibinfo {author} {\bibfnamefont {D.~P.}\ \bibnamefont {Schneider}}, \bibinfo {author} {\bibfnamefont {G.}~\bibnamefont {Chartas}},\ and\ \bibinfo {author} {\bibfnamefont {G.~P.}\ \bibnamefont {Garmire}},\ }\bibfield  {title} {\bibinfo {title} {{The X-ray Properties of the Most-Luminous Quasars from the Sloan Digital Sky Survey}},\ }\href {https://doi.org/10.1086/519990} {\bibfield  {journal} {\bibinfo  {journal} {Astrophys. J.}\ }\textbf {\bibinfo {volume} {665}},\ \bibinfo {pages} {1004} (\bibinfo {year} {2007})},\ \Eprint {https://arxiv.org/abs/0705.3059} {arXiv:0705.3059 [astro-ph]} \BibitemShut {NoStop}%
\bibitem [{\citenamefont {Lusso}\ \emph {et~al.}(2010)\citenamefont {Lusso}, \citenamefont {Comastri}, \citenamefont {Vignali}, \citenamefont {Zamorani}, \citenamefont {Brusa}, \citenamefont {Gilli}, \citenamefont {Iwasawa}, \citenamefont {Salvato}, \citenamefont {Civano}, \citenamefont {Elvis}, \citenamefont {Merloni}, \citenamefont {Bongiorno}, \citenamefont {Trump}, \citenamefont {Koekemoer}, \citenamefont {Schinnerer}, \citenamefont {Le~Floc’h}, \citenamefont {Cappelluti}, \citenamefont {Jahnke}, \citenamefont {Sargent}, \citenamefont {Silverman}, \citenamefont {Mainieri}, \citenamefont {Fiore}, \citenamefont {Bolzonella}, \citenamefont {Le~Fèvre}, \citenamefont {Garilli}, \citenamefont {Iovino}, \citenamefont {Kneib}, \citenamefont {Lamareille}, \citenamefont {Lilly}, \citenamefont {Mignoli}, \citenamefont {Scodeggio},\ and\ \citenamefont {Vergani}}]{Lusso_2010}%
  \BibitemOpen
  \bibfield  {author} {\bibinfo {author} {\bibfnamefont {E.}~\bibnamefont {Lusso}}, \bibinfo {author} {\bibfnamefont {A.}~\bibnamefont {Comastri}}, \bibinfo {author} {\bibfnamefont {C.}~\bibnamefont {Vignali}}, \bibinfo {author} {\bibfnamefont {G.}~\bibnamefont {Zamorani}}, \bibinfo {author} {\bibfnamefont {M.}~\bibnamefont {Brusa}}, \bibinfo {author} {\bibfnamefont {R.}~\bibnamefont {Gilli}}, \bibinfo {author} {\bibfnamefont {K.}~\bibnamefont {Iwasawa}}, \bibinfo {author} {\bibfnamefont {M.}~\bibnamefont {Salvato}}, \bibinfo {author} {\bibfnamefont {F.}~\bibnamefont {Civano}}, \bibinfo {author} {\bibfnamefont {M.}~\bibnamefont {Elvis}}, \bibinfo {author} {\bibfnamefont {A.}~\bibnamefont {Merloni}}, \bibinfo {author} {\bibfnamefont {A.}~\bibnamefont {Bongiorno}}, \bibinfo {author} {\bibfnamefont {J.~R.}\ \bibnamefont {Trump}}, \bibinfo {author} {\bibfnamefont {A.~M.}\ \bibnamefont {Koekemoer}}, \bibinfo {author} {\bibfnamefont {E.}~\bibnamefont {Schinnerer}}, \bibinfo {author} {\bibfnamefont {E.}~\bibnamefont
  {Le~Floc’h}}, \bibinfo {author} {\bibfnamefont {N.}~\bibnamefont {Cappelluti}}, \bibinfo {author} {\bibfnamefont {K.}~\bibnamefont {Jahnke}}, \bibinfo {author} {\bibfnamefont {M.}~\bibnamefont {Sargent}}, \bibinfo {author} {\bibfnamefont {J.}~\bibnamefont {Silverman}}, \bibinfo {author} {\bibfnamefont {V.}~\bibnamefont {Mainieri}}, \bibinfo {author} {\bibfnamefont {F.}~\bibnamefont {Fiore}}, \bibinfo {author} {\bibfnamefont {M.}~\bibnamefont {Bolzonella}}, \bibinfo {author} {\bibfnamefont {O.}~\bibnamefont {Le~Fèvre}}, \bibinfo {author} {\bibfnamefont {B.}~\bibnamefont {Garilli}}, \bibinfo {author} {\bibfnamefont {A.}~\bibnamefont {Iovino}}, \bibinfo {author} {\bibfnamefont {J.~P.}\ \bibnamefont {Kneib}}, \bibinfo {author} {\bibfnamefont {F.}~\bibnamefont {Lamareille}}, \bibinfo {author} {\bibfnamefont {S.}~\bibnamefont {Lilly}}, \bibinfo {author} {\bibfnamefont {M.}~\bibnamefont {Mignoli}}, \bibinfo {author} {\bibfnamefont {M.}~\bibnamefont {Scodeggio}},\ and\ \bibinfo {author} {\bibfnamefont
  {D.}~\bibnamefont {Vergani}},\ }\bibfield  {title} {\bibinfo {title} {The x-ray to optical-uv luminosity ratio of x-ray selected type 1 agn in xmm-cosmos},\ }\href {https://doi.org/10.1051/0004-6361/200913298} {\bibfield  {journal} {\bibinfo  {journal} {Astronomy and Astrophysics}\ }\textbf {\bibinfo {volume} {512}},\ \bibinfo {pages} {A34} (\bibinfo {year} {2010})}\BibitemShut {NoStop}%
\bibitem [{\citenamefont {Salvestrini}\ \emph {et~al.}(2019)\citenamefont {Salvestrini}, \citenamefont {Risaliti}, \citenamefont {Bisogni}, \citenamefont {Lusso},\ and\ \citenamefont {Vignali}}]{Salvestrini:2019thn}%
  \BibitemOpen
  \bibfield  {author} {\bibinfo {author} {\bibfnamefont {F.}~\bibnamefont {Salvestrini}}, \bibinfo {author} {\bibfnamefont {G.}~\bibnamefont {Risaliti}}, \bibinfo {author} {\bibfnamefont {S.}~\bibnamefont {Bisogni}}, \bibinfo {author} {\bibfnamefont {E.}~\bibnamefont {Lusso}},\ and\ \bibinfo {author} {\bibfnamefont {C.}~\bibnamefont {Vignali}},\ }\bibfield  {title} {\bibinfo {title} {{Quasars as standard candles II: The non linear relation between UV and X-ray emission at high redshifts}},\ }\href {https://doi.org/10.1051/0004-6361/201935491} {\bibfield  {journal} {\bibinfo  {journal} {Astron. Astrophys.}\ }\textbf {\bibinfo {volume} {631}},\ \bibinfo {pages} {A120} (\bibinfo {year} {2019})},\ \Eprint {https://arxiv.org/abs/1909.12309} {arXiv:1909.12309 [astro-ph.GA]} \BibitemShut {NoStop}%
\bibitem [{\citenamefont {Bisogni}\ \emph {et~al.}(2021)\citenamefont {Bisogni}, \citenamefont {Lusso}, \citenamefont {Civano}, \citenamefont {Nardini}, \citenamefont {Risaliti}, \citenamefont {Elvis},\ and\ \citenamefont {Fabbiano}}]{Bisogni:2021hue}%
  \BibitemOpen
  \bibfield  {author} {\bibinfo {author} {\bibfnamefont {S.}~\bibnamefont {Bisogni}}, \bibinfo {author} {\bibfnamefont {E.}~\bibnamefont {Lusso}}, \bibinfo {author} {\bibfnamefont {F.}~\bibnamefont {Civano}}, \bibinfo {author} {\bibfnamefont {E.}~\bibnamefont {Nardini}}, \bibinfo {author} {\bibfnamefont {G.}~\bibnamefont {Risaliti}}, \bibinfo {author} {\bibfnamefont {M.}~\bibnamefont {Elvis}},\ and\ \bibinfo {author} {\bibfnamefont {G.}~\bibnamefont {Fabbiano}},\ }\bibfield  {title} {\bibinfo {title} {{The Chandra view of the relation between X-ray and UV emission in quasars}},\ }\href {https://doi.org/10.1051/0004-6361/202140852} {\bibfield  {journal} {\bibinfo  {journal} {Astron. Astrophys.}\ }\textbf {\bibinfo {volume} {655}},\ \bibinfo {pages} {A109} (\bibinfo {year} {2021})},\ \Eprint {https://arxiv.org/abs/2109.03252} {arXiv:2109.03252 [astro-ph.GA]} \BibitemShut {NoStop}%
\bibitem [{\citenamefont {Luongo}\ \emph {et~al.}(2022)\citenamefont {Luongo}, \citenamefont {Muccino}, \citenamefont {\'O~Colg\'ain}, \citenamefont {Sheikh-Jabbari},\ and\ \citenamefont {Yin}}]{Luongo:2021nqh}%
  \BibitemOpen
  \bibfield  {author} {\bibinfo {author} {\bibfnamefont {O.}~\bibnamefont {Luongo}}, \bibinfo {author} {\bibfnamefont {M.}~\bibnamefont {Muccino}}, \bibinfo {author} {\bibfnamefont {E.}~\bibnamefont {\'O~Colg\'ain}}, \bibinfo {author} {\bibfnamefont {M.~M.}\ \bibnamefont {Sheikh-Jabbari}},\ and\ \bibinfo {author} {\bibfnamefont {L.}~\bibnamefont {Yin}},\ }\bibfield  {title} {\bibinfo {title} {{Larger H0 values in the CMB dipole direction}},\ }\href {https://doi.org/10.1103/PhysRevD.105.103510} {\bibfield  {journal} {\bibinfo  {journal} {Phys. Rev. D}\ }\textbf {\bibinfo {volume} {105}},\ \bibinfo {pages} {103510} (\bibinfo {year} {2022})},\ \Eprint {https://arxiv.org/abs/2108.13228} {arXiv:2108.13228 [astro-ph.CO]} \BibitemShut {NoStop}%
\bibitem [{\citenamefont {Jia}\ \emph {et~al.}(2022)\citenamefont {Jia}, \citenamefont {Hu}, \citenamefont {Yang}, \citenamefont {Zhang},\ and\ \citenamefont {Wang}}]{Jia:2022mwb}%
  \BibitemOpen
  \bibfield  {author} {\bibinfo {author} {\bibfnamefont {X.~D.}\ \bibnamefont {Jia}}, \bibinfo {author} {\bibfnamefont {J.~P.}\ \bibnamefont {Hu}}, \bibinfo {author} {\bibfnamefont {J.}~\bibnamefont {Yang}}, \bibinfo {author} {\bibfnamefont {B.~B.}\ \bibnamefont {Zhang}},\ and\ \bibinfo {author} {\bibfnamefont {F.~Y.}\ \bibnamefont {Wang}},\ }\bibfield  {title} {\bibinfo {title} {{Eiso\textendash{}Ep correlation of gamma-ray bursts: calibration and cosmological applications}},\ }\href {https://doi.org/10.1093/mnras/stac2356} {\bibfield  {journal} {\bibinfo  {journal} {Mon. Not. Roy. Astron. Soc.}\ }\textbf {\bibinfo {volume} {516}},\ \bibinfo {pages} {2575} (\bibinfo {year} {2022})},\ \Eprint {https://arxiv.org/abs/2208.09272} {arXiv:2208.09272 [astro-ph.HE]} \BibitemShut {NoStop}%
\bibitem [{\citenamefont {Khadka}\ \emph {et~al.}(2021)\citenamefont {Khadka}, \citenamefont {Luongo}, \citenamefont {Muccino},\ and\ \citenamefont {Ratra}}]{Khadka:2021vqa}%
  \BibitemOpen
  \bibfield  {author} {\bibinfo {author} {\bibfnamefont {N.}~\bibnamefont {Khadka}}, \bibinfo {author} {\bibfnamefont {O.}~\bibnamefont {Luongo}}, \bibinfo {author} {\bibfnamefont {M.}~\bibnamefont {Muccino}},\ and\ \bibinfo {author} {\bibfnamefont {B.}~\bibnamefont {Ratra}},\ }\bibfield  {title} {\bibinfo {title} {Do gamma-ray burst measurements provide a useful test of cosmological models?},\ }\href {https://doi.org/10.1088/1475-7516/2021/09/042} {\bibfield  {journal} {\bibinfo  {journal} {JCAP}\ }\textbf {\bibinfo {volume} {09}},\ \bibinfo {pages} {042}},\ \Eprint {https://arxiv.org/abs/2105.12692} {arXiv:2105.12692 [astro-ph.CO]} \BibitemShut {NoStop}%
\bibitem [{\citenamefont {Amati}\ \emph {et~al.}(2002)\citenamefont {Amati} \emph {et~al.}}]{Amati:2002ny}%
  \BibitemOpen
  \bibfield  {author} {\bibinfo {author} {\bibfnamefont {L.}~\bibnamefont {Amati}} \emph {et~al.},\ }\bibfield  {title} {\bibinfo {title} {{Intrinsic spectra and energetics of BeppoSAX gamma-ray bursts with known redshifts}},\ }\href {https://doi.org/10.1051/0004-6361:20020722} {\bibfield  {journal} {\bibinfo  {journal} {Astron. Astrophys.}\ }\textbf {\bibinfo {volume} {390}},\ \bibinfo {pages} {81} (\bibinfo {year} {2002})},\ \Eprint {https://arxiv.org/abs/astro-ph/0205230} {arXiv:astro-ph/0205230} \BibitemShut {NoStop}%
\bibitem [{\citenamefont {Velten}\ and\ \citenamefont {Gomes}(2020)}]{Velten:2019vwo}%
  \BibitemOpen
  \bibfield  {author} {\bibinfo {author} {\bibfnamefont {H.}~\bibnamefont {Velten}}\ and\ \bibinfo {author} {\bibfnamefont {S.}~\bibnamefont {Gomes}},\ }\bibfield  {title} {\bibinfo {title} {{Is the Hubble diagram of quasars in tension with concordance cosmology?}},\ }\href {https://doi.org/10.1103/PhysRevD.101.043502} {\bibfield  {journal} {\bibinfo  {journal} {Phys. Rev. D}\ }\textbf {\bibinfo {volume} {101}},\ \bibinfo {pages} {043502} (\bibinfo {year} {2020})},\ \Eprint {https://arxiv.org/abs/1911.11848} {arXiv:1911.11848 [astro-ph.CO]} \BibitemShut {NoStop}%
\bibitem [{\citenamefont {Holm}\ \emph {et~al.}(2023{\natexlab{a}})\citenamefont {Holm}, \citenamefont {Herold}, \citenamefont {Simon}, \citenamefont {Ferreira}, \citenamefont {Hannestad}, \citenamefont {Poulin},\ and\ \citenamefont {Tram}}]{Holm:2023laa}%
  \BibitemOpen
  \bibfield  {author} {\bibinfo {author} {\bibfnamefont {E.~B.}\ \bibnamefont {Holm}}, \bibinfo {author} {\bibfnamefont {L.}~\bibnamefont {Herold}}, \bibinfo {author} {\bibfnamefont {T.}~\bibnamefont {Simon}}, \bibinfo {author} {\bibfnamefont {E.~G.~M.}\ \bibnamefont {Ferreira}}, \bibinfo {author} {\bibfnamefont {S.}~\bibnamefont {Hannestad}}, \bibinfo {author} {\bibfnamefont {V.}~\bibnamefont {Poulin}},\ and\ \bibinfo {author} {\bibfnamefont {T.}~\bibnamefont {Tram}},\ }\bibfield  {title} {\bibinfo {title} {{Bayesian and frequentist investigation of prior effects in EFT of LSS analyses of full-shape BOSS and eBOSS data}},\ }\href {https://doi.org/10.1103/PhysRevD.108.123514} {\bibfield  {journal} {\bibinfo  {journal} {Phys. Rev. D}\ }\textbf {\bibinfo {volume} {108}},\ \bibinfo {pages} {123514} (\bibinfo {year} {2023}{\natexlab{a}})},\ \Eprint {https://arxiv.org/abs/2309.04468} {arXiv:2309.04468 [astro-ph.CO]} \BibitemShut {NoStop}%
\bibitem [{\citenamefont {Holm}\ \emph {et~al.}(2023{\natexlab{b}})\citenamefont {Holm}, \citenamefont {Nygaard}, \citenamefont {Dakin}, \citenamefont {Hannestad},\ and\ \citenamefont {Tram}}]{holm2023prospect}%
  \BibitemOpen
  \bibfield  {author} {\bibinfo {author} {\bibfnamefont {E.~B.}\ \bibnamefont {Holm}}, \bibinfo {author} {\bibfnamefont {A.}~\bibnamefont {Nygaard}}, \bibinfo {author} {\bibfnamefont {J.}~\bibnamefont {Dakin}}, \bibinfo {author} {\bibfnamefont {S.}~\bibnamefont {Hannestad}},\ and\ \bibinfo {author} {\bibfnamefont {T.}~\bibnamefont {Tram}},\ }\href@noop {} {\bibinfo {title} {Prospect: A profile likelihood code for frequentist cosmological parameter inference}} (\bibinfo {year} {2023}{\natexlab{b}}),\ \Eprint {https://arxiv.org/abs/2312.02972} {arXiv:2312.02972 [astro-ph.CO]} \BibitemShut {NoStop}%
\bibitem [{\citenamefont {Demianski}\ \emph {et~al.}(2017{\natexlab{a}})\citenamefont {Demianski}, \citenamefont {Piedipalumbo}, \citenamefont {Sawant},\ and\ \citenamefont {Amati}}]{Demianski:2016zxi}%
  \BibitemOpen
  \bibfield  {author} {\bibinfo {author} {\bibfnamefont {M.}~\bibnamefont {Demianski}}, \bibinfo {author} {\bibfnamefont {E.}~\bibnamefont {Piedipalumbo}}, \bibinfo {author} {\bibfnamefont {D.}~\bibnamefont {Sawant}},\ and\ \bibinfo {author} {\bibfnamefont {L.}~\bibnamefont {Amati}},\ }\bibfield  {title} {\bibinfo {title} {{Cosmology with gamma-ray bursts: I. The Hubble diagram through the calibrated $E_{\rm p,i}$ - $E_{\rm iso}$ correlation}},\ }\href {https://doi.org/10.1051/0004-6361/201628909} {\bibfield  {journal} {\bibinfo  {journal} {Astron. Astrophys.}\ }\textbf {\bibinfo {volume} {598}},\ \bibinfo {pages} {A112} (\bibinfo {year} {2017}{\natexlab{a}})},\ \Eprint {https://arxiv.org/abs/1610.00854} {arXiv:1610.00854 [astro-ph.CO]} \BibitemShut {NoStop}%
\bibitem [{\citenamefont {Demianski}\ \emph {et~al.}(2017{\natexlab{b}})\citenamefont {Demianski}, \citenamefont {Piedipalumbo}, \citenamefont {Sawant},\ and\ \citenamefont {Amati}}]{Demianski:2016dsa}%
  \BibitemOpen
  \bibfield  {author} {\bibinfo {author} {\bibfnamefont {M.}~\bibnamefont {Demianski}}, \bibinfo {author} {\bibfnamefont {E.}~\bibnamefont {Piedipalumbo}}, \bibinfo {author} {\bibfnamefont {D.}~\bibnamefont {Sawant}},\ and\ \bibinfo {author} {\bibfnamefont {L.}~\bibnamefont {Amati}},\ }\bibfield  {title} {\bibinfo {title} {{Cosmology with gamma-ray bursts: II Cosmography challenges and cosmological scenarios for the accelerated Universe}},\ }\href {https://doi.org/10.1051/0004-6361/201628911} {\bibfield  {journal} {\bibinfo  {journal} {Astron. Astrophys.}\ }\textbf {\bibinfo {volume} {598}},\ \bibinfo {pages} {A113} (\bibinfo {year} {2017}{\natexlab{b}})},\ \Eprint {https://arxiv.org/abs/1609.09631} {arXiv:1609.09631 [astro-ph.CO]} \BibitemShut {NoStop}%
\bibitem [{\citenamefont {Luongo}\ and\ \citenamefont {Muccino}(2021)}]{Luongo:2020hyk}%
  \BibitemOpen
  \bibfield  {author} {\bibinfo {author} {\bibfnamefont {O.}~\bibnamefont {Luongo}}\ and\ \bibinfo {author} {\bibfnamefont {M.}~\bibnamefont {Muccino}},\ }\bibfield  {title} {\bibinfo {title} {{Model-independent calibrations of gamma-ray bursts using machine learning}},\ }\href {https://doi.org/10.1093/mnras/stab795} {\bibfield  {journal} {\bibinfo  {journal} {Mon. Not. Roy. Astron. Soc.}\ }\textbf {\bibinfo {volume} {503}},\ \bibinfo {pages} {4581} (\bibinfo {year} {2021})},\ \Eprint {https://arxiv.org/abs/2011.13590} {arXiv:2011.13590 [astro-ph.CO]} \BibitemShut {NoStop}%
\bibitem [{\citenamefont {Alfano}\ \emph {et~al.}(2024)\citenamefont {Alfano}, \citenamefont {Capozziello}, \citenamefont {Luongo},\ and\ \citenamefont {Muccino}}]{Alfano:2024ukk}%
  \BibitemOpen
  \bibfield  {author} {\bibinfo {author} {\bibfnamefont {A.~C.}\ \bibnamefont {Alfano}}, \bibinfo {author} {\bibfnamefont {S.}~\bibnamefont {Capozziello}}, \bibinfo {author} {\bibfnamefont {O.}~\bibnamefont {Luongo}},\ and\ \bibinfo {author} {\bibfnamefont {M.}~\bibnamefont {Muccino}},\ }\href@noop {} {\bibinfo {title} {Cosmological transition epoch from gamma-ray burst correlations}} (\bibinfo {year} {2024}),\ \Eprint {https://arxiv.org/abs/2402.18967} {arXiv:2402.18967 [astro-ph.CO]} \BibitemShut {NoStop}%
\bibitem [{\citenamefont {Wang}\ and\ \citenamefont {Liang}(2024)}]{wang2024cosmological}%
  \BibitemOpen
  \bibfield  {author} {\bibinfo {author} {\bibfnamefont {H.}~\bibnamefont {Wang}}\ and\ \bibinfo {author} {\bibfnamefont {N.}~\bibnamefont {Liang}},\ }\href@noop {} {\bibinfo {title} {Cosmological constraints on long gamma-ray bursts from fermi observations}} (\bibinfo {year} {2024}),\ \Eprint {https://arxiv.org/abs/2405.14357} {arXiv:2405.14357 [astro-ph.CO]} \BibitemShut {NoStop}%
\bibitem [{\citenamefont {Rubin}\ \emph {et~al.}(2023)\citenamefont {Rubin}, \citenamefont {Aldering}, \citenamefont {Betoule}, \citenamefont {Fruchter}, \citenamefont {Huang}, \citenamefont {Kim}, \citenamefont {Lidman}, \citenamefont {Linder}, \citenamefont {Perlmutter}, \citenamefont {Ruiz-Lapuente},\ and\ \citenamefont {Suzuki}}]{Rubin:2023ovl}%
  \BibitemOpen
  \bibfield  {author} {\bibinfo {author} {\bibfnamefont {D.}~\bibnamefont {Rubin}}, \bibinfo {author} {\bibfnamefont {G.}~\bibnamefont {Aldering}}, \bibinfo {author} {\bibfnamefont {M.}~\bibnamefont {Betoule}}, \bibinfo {author} {\bibfnamefont {A.}~\bibnamefont {Fruchter}}, \bibinfo {author} {\bibfnamefont {X.}~\bibnamefont {Huang}}, \bibinfo {author} {\bibfnamefont {A.~G.}\ \bibnamefont {Kim}}, \bibinfo {author} {\bibfnamefont {C.}~\bibnamefont {Lidman}}, \bibinfo {author} {\bibfnamefont {E.}~\bibnamefont {Linder}}, \bibinfo {author} {\bibfnamefont {S.}~\bibnamefont {Perlmutter}}, \bibinfo {author} {\bibfnamefont {P.}~\bibnamefont {Ruiz-Lapuente}},\ and\ \bibinfo {author} {\bibfnamefont {N.}~\bibnamefont {Suzuki}},\ }\href@noop {} {\bibinfo {title} {Union through unity: Cosmology with 2,000 sne using a unified bayesian framework}} (\bibinfo {year} {2023}),\ \Eprint {https://arxiv.org/abs/2311.12098} {arXiv:2311.12098 [astro-ph.CO]} \BibitemShut {NoStop}%
\bibitem [{\citenamefont {Abbott}\ \emph {et~al.}(2024)\citenamefont {Abbott} \emph {et~al.}}]{DES:2024jxu}%
  \BibitemOpen
  \bibfield  {author} {\bibinfo {author} {\bibfnamefont {T.~M.~C.}\ \bibnamefont {Abbott}} \emph {et~al.} (\bibinfo {collaboration} {DES}),\ }\bibfield  {title} {\bibinfo {title} {{The Dark Energy Survey: Cosmology Results with \ensuremath{\sim}1500 New High-redshift Type Ia Supernovae Using the Full 5 yr Data Set}},\ }\href {https://doi.org/10.3847/2041-8213/ad6f9f} {\bibfield  {journal} {\bibinfo  {journal} {Astrophys. J. Lett.}\ }\textbf {\bibinfo {volume} {973}},\ \bibinfo {pages} {L14} (\bibinfo {year} {2024})},\ \Eprint {https://arxiv.org/abs/2401.02929} {arXiv:2401.02929 [astro-ph.CO]} \BibitemShut {NoStop}%
\bibitem [{\citenamefont {\'O~Colg\'ain}\ \emph {et~al.}(2025)\citenamefont {\'O~Colg\'ain}, \citenamefont {Pourojaghi},\ and\ \citenamefont {Sheikh-Jabbari}}]{Colgain:2024ksa}%
  \BibitemOpen
  \bibfield  {author} {\bibinfo {author} {\bibfnamefont {E.}~\bibnamefont {\'O~Colg\'ain}}, \bibinfo {author} {\bibfnamefont {S.}~\bibnamefont {Pourojaghi}},\ and\ \bibinfo {author} {\bibfnamefont {M.~M.}\ \bibnamefont {Sheikh-Jabbari}},\ }\bibfield  {title} {\bibinfo {title} {{Implications of DES 5YR SNe Dataset for $\Lambda $CDM}},\ }\href {https://doi.org/10.1140/epjc/s10052-025-13995-4} {\bibfield  {journal} {\bibinfo  {journal} {Eur. Phys. J. C}\ }\textbf {\bibinfo {volume} {85}},\ \bibinfo {pages} {286} (\bibinfo {year} {2025})},\ \Eprint {https://arxiv.org/abs/2406.06389} {arXiv:2406.06389 [astro-ph.CO]} \BibitemShut {NoStop}%
\bibitem [{\citenamefont {Adame}\ \emph {et~al.}(2024{\natexlab{a}})\citenamefont {Adame} \emph {et~al.}}]{DESI:2024hhd}%
  \BibitemOpen
  \bibfield  {author} {\bibinfo {author} {\bibfnamefont {A.~G.}\ \bibnamefont {Adame}} \emph {et~al.} (\bibinfo {collaboration} {DESI}),\ }\href@noop {} {\bibinfo {title} {{DESI 2024 VII: Cosmological Constraints from the Full-Shape Modeling of Clustering Measurements}}} (\bibinfo {year} {2024}{\natexlab{a}}),\ \Eprint {https://arxiv.org/abs/2411.12022} {arXiv:2411.12022 [astro-ph.CO]} \BibitemShut {NoStop}%
\bibitem [{\citenamefont {Adame}\ \emph {et~al.}(2024{\natexlab{b}})\citenamefont {Adame} \emph {et~al.}}]{DESI:2024jis}%
  \BibitemOpen
  \bibfield  {author} {\bibinfo {author} {\bibfnamefont {A.~G.}\ \bibnamefont {Adame}} \emph {et~al.},\ }\href@noop {} {\bibinfo {title} {{DESI 2024 V: Full-Shape Galaxy Clustering from Galaxies and Quasars}}} (\bibinfo {year} {2024}{\natexlab{b}}),\ \Eprint {https://arxiv.org/abs/2411.12021} {arXiv:2411.12021 [astro-ph.CO]} \BibitemShut {NoStop}%
\bibitem [{\citenamefont {Adil}\ \emph {et~al.}(2023{\natexlab{b}})\citenamefont {Adil}, \citenamefont {Akarsu}, \citenamefont {Malekjani}, \citenamefont {Colg\'ain}, \citenamefont {Pourojaghi}, \citenamefont {Sen},\ and\ \citenamefont {Sheikh-Jabbari}}]{Adil:2023jtu}%
  \BibitemOpen
  \bibfield  {author} {\bibinfo {author} {\bibfnamefont {S.~A.}\ \bibnamefont {Adil}}, \bibinfo {author} {\bibfnamefont {O.}~\bibnamefont {Akarsu}}, \bibinfo {author} {\bibfnamefont {M.}~\bibnamefont {Malekjani}}, \bibinfo {author} {\bibfnamefont {E.~O.}\ \bibnamefont {Colg\'ain}}, \bibinfo {author} {\bibfnamefont {S.}~\bibnamefont {Pourojaghi}}, \bibinfo {author} {\bibfnamefont {A.~A.}\ \bibnamefont {Sen}},\ and\ \bibinfo {author} {\bibfnamefont {M.~M.}\ \bibnamefont {Sheikh-Jabbari}},\ }\bibfield  {title} {\bibinfo {title} {{S8 increases with effective redshift in \ensuremath{\Lambda}CDM cosmology}},\ }\href {https://doi.org/10.1093/mnrasl/slad165} {\bibfield  {journal} {\bibinfo  {journal} {Mon. Not. Roy. Astron. Soc.}\ }\textbf {\bibinfo {volume} {528}},\ \bibinfo {pages} {L20} (\bibinfo {year} {2023}{\natexlab{b}})},\ \Eprint {https://arxiv.org/abs/2303.06928} {arXiv:2303.06928 [astro-ph.CO]} \BibitemShut {NoStop}%
\bibitem [{\citenamefont {Akarsu}\ \emph {et~al.}(2024{\natexlab{b}})\citenamefont {Akarsu}, \citenamefont {\'O~Colg\'ain}, \citenamefont {Sen},\ and\ \citenamefont {Sheikh-Jabbari}}]{Akarsu:2024hsu}%
  \BibitemOpen
  \bibfield  {author} {\bibinfo {author} {\bibfnamefont {O.}~\bibnamefont {Akarsu}}, \bibinfo {author} {\bibfnamefont {E.}~\bibnamefont {\'O~Colg\'ain}}, \bibinfo {author} {\bibfnamefont {A.~A.}\ \bibnamefont {Sen}},\ and\ \bibinfo {author} {\bibfnamefont {M.~M.}\ \bibnamefont {Sheikh-Jabbari}},\ }\href@noop {} {\bibinfo {title} {{Further support for $S_8$ increasing with effective redshift}}} (\bibinfo {year} {2024}{\natexlab{b}}),\ \Eprint {https://arxiv.org/abs/2410.23134} {arXiv:2410.23134 [astro-ph.CO]} \BibitemShut {NoStop}%
\bibitem [{\citenamefont {Foreman-Mackey}\ \emph {et~al.}(2013)\citenamefont {Foreman-Mackey}, \citenamefont {Hogg}, \citenamefont {Lang},\ and\ \citenamefont {Goodman}}]{Foreman-Mackey:2012any}%
  \BibitemOpen
  \bibfield  {author} {\bibinfo {author} {\bibfnamefont {D.}~\bibnamefont {Foreman-Mackey}}, \bibinfo {author} {\bibfnamefont {D.~W.}\ \bibnamefont {Hogg}}, \bibinfo {author} {\bibfnamefont {D.}~\bibnamefont {Lang}},\ and\ \bibinfo {author} {\bibfnamefont {J.}~\bibnamefont {Goodman}},\ }\bibfield  {title} {\bibinfo {title} {{emcee: The MCMC Hammer}},\ }\href {https://doi.org/10.1086/670067} {\bibfield  {journal} {\bibinfo  {journal} {Publ. Astron. Soc. Pac.}\ }\textbf {\bibinfo {volume} {125}},\ \bibinfo {pages} {306} (\bibinfo {year} {2013})},\ \Eprint {https://arxiv.org/abs/1202.3665} {arXiv:1202.3665 [astro-ph.IM]} \BibitemShut {NoStop}%
\bibitem [{\citenamefont {Lewis}(2019)}]{Lewis:2019xzd}%
  \BibitemOpen
  \bibfield  {author} {\bibinfo {author} {\bibfnamefont {A.}~\bibnamefont {Lewis}},\ }\href@noop {} {\bibinfo {title} {Getdist: a python package for analysing monte carlo samples}} (\bibinfo {year} {2019}),\ \Eprint {https://arxiv.org/abs/1910.13970} {arXiv:1910.13970 [astro-ph.IM]} \BibitemShut {NoStop}%
\bibitem [{\citenamefont {Alam}\ \emph {et~al.}(2021)\citenamefont {Alam} \emph {et~al.}}]{eBOSS:2020yzd}%
  \BibitemOpen
  \bibfield  {author} {\bibinfo {author} {\bibfnamefont {S.}~\bibnamefont {Alam}} \emph {et~al.} (\bibinfo {collaboration} {eBOSS}),\ }\bibfield  {title} {\bibinfo {title} {{Completed SDSS-IV extended Baryon Oscillation Spectroscopic Survey: Cosmological implications from two decades of spectroscopic surveys at the Apache Point Observatory}},\ }\href {https://doi.org/10.1103/PhysRevD.103.083533} {\bibfield  {journal} {\bibinfo  {journal} {Phys. Rev. D}\ }\textbf {\bibinfo {volume} {103}},\ \bibinfo {pages} {083533} (\bibinfo {year} {2021})},\ \Eprint {https://arxiv.org/abs/2007.08991} {arXiv:2007.08991 [astro-ph.CO]} \BibitemShut {NoStop}%
\bibitem [{\citenamefont {Craig}\ \emph {et~al.}(2024)\citenamefont {Craig}, \citenamefont {Green}, \citenamefont {Meyers},\ and\ \citenamefont {Rajendran}}]{craig2024}%
  \BibitemOpen
  \bibfield  {author} {\bibinfo {author} {\bibfnamefont {N.}~\bibnamefont {Craig}}, \bibinfo {author} {\bibfnamefont {D.}~\bibnamefont {Green}}, \bibinfo {author} {\bibfnamefont {J.}~\bibnamefont {Meyers}},\ and\ \bibinfo {author} {\bibfnamefont {S.}~\bibnamefont {Rajendran}},\ }\href {https://arxiv.org/abs/2405.00836} {\bibinfo {title} {No $\nu$s is good news}} (\bibinfo {year} {2024}),\ \Eprint {https://arxiv.org/abs/2405.00836} {arXiv:2405.00836 [astro-ph.CO]} \BibitemShut {NoStop}%
\bibitem [{\citenamefont {Naredo-Tuero}\ \emph {et~al.}(2024)\citenamefont {Naredo-Tuero}, \citenamefont {Escudero}, \citenamefont {Fernández-Martínez}, \citenamefont {Marcano},\ and\ \citenamefont {Poulin}}]{naredotuero2024}%
  \BibitemOpen
  \bibfield  {author} {\bibinfo {author} {\bibfnamefont {D.}~\bibnamefont {Naredo-Tuero}}, \bibinfo {author} {\bibfnamefont {M.}~\bibnamefont {Escudero}}, \bibinfo {author} {\bibfnamefont {E.}~\bibnamefont {Fernández-Martínez}}, \bibinfo {author} {\bibfnamefont {X.}~\bibnamefont {Marcano}},\ and\ \bibinfo {author} {\bibfnamefont {V.}~\bibnamefont {Poulin}},\ }\href {https://arxiv.org/abs/2407.13831} {\bibinfo {title} {Living at the edge: A critical look at the cosmological neutrino mass bound}} (\bibinfo {year} {2024}),\ \Eprint {https://arxiv.org/abs/2407.13831} {arXiv:2407.13831 [astro-ph.CO]} \BibitemShut {NoStop}%
\end{thebibliography}%

\end{document}